%% file: ges.tex
\documentclass{aa}  
\usepackage{graphicx}
\DeclareGraphicsExtensions{.pdf,.png,.jpg,.ps,.eps}
\usepackage{xcolor}
\usepackage{natbib}
\bibpunct{(}{)}{;}{a}{}{,}
\usepackage{ltablex} 
\usepackage{placeins} 
\usepackage{lscape}
\usepackage{color}
\usepackage{textgreek}
\usepackage{upgreek}
\usepackage[varg]{txfonts}
\usepackage{siunitx}

\newcommand{\GG}{\mbox{$G_3$}}
\newcommand{\GGc}{\mbox{$G^\prime_3$}}
\newcommand{\BP}{\mbox{$G_{\rm BP3}$}}
\newcommand{\RP}{\mbox{$G_{\rm RP3}$}}
\newcommand{\BPRP}{\mbox{$G_{\rm BP3}-G_{\rm RP3}$}}
\newcommand{\pic}{\mbox{$\varpi_{\rm c}$}}
\newcommand{\mci}[1]{\multicolumn{1}{c}{#1}}
\newcommand{\VO}[1]{Villafranca~O-{#1}}
\newcommand{\EBV}{\mbox{$E(4405-5495)$}}
\newcommand{\RV}{\mbox{$R_{5495}$}}
\newcommand{\lili}{\mbox{LiLiMaRlin}}
\newcommand{\kms}{km s$^{-1}$}
\def\Teff{\mbox{$T_{\rm eff}$}}

\usepackage[breaklinks=true, colorlinks=true, linkcolor=blue,citecolor=blue]{hyperref}

\begin{document} 

   \title{\textbf{\textit{Gaia}}-ESO Survey: massive stars in the Carina Nebula }

   \subtitle{I. A new census of OB stars}

\titlerunning{Massive stars in the Carina Nebula}
\authorrunning{Berlanas et al.}

   \author{S. R. Berlanas\inst{1,2}
          \and
         J. Maíz Apellániz\inst{3}
           \and
          A. Herrero\inst{4,5} 
          \and
           L. Mahy\inst{6} 
          \and          
          R. Blomme\inst{6}
          \and
          I. Negueruela\inst{1}            
          \and
          R. Dorda\inst{1,7} 
          \and \linebreak
           F. Comerón\inst{8}
           \and
          E. Gosset\inst{9} 
           \and
          M. Pantaleoni Gonz\'alez\inst{3,10}
           \and
          J. A. Molina Lera\inst{11}
           \and
          A. Sota\inst{12}
          \and
          T. Furst\inst{6,9}
           \and     \linebreak     
          E. J. Alfaro\inst{12}
          \and
          M. Bergemann\inst{13,14}
           \and
         G. Carraro\inst{15}
         \and 
          J. E. Drew\inst{16}
          \and
          L. Morbidelli\inst{17}
          \and
          J. S. Vink\inst{18}}

   \institute{Departamento de Física Aplicada, Universidad de Alicante, E-\num[detect-all]{03690}, San Vicente del Raspeig, Alicante, Spain\
    \and Astrophysics Group, Keele University, Keele ST5 5BG, Staffordshire,  UK\
    \and   Centro de Astrobiología (CAB), CSIC-INTA, Campus ESAC, E-\num[detect-all]{28692} Villanueva de la Cañada, Madrid, Spain\    
    \and Instituto de Astrofísica de Canarias, E-\num[detect-all]{38200} La Laguna, Tenerife, Spain\   		
    \and Departamento de Astrofísica, Universidad de La Laguna, E-\num[detect-all]{38205} La Laguna, Tenerife, Spain\  
    \and  Royal Observatory of Belgium, Ringlaan 3, 1180 Brussels, Belgium\  
    \and School of Architecture, Universidad Europea de Canarias, Tenerife, Spain \    
    \and ESO, Karl-Schwarzschild-Strasse 2, \num[detect-all]{85748} Garching bei M\"unchen, Germany\
    \and Space Sciences, Technologies and Astrophysics Research (STAR) Institute, Université de Liège, Allée du 6 Août, 19c, Bât B5c, 4000 Liège, Belgium\
    \and Departamento de Astrof{\'\i}sica y F{\'\i}sica de la Atm\'osfera, Universidad Complutense de Madrid. E-\num[detect-all]{28040} Madrid, Spain\
    \and Instituto de Astronom\'{\i}a y F\'{\i}sica del Espacio, UBA-CONICET. CC 67, Suc. 28, 1428 Buenos Aires, Argentina\
    \and Instituto de Astrof\'{\i}sica de Andaluc\'{\i}a (IAA), CSIC. Glorieta de la Astronom\'{\i}a s/n. E-\num[detect-all]{18008} Granada, Spain\
    \and Max Planck Institute for Astronomy, K\"onigstuhl 17, 69117, Heidelberg, Germany\
    \and Niels Bohr International Academy, Niels Bohr Institute, University of Copenhagen Blegdamsvej 17, DK-2100 Copenhagen, Denmark\
     \and Dipartimento di Fisica e Astronomia Galileo Galilei, Universitá di Padova, Vicolo Osservatorio 3, I-35122, Padova, Italy\
     \and  Department of Physics \& Astronomy, University College London, Gower Street, London, WC1E 6BT, UK\  
     \and INAF - Osservatorio Astrofisico di Arcetri, Largo E. Fermi 5, 50125 Florence, Italy\
    \and Armagh Observatory and Planetarium, College Hill, Armagh BT61 9DG, N. Ireland\
    }

   \date{Received month day, year; accepted month day, year}

 
  \abstract
   {The Carina Nebula is one of the major massive star-forming regions in the Galaxy. Its relatively nearby distance (2.35~kpc) makes it an ideal laboratory for the study of massive star formation, structure and evolution, both for individual stars and stellar systems. 
   Thanks to the high-quality spectra provided by \textit{Gaia}-ESO survey and the LiLiMaRlin library, as well as \textit{Gaia}~EDR3 astrometry, a detailed and homogeneous spectroscopic characterization of its massive stellar content can be carried out. }
   {Our main objective is to spectroscopically characterize all massive members of the Carina Nebula in the \textit{Gaia}-ESO survey footprint to provide an updated census of massive stars in the region and an updated estimate of the binary fraction of O stars.}
   {We perform accurate spectral classification by using an interactive code that compares spectra with spectral libraries of OB standards, as well as line-based classic methods. Membership is calculated using our own algorithm based on \textit{Gaia}~EDR3 astrometry. To check the correlation between the spectroscopic n-qualifier and the rotational velocity, we use the semi-automated tool for the line-broadening characterization of OB stars  which is based on a combined Fourier Transform and Goodness-of-fit methodology.  }
   {The \textit{Gaia}-ESO survey sample of massive OB stars in the Carina Nebula consists of 234 stars. The addition of brighter sources from the Galactic O-Star Spectroscopic Survey and additional sources from the literature allows us to create the most complete census of massive OB stars done so far in the region. It contains a total of 
   316 stars, being 18 of them in the background and four in the foreground. Of the 294 stellar systems in Car OB1, 74 are of O type, 214 are of non-supergiant B type and 6 are of WR or non-O supergiant (II to Ia) spectral class. We identify 20 spectroscopic binary systems with an O-star primary, of which 6 are reported for the first time, and another 18 with a B-star primary, of which 13 are new detections. The average observed double-lined binary fraction of O-type stars in the surveyed region is 0.35, which represents a lower limit. We find a good correlation between the spectroscopic n-qualifier and the projected rotational velocity of the stars. The fraction of candidate runaways among the stars with and without the n-qualifier is 4.4$\%$ and 2.4$\%$, respectively, although non resolved double-lined binaries can be contaminating the fast rotator sample.}
  {}


   \maketitle
%
\section{Introduction}

The \textit{Gaia}-ESO Large Public Spectroscopic Survey \citep[GES,][]{gilmore22, randich22} has obtained high quality spectra of $\sim$10$^{5}$ stars in our Galaxy using FLAMES at the Very Large Telescope (VLT) with its high-resolution UVES and its intermediate-resolution GIRAFFE spectrographs.
GES has systematically covered all the major components of the Milky Way,  providing an homogeneous and unique overview of the kinematics, chemical composition, formation history, and evolution of young, mature and ancient Galactic populations. Open clusters are useful tools for this aim, where it is possible to study stellar populations of different ages in different evolutionary stages \citep[see][]{bragaglia22}.

Numerous spectroscopic studies of massive stars have been carried out in Galactic young stellar clusters and OB associations, the most extensive to date being the Galactic O-Star Spectroscopic Survey \citep[GOSSS,][]{Maizetal11}. As some examples of such studies,  \citet{figer05} determined the upper mass limit of the Initial Mass Function (IMF) in the Arches Cluster, a result that was later challenged by studies in R136 \citep{crowther10}. Other examples are the determination of the chemical composition of stars in Orion \citep{ssimon10}; the membership, chemical and stellar parameter determination studies in Cygnus OB2 \citep{berlanas18a, berlanas18b, berlanas20}; the characterization of very massive obscured clusters in the Milky Way like Westerlund 1 \citep{clark05, negueruela10, negueruela22}; and the analysis of the multiplicity of massive stars in clusters \citep{debecker04,debecker06, mahy09,sana11, mahy13,  banyard22} and in the whole northern hemisphere \citep{Maizetal19b,Trigetal21,mahy22}. Outside the Milky Way, the most thorough analysis is that of the many papers\footnote{\url{https://www.roe.ac.uk/~cje/tarantula/f2-pubs.html}.} published by the VLT-FLAMES Tarantula Survey collaboration (VFTS, \citealt{Evanetal11a}).

The Carina Nebula complex consists of several stellar groups, some bound and some not, immersed in the Car~OB1 association (\citealt{Maizetal20b,Maizetal22a}, from now on Villafranca~I~and~II, respectively, and references therein). It represents a unique region to study Galactic massive stars with FLAMES since it contains a large number of O-type stars \citep{Walb72, Walb73a, Walb82a, LevaMala82, Morretal88, Sotaetal14, Maizetal16, Alexetal16, berlanas17, mohr17}. It is the most massive star-forming region within 3 kpc of the Sun. The distance to its most famous member, $\eta$~Car, was geometrically determined with excellent precision  to be 2.35$\pm$0.05~kpc by \citet{smith06b}. The recent {\it Gaia}~EDR3 \citep{Browetal21} analysis in Villafranca~I+II has not only confirmed that value but has also found that there are little distance variations between at least Trumpler~14, Trumpler~16~W, and Trumpler~16~E, three of the stellar groups in the complex. In a new installment of the series (Villafranca~III, Molina Lera et al. in preparation) the authors show that those distance variations are still small when including other stellar groups in Car~OB1. Even though the Carina Nebula harbors hundreds of massive stars, there is no systematic spectroscopic analysis of its early-type members. Thanks to the high-quality spectra provided by GES and astrometry by {\it Gaia}~EDR3, a detailed and homogeneous spectroscopic study of its massive stellar content can be carried out.
The analysis of the Carina massive stellar population will be highly relevant for problems like the initial mass function (IMF, \citealt{crowther10}), the chemical composition, rotation and internal mixing \citep{maynet00, ragudelo13, ssimon14a, herrero16, holgado22}, or the stellar multiplicity of massive stars \citep[see][]{langer12,sana12,Sotaetal14,demink14}. In particular, although \cite{sana11} and \cite{sana17} quote fractions of binary systems in excess of 0.5 for the O-type star population in the Milky Way, the former authors give a null fraction in a cluster like Trumpler 14, making clear the need for a systematic survey in the region. In addition, binarity may be the origin of fast rotating and runaway stars \citep[e.g.][]{demink13, demink14,mahy20a, holgado22} by ejecting stars that have gained mass and angular momentum from the binary system after the explosion of the primary as supernova. The Carina region, containing a large number of massive stars at a relatively nearby distance, is an ideal place to test the theories of massive star evolution.

As a first step this work focuses on the creation of the most complete to date census of massive stars and the identification of double-lined spectroscopic binaries (SB2) in Car OB1. It is organized as follows.
In Section~\ref{data} we describe how we have obtained our spectroscopy, compiled our spectral types, and used {\it Gaia} to determine the distances. In Section~\ref{census} we present our census of massive stars in the central part of the Carina Nebula. We discuss the results in Sect.~\ref{discussion}, where we explore the completeness of the census,  determine the binary fraction of OB stars and investigate the correlation between the n spectroscopic qualifier, the projected rotational velocity and the runaway status. Finally, we summarize the conclusions in Sect.~\ref{conclusion}.

\section{Data and methods}\label{data}

\subsection{GES strategy and spectroscopy}\label{ges}

\begin{table}
\caption{Wavelength range and resolving power of the GES spectra obtained with different gratings.}
\label{table_setups}
\centering
\begin{tabular}{lllllllll}
\hline\hline \\[-1.5ex]
& \multicolumn{1}{c}{Grating} & \multicolumn{1}{c}{Wavelength} & \multicolumn{1}{c}{Resolving} & \multicolumn{1}{c}{Data release} \\
&         & \multicolumn{1}{c}{range (\AA)} & \multicolumn{1}{c}{power} \\
\hline \\[-1.5ex]
\multicolumn{3}{l}{GIRAFFE} \\ \\[-1.5ex]
& HR03  & $4033-4201$ & \num{24800} &  iDR3-6\\
& HR04  & $4188-4392$ & \num{20350} &  iDR3-6\\
& HR05A & $4340-4587$ & \num{18470} &  iDR5-6 \\
& HR06  & $4538-4759$ & \num{20350} &  iDR3-6\\
& HR14A & $6308-6701$ & \num{17740} &  iDR3-6\\ \\[-1.5ex]
\multicolumn{3}{l}{UVES} \\  \\[-1.5ex]
& 520   & $4140-6210$ & \num{47000} &  iDR3-6\\ 
& 580   & $4760-6840$ & \num{47000} &  iDR5-6\\
\hline
\end{tabular}
\end{table}

GES spectroscopic data for hot stars were obtained using the FLAMES intermediate-resolution (R$\sim$\num{20000}) GIRAFFE and the high-resolution (R$\sim$\num{47000}) UVES spectrographs on the Very Large Telescope (VLT). See \cite{blomme22} for further details on the analysis of GES hot stars and Table~\ref{table_setups} for the wavelength range covered by each of the setups. In the rest of this subsection we present the aspects that are more relevant to the Carina Nebula GES data set. A previous GES paper on the Carina Nebula \citep{Damietal17a} used a different data set and concentrated on stars of lower mass than the ones analyzed here.

\begin{figure*}[t!]
\centering
\includegraphics[width=18cm, trim={0 0 0 0.8cm},clip]{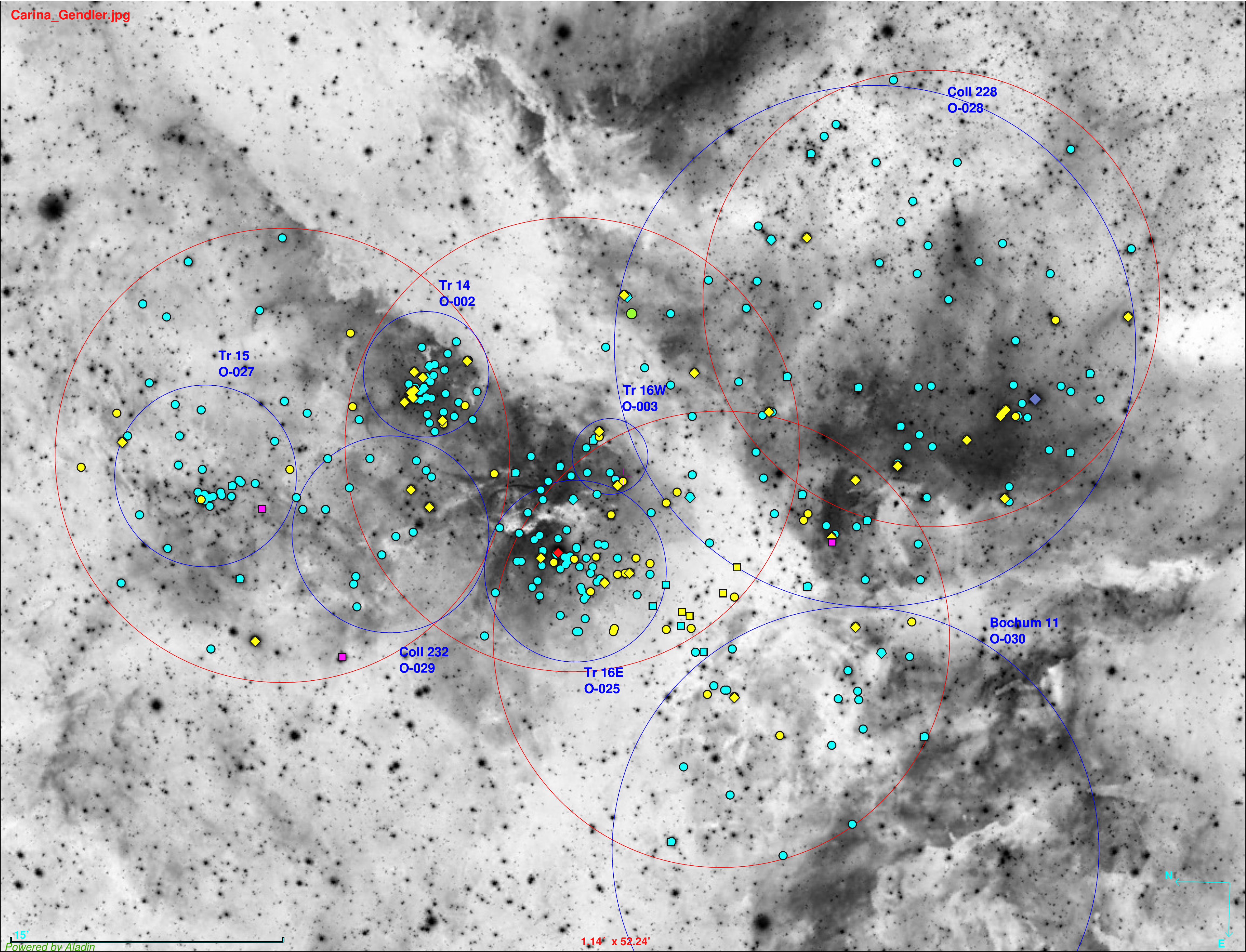}
\caption{Negative image of the Great Carina Nebula by Robert Gendler and Stephane Guisard showing the location of the whole census of massive stars in the GES surveyed area presented in this work. Yellow and cyan colors indicate O and B-type stars, respectively. Green, red, purple and pink colors have been used to represent the sdO, LBV,  WR and RSG stars, respectively. Small filled-circles refer to the GES sample while rhombuses and squares refer to stars from GOSSS/LiLiMarlin and other works \citep[][]{smith06a,Alexetal16,Preietal21} not present in GES, respectively. Red circles indicate the observing GES pointings while the blue ones indicate the Villafranca groups: O-002 (Trumpler~14), O-003 (Trumpler~16~W), O-025 (Trumpler~16~E), O-027 (Trumpler~15), O-028 (Collinder~228), O-029 (Collinder~232), and O-030 (Bochum~11). The V-shaped extinction lane that dominates the appearance of the nebula is clearly seen crossing the image from top to bottom.}
\label{carina}
\end{figure*}

The central part of the Carina Nebula can be divided into six stellar groups: Trumpler~14, Trumpler~15, Trumpler~16~W, Trumpler~16~E, Collinder 228, and Collinder~232 (\citealt{Walb95,smith06a}; Villafranca~I+II+III). Of those stellar groups, only Trumpler~14 and Trumpler~15 and possibly Trumpler~16~E appear to be real bound clusters, with the rest being parts of the association defined by (apparent or real) structures seen in the stellar distribution and nebulosity (e.g. the separation of Collinder~228 from the other groups likely originates in the prominent V-shaped dust lane that crosses the H\,{\sc ii}~region). Other stellar groups farther away from the central region but likely members of the Car~OB1 association include NGC~3293 \citep[see][]{morel22}, NGC~3324, Bochum~10, Bochum~11, Loden~153, IC~2581, Ruprecht~90, and ASCC~62. Given the large size of the nebula and the high stellar density in some regions (e.g. the core of Trumpler~14), four different GIRAFFE+UVES pointings were needed to cover a substantial fraction of the massive stars in the six central groups and part of those in Bochum~11 (see Fig.~\ref{carina}). The four pointings are centered at RA+$\delta$ J2000 coordinates (161.10,$-$59.430), (161.07,$-$59.695), (161.42,$-$59.835), and (160.79,$-$60.020), respectively, with the values expressed in degrees. The sample selection was done by compiling the available spectroscopic and photometric information at the time of the survey design (e.g. the Galactic O-Star Catalog, GOSSS, \citealt{Maizetal04b,Sotaetal08}) but, of course, no {\it Gaia} data existed back then. For that reason, we complemented our spectroscopy with GOSSS data and we have to evaluate the completeness of our sample (see below for both).

One important difference between this one and most of the other GES data sets is the existence of a significant nebulosity in the region. Furthermore, the nebular Balmer and He\,{\sc i} emission lines not only are strong but they are placed on top of important diagnostic stellar absorption lines. For that reason, we devised a specific strategy to eliminate or at least mitigate their influence when we prepared these observations. Each of the four pointings was divided into two subpointings (for a total of eight) with half of the fibers dedicated to stars and the other half to nebulosity. Each of the two subpointings within a given pointing have identical fiber configurations but the field center is displaced by 10\arcsec\ between the two of them. In that way, each star observed in a given subpointing has a nebular counterpart 10\arcsec\ away observed in the other subpointing. One of us (JMA) wrote an IDL code to manually review each star/nebulosity spectral pair and use the second one to subtract the nebular contribution from the stellar fiber. This strategy is the best possible one given the limitations of the observational setup, but it is not ideal, as in some cases nebular emission can change substantially in scales smaller than 10\arcsec. This is one of the advantages of long-slit spectroscopy (such as that obtained by GOSSS) over its fiber-fed alternative, as the former allows a sampling of nebular emission closer to the target and at two different locations with respect to the star. In practical terms, the issue is significative only for faint stars at H$\alpha$, as the nebular contribution for bright stars and other relevant lines is usually small.

We examined all of the spectra in our GES datasets to identify OB massive stars (B2 or earlier for dwarfs, B5 or earlier for giants, and all B subtypes for supergiants) and obtained a sample of 234 objects, 18 of which were observed with FLAMES-UVES  and 216 with FLAMES-GIRAFFE. The spectrograms are shown in Figs.~\ref{UVES_spectra}~and~\ref{GIRAFFE_spectra}, in the first case at the original \num{47000} spectral resolution and in the second case at the \num{2500} spectral resolution used for spectral classification.

\subsection{GOSSS spectroscopy}\label{GOSSS}

The GOSSS project was born a decade and a half ago with the idea of obtaining mid-low resolution ($R\sim 2500$) blue-violet spectroscopy of any optically-accessible Galactic object that had ever been classified as an O star to confirm its nature and to provide homogeneous spectral classifications for the whole sample. While doing that, GOSSS managed not only to discover a sizable number of new O-type stars but also to reject quite a number of them as being of B type (or, more egregiously, of even later types) and to obtain good-quality spectroscopy of several thousands of other early-type stars. In the first three major papers, \citet{Sotaetal11a} or GOSSS~I, \citet{Sotaetal14} or GOSSS~II, and \citet{Maizetal16} or GOSSS~III, GOSSS published spectra for 590 O-type stars and for a few later-type objects. Since that time, GOSSS has collected a large number of new spectra, some of them in the Carina Nebula. Of those, eight new GOSSS spectra for O-type stars will appear in the fourth major installment of the project (GOSSS~IV, Ma{\'\i}z Apell\'aniz et al. in preparation) but the spectral classifications are already listed here in Table~\ref{table_spclas}. In this paper we also present GOSSS spectra for one Wolf-Rayet and 17 early-B stars in Fig.~\ref{GOSSS_spectra} as a complement to the GES data.

\subsection{Spectral classifications}\label{classif}

We obtained the spectral classifications using the MGB tool \citep{Maizetal12,Maizetal15b}, which compares the observed spectra with a standard library of OB stars (in this case the GOSSS library, see GOSSS~III+IV). This interactive software allows us to vary the spectral subtype, luminosity class, line broadening, and spectral resolving power of the standard spectrum until we obtain the  best match. In addition, it also allows us to combine two standard spectra (with different velocities and flux fractions) to fit SB2 systems. The spectral classification was performed for the three types of spectroscopic data (UVES, GIRAFFE, and GOSSS) at the same spectral resolution of the GOSSS library, 2500. The spectral classifications are given in Table~\ref{table_spclas}.

There is one specific issue with GIRAFFE spectra and spectroscopic binaries that needs to be discussed. In general, each grating was observed at a different epoch and this generates a problem for spectroscopic binaries, as different lines of the same ion may be at different velocities. We have dealt with this issue on a case by case basis but in some we are only able to provide a poor-quality spectral classification. See subsection~\ref{individual} for some examples. 

\subsection{Spectral types from other sources and cataloguing}\label{gosc}

In addition to those from GES and GOSSS, in Table~\ref{table_spclas} we give spectral types from other further sources. The first one is \lili\ ({\bf Li}brary of {\bf Li}braries of {\bf Ma}ssive-star High-{\bf R}eso{\bf l}ut{\bf i}o{\bf n} Spectra, \citealt{Maizetal19a}) that is collecting multi-epoch high-resolution optical+NIR spectra of massive stars, with over \num{60000} epochs to date. For the case of the Carina Nebula, the library currently has FEROS, UVES, and HARPS spectra. \lili\ is especially useful for the analysis of SB2 and SB3 systems, where finding the right epoch is usually necessary to separate the different components in velocity. One of the \lili\ spectral types had appeared before in Villafranca~I but there are also nine whose spectra will appear in GOSSS~IV and another 16 whose spectra will appear in Villafranca~III. For the two latter papers, those spectral types are listed here for the first time.

Multi-epoch high-resolution spectroscopy such as that from \lili\ can be used to separate in velocity spectroscopic binaries. If one wants to spatially separate close visual binaries, then what is needed is the combination of high-spatial resolution with spectroscopy, either from the ground \citep{Maizetal18a,Maizetal21b} or from space \citep{MaizBarb20}. For the case of the Carina Nebula, we give in Table~\ref{table_spclas} the spatially resolved spectral types for HD~\num{93129}~Aa,Ab, one of the most massive systems in the region, obtained with STIS/HST \citep{Maizetal17a}. 

Another source is the already mentioned GOSC, which is a catalog that compiles information about massive stars (with an emphasis on O stars) from different sources. GOSC has a private and a (increasingly growing) public version, which will be heavily updated after GOSSS~IV is published. Here we have used the private version of GOSC to search for additional massive stars in the region of interest and provide spectral types. In particular, we have included the results from  \citet{smith06a}, a previous census of the massive stars in the Carina Nebula, from \citet{Alexetal16}, a spectroscopic survey of the region, and from \citet{Preietal21}, a near-infrared spectroscopy survey to identify obscured OB stars.

Besides the flow of information from GOSC to this paper mentioned in the previous paragraph, there will be a flow of information in the opposite direction, as the spectral types here will be included in GOSC. In addition to GOSC, these spectral types will be used to update the Alma Luminous Star Catalog (ALS, \citealt{Reed03}), a compilation of (originally)  photometric and spectroscopic information for Galactic OB stars. In \citet{Pantetal21}, the original ALS catalog was cross-matched with {\it Gaia}~DR2 to eliminate the many misidentifications and duplicates present and to provide astrometric information. In a soon-to-be submitted third paper, the cross-match will be revised with {\it Gaia}~DR3 information and the catalog will be expanded with new information, such as the one in this paper.

\subsection{Gaia~EDR3 data}\label{gaia}

We have searched the {\it Gaia}~EDR3 archive \citep{Browetal21} for the astrometric and photometric information of the sample in the paper. {\it Gaia}~EDR3 parallaxes, $\varpi$, have a zero point, $Z_{\rm EDR3}$ \citep{Lindetal21b}, that needs to be applied to yield corrected parallaxes, \pic. Furthermore, the internal parallax uncertainties are underestimated and have to be converted into external (or true) uncertainties. Here we follow the procedure outlined in \citet{Maizetal21c} and \cite{Maiz22} to list the corrected parallaxes with their external uncertainties in Table~\ref{table_main}. We also list there the membership of each star to a foreground or background population according to its parallax (see Appendix~\ref{ap_list} for details).

As the inverse of the parallax is a biased estimator of the distance \citep{LutzKelk73}, one has to do a proper estimation of the distance involving a prior. The prior depends on the analyzed population  itself: notoriously, OB stars do not follow the same spatial distribution in the Milky Way compared to its older populations. 
Here we use the thin disk model and prior of \citet{Maiz01a,Maiz05c} updated with the parameters of \citet{Maizetal08a} to calculate the distances and uncertainties of individual stars. See \citet{Pantetal21} for a comparison among different distance estimates to OB stars using {\it Gaia} parallaxes.  

{\it Gaia}~EDR3 provides photometry in three bands, \GG, \BP, and \RP, with the two last bands being actually the result of integrating spectrophotometry in the wavelength direction. The analysis of previous {\it Gaia} data releases \citep{Maiz17a,MaizWeil18} revealed that the sensitivity curves of the {\it Gaia} instrument change with time, leading to slightly different intrinsic photometric values between data releases (each being an average over different time frames). Furthermore, in some cases the processing introduces small trends and artifacts in the published magnitudes that require corrections. In a paper that will be submitted soon, a team that includes some of us have computed such an analysis for \GG, leading to a corrected value \GGc. In Table~\ref{table_main} we list the \GGc\ and \BPRP\ values for our sample.

\section{Census}\label{census}

Here we present the new census of massive stars in the central region of the Carina Nebula and discuss some individual stars of interest, especially if they have received little or no attention before. The census itself is presented in two tables in the Appendix already introduced in the previous section. Table~\ref{table_main} lists the star identifications and coordinates, {\it Gaia}~EDR3 corrected photometry and parallaxes, and the group identification (see previous section and Fig.~\ref{carina}). Table~\ref{table_spclas} gives the spectral classifications from different sources. Disagreements between spectral classifications are sometimes attributable to the nature of spectroscopic binaries caught in different orbital phases but in other cases they are due to differences in data quality (e.g. wavelength range, S/N, uncorrected artifacts) or classification criteria (e.g. choice of lines for classification, standards used, consideration of line broadening). When in doubt, one should consult the published spectrograms, not the spectral types themselves. That is the reason for publishing long appendices with figures such as the one here.

\subsection{Overall properties}

The resulting census of stars presented in this work contains 316 massive stars\footnote{Note that this number does not distinguish between single and binary or multiple stars, so hereinafter we refer to stellar systems when both types are included in the statistics.}. Note that, by definition and as stated before, massive OB stars include all O-types and those B2-types or earlier for dwarfs, B5-types or earlier for giants, and all B subtypes for supergiants (I or II luminosity classes). Red supergiants (RSGs), Wolf-Rayet (WR) stars, and some B subtypes close to the OB-star limit (e.g. B2.5~V) are also included in the census for completeness. We have separated stars with distances compatible with Car~OB1 from those in the foreground and background, finding four systems in the foreground (one RSG, two B dwarfs and one sdO) and 18 in the background (two O stars, five B supergiants and 11 B non-supergiants). These systems are listed in Table~\ref{non-members}.

Of the 294 stellar systems in our census in Car~OB1, 74 are of O type, 214 are of non-supergiant B type and 6 are of WR or non-O supergiant (II to Ia) spectral class (they are listed in Table~\ref{six-members}). Note that other WR stars in Car~OB1 fall outside the surveyed area (WR~22, WR~23 and WR~27). Compared to the previous census of the massive stars in the Carina Nebula by \cite{smith06a} we have significantly increased the content of known OB stars in the region. Considering only the area surveyed in this work, the number of 105 OB stellar systems with spectral types as late as B2 reported by \cite{smith06a} has been increased by a factor of 2.8.  

There are three RSGs in the field of view: HD~\num{93420}, HD~\num{93281}, and HDE~\num{303310} (=~RT~Car), all of them included in the study of \citet{Humpetal72}. They are the second, fifth, and sixth \GGc\ brightest sources, as the bolometric correction in that photometric band is significantly lower for RSGs than for O-type and WR stars. $\eta$~Car, of course, is in a different luminosity category and is almost two magnitudes brighter in \GGc\ than the brightest RSG. HD~\num{93420} is three sigmas\footnote{The external parallax uncertainty for such a bright star is much larger than the internal uncertainty \citep{Maiz22}, so the distance in sigmas would be also larger if we were to use the second one.} closer to us in parallax than 0.44~mas, the limit we are using to include a star in Car~OB1, and that places it in the foreground (but closer to Car~OB1 than to us). The other two stars have parallaxes compatible with being in Car~OB1, HD~\num{93281} in \VO{028} (Collinder~228) and HDE~\num{303310} in \VO{029} (Trumpler~15). We list in this paper their new spectral classifications from Villafranca~III, derived from recently obtained FEROS spectra. The classification for HD~\num{93420} is identical to that of \citet{Humpetal72} but the other two are of slightly later type, with HDE~\num{303310} at M3~Iab and HD~\num{93281} at M1.5~Iab. We see no sign of the alleged B-type companion for HD~\num{93281} \citep[see][]{Humpetal72} other than the strong H$\alpha$ emission.

The three RSGs are not the only sign of the existence of previous generations of massive-star formation in Car~OB1 and its immediate foreground. We also find in our sample evolved B stars such as HDE~\num{305535}, HDE~\num{305452}, CPD~$-$58~2605, CPD~$-$59~2469, and CPD~$-$59~2504. We find only one B-supergiant of luminosity class I, HDE~\num{305530}, at the distance of Car~OB1 in the footprint of this paper (the two stars by \cite{Damietal17a} classified as B I, 2MASS J10440384$-$5934344 and 2MASS J10452875$-$5930037, are classified here as B2V and B1.5Vp, respectively), but a number of B and later-type supergiants are observed in its vicinity (Villafranca~III). All of this establishes the existence not only of those older massive stars but also of the supernova explosions associated to those star-formation episodes. It has been known for a long time that the gas in the foreground of some of the OB stars in Carina shows the most complex kinematics in any Galactic sightline \citep{WalbHess75,Walb82d,Walbetal02c}, with up to 26 individual components and a range of velocities between $-$388 km/s and $+$127~km/s. Those components must have been produced by supernova explosions whose progenitors were evolved massive stars. The remaining three RSGs and the evolved B stars must be just the tip of the iceberg of the previous massive populations. Those complex kinematics are the main reason why the interstellar lines present in the spectra of the Carina OB stars are so strong \citep{Penaetal11,Penaetal13}, as the spread in velocity yields a more advantageous curve of growth. Due to the additional  Routly-Spitzer effect \citep{RoutSpit51,RoutSpit52} the Ca\,{\sc ii}~H+K lines are especially strong for these stars, making them deviate strongly from the relationship between extinction and their EW derived from other sightlines.

\subsection{Individual stars}\label{individual}

The Carina Nebula field has a large number of interesting stars, starting with $\eta$~Car, that have been analyzed in the past \citep[see, e.g.,][]{damineli00, damineli08, iping05}. Our goal in this subsection is not to discuss such objects per se but to present new interesting objects that have received little or no attention in the past or new aspects of old objects that are mentioned for the first time.

\paragraph{QZ Car~Aa,Ac.} This complex system \citep{SanBetal17,rainot20} is the brightest of O-type in the Carina Nebula. \citet{Mayeetal01} identified it as an SB1E+SB1 system and measured the two periods as \num{5.991}~d (eclipsing) and \num{20.73596}~d (non-eclipsing). In GOSSS~II the system was classified as O9.7~Ibn with no resolved components\footnote{Some papers quote spectral types for the four components but these are estimates: to our knowledge both spectroscopic binaries are still SB1 and no resolved (spatially or kinematically) spectral types have been determined.}. In GOSSS~IV the system is determined to be O9.7~Ib~+~O9~II: using \lili\ but the authors note that the secondary luminosity class is poorly determined, possibly as the result of contamination by one of the additional stars. The high luminosity of the two primaries coupled with the smaller contribution of the secondaries explains why this system seats at the top of the optical-luminosity food chain of the O-type stars in the Carina Nebula. The {\it Gaia}~EDR3 parallax uncertainty is quite large. 

\paragraph{HD~\num{93129}~Aa,Ab.} This system was spatially resolved by \citet{Maizetal17a} using HST/STIS and determined to be composed of two O2~If* stars, with one of them having a companion in a tight orbit, likely a late-O star. The orbit is highly eccentric and passed though periastron in $2018.70^{+0.22}_{-0.12}$ \citep{delPetal20}. Given that the periastron took place at a 3-D separation of just 18.6$\pm$1.0~AU (when the system was first spatially resolved in 1996 it was $\sim$375~AU), an order of magnitude (or even less) smaller than the expected semi-major axis of the inner orbit, and the high eccentricity, it is possible that the system has transitioned from an elliptic orbit to a hyperbolic trajectory and a possible ejection from \VO{002} (Trumpler~14). If that had happened, this could be another example of an orphan cluster where the most (in this case, two) massive stars of a cluster are ejected through a dynamical interaction \citep{Maizetal22b}. Further observations are needed, especially with HST/STIS later in this decade (if it is still operational) when Aa~and~Ab are expected to reach plane-of-the-sky separations of $\sim$40~mas.

\paragraph{HD~\num{93403}.} \citet{Rauwetal00} classified this SB2 system as O5.5~I~+~O7~V. In GOSSS~II the two components could not be resolved and it received a classification of O5.5~III(fc)~var. In GOSSS~IV it is now kinematically resolved and classified as O5~Ifc~+~O7.5~V using either GOSSS or \lili\ data, that is, the primary is slightly earlier and the secondary slightly later compared to \citet{Rauwetal00}. The {\it Gaia}~EDR3 parallax uncertainty is quite large. 

\paragraph{HDE~\num{305520}.} \citet{Alexetal16} classified this system as B1~Ia. In Villafranca~III we use \lili\ data to reclassify it as B0.7~Iab. This \VO{028} object is the only B~supergiant at the distance of Car~OB1 in our sample. 

\paragraph{V572 Car.} \citet{Rauwetal01a} classified this SB3 system in \VO{025} (Trumpler~16~E) as composed by an O7~V~+~O9.5~V inner eclipsing binary and an outer B0.2~IV star. In GOSSS~III only two components were seen and received an O7.5~V(n)~+~B0~V(n) classification. With the new data we now detect the system as an SB3: O6.5~Vz~+~B0~V~+~B0.2~V in \lili\ data in GOSSS~IV and O6.5~Vz~+~B0~V~+~B0.5:~V in UVES. As it happened with HD~\num{93403}, the primary is slightly earlier and the secondary slightly later compared to the original classification. The outer star has been detected in NIR Long-Baseline Interferometry and is currently further monitored \citep[see][]{gosset14}.

\paragraph{CPD~$-$59~2554.} This system was classified as O9.5~IV in GOSSS~II. Using \lili\ in GOSSS~IV and UVES here it is now found to be an SB2. In both cases the spectral classification is O9.2~V~+~B1:~V.

\paragraph{HD~\num{93342}.} This object was considered as a \VO{027} (Trumpler~15) member by \citet{smith06a}, where it received a classification as O9~III. \citet{Alexetal16}, on the other hand, classified it as B1~Ia and in Villafranca~III we reclassify it as B1.5~Ib using \lili\ data, confirming it is a B-type supergiant and not an O star. Its {\it Gaia}~EDR3 distance is $3.58^{+0.41}_{-0.33}$~kpc, placing it beyond Car~OB1, something that is consistent with its red color (it is the brightest OB star in our sample with $\BPRP > 1.0$). 

\paragraph{HD~\num{93056}.} \citet{Alexetal16} classified this system as O9~V~+~B2~V. In Villafranca~III we use \lili\ data to reclassify it as B1:~V:n. Furthermore, in the UVES data no He\,{\sc ii} is detected (either 4542, 4686, or 5412), as it should in an SB2 system composed of a late-O and an early-B stars even if caught at a disfavorable phase. 

\paragraph{HD~\num{93501}.} \citet{Alexetal16} classified this system as B0~V. In Villafranca~III we use \lili\ data to reclassify it as B1.5:~III:(n)e. Its {\it Gaia}~EDR3 distance is $1.87^{+0.12}_{-0.11}$~kpc, placing it in the foreground.

\paragraph{CPD~$-$59~2592.} \citet{Alexetal16} classified this object as B1~Ib. In Villafranca~III we use \lili\ data to reclassify it as B2.5~Ia. Its {\it Gaia}~EDR3 distance is $4.71^{+0.69}_{-0.53}$~kpc, placing it beyond Car~OB1, something that is consistent with its red color (it is the brightest OB star in our sample with $\BPRP > 1.2$). 

\paragraph{HDE~\num{305439}~A,B.} With GIRAFFE data we classify the A component as B0~Ia. Its {\it Gaia}~EDR3 distance is $4.48^{+0.54}_{-0.44}$~kpc, placing it beyond Car~OB1, something that is consistent with its moderately red color (it is the fourth brightest OB star in our sample with $\BPRP > 0.7$). In Villafranca~III we use \lili\ data to classify the B component, located 3\farcs7 away, as B0.7~Ib. Its parallax is consistent with being at the same distance, making the system a likely pair of B supergiants.

\paragraph{HDE~\num{305535}.} This object was classified as B2.5~V by \citet{Alexetal16}. Here we derive a classification of B4~III(n) from UVES data, which leads to an absolute magnitude more consistent with its spectral type and low extinction.

\paragraph{HD~\num{93343}.} \citet{Rauwetal09} classified this SB2 system as O7-8.5~+~O8. In GOSSS~III the two components could not be resolved and it received a classification of O8~V. In GOSSS~IV it is now kinematically resolved and classified as O7.5~Vz~+~O7.5:~V(n).

\paragraph{CPD~$-$59~2636~A,B.} This system is a visual binary with a 0\farcs3 separation and $\Delta m$~=~0.6~mag in which both components are spectroscopic binaries \citep{Albaetal02}: A (A+B in \citealt{Albaetal02}) is an SB2 with a 3.6284~d period and B (C in \citealt{Albaetal02}) is an SB1 with a 5.034~d period. Those authors gave a spectral classification of O7~V~+~O8~V to A and of O9~V to B. In GOSSS~II, the authors were only able to give two spectral types as O8~V~+~O8~V but with GES we are able to see the three components in an UVES single epoch and derive spectral types of O7.5~V~+~O8~V~+~O8~V. Further epochs are needed to solve the small discrepancies with the \citet{Albaetal02} classification. {\it Gaia}~EDR3 does not provide a parallax for CPD~$-$59~2636~A,B, which is common for a visual binary of this separation and magnitude difference, but it is a likely member of \VO{025} (Trumpler~16~E).

\paragraph{HDE~\num{305534}.} \citet{Alexetal16} identified this system as a spectroscopic binary and classified it as B0~V~+~B0~V. In Villafranca~III we use \lili\ data to confirm it is an SB2 and reclassify it as B0~V~+~B1:~V. 

\paragraph{HDE~\num{305543}.} \citet{Gagnetal11} identified this system as a spectroscopic binary and classified it as B0~V~+~B0~V. In Villafranca~III we use \lili\ data to confirm it is an SB2 and reclassify it as B0.2~V(n)~+~B1:~V(n).

\paragraph{HDE~\num{303312}.} We detect this object as a SB2 for the first time with GIRAFFE and assign it spectral types O9.5~III~+~B0.5:~V. In GOSSS~II, where it was likely caught at a disfavorable phase, it had received the intermediate type O9.7~IV. It was already known to be an eclipsing binary with a 9.4109~d period  \citep{Oter06}.

\paragraph{CPD~$-$58~2649~A.} We classify this system as an SB2 with spectral types O9.7~III:~+B0:~V in GOSSS~IV with GOSSS data. With GIRAFFE data, we can only give a poorer classification of O9.5:~+~B0: due to the different phases in each grating, but in any case both components are clearly later than the O7~V~+~O8~V of \citet{Alexetal16}. There is a visual companion detected in {\it Gaia}~EDR3 with a separation of 1\farcs2 that, though relatively weak, may contaminate the GOSSS and GES spectra.

\paragraph{ALS~\num{15860}.} This object was considered as a \VO{027} (Trumpler~15) member by \citet{smith06a}, where it received a classification as O9~I-II. Using either the GOSSS or GES data here we classify it as B1~Iab. Its {\it Gaia}~EDR3 distance is $3.31^{+0.23}_{-0.20}$~kpc, placing it beyond Car~OB1, consistent with its red color (it is the brightest OB star in our sample with $\BPRP > 1.8$). 

\paragraph{CPD~$-$58~2634.} We classify this object as B1.5~V using GIRAFFE data. Its {\it Gaia}~EDR3 distance is $1.869^{+0.077}_{-0.071}$~kpc, placing it in the foreground. Its parallax is consistent with being at the same distance as HD~\num{93501}.

\paragraph{CPD~$-$59~2591.} This system in \VO{028} (Collinder~228) was classified as an SB2 with spectral types O8~Vz~+~B0.5:~V both in GOSSS~III using GOSSS spectroscopy and in GOSSS~IV using \lili\ data. Here it is seen as SB2 but the classification is of poorer quality due to the multiple epochs of the GIRAFFE data.

\paragraph{CPD~$-$59~2535.} This system was classified as B2~V by \citet{Alexetal16} but there is no GES, GOSSS, or \lili\ data. Its {\it Gaia}~EDR3 distance is $3.16^{+0.23}_{-0.20}$~kpc, placing it in the background.

\paragraph{2MASS~J10424476$-$6005020.} A GIRAFFE spectrum is used to identify this system as an SB2 with the spectral types B0.2~V~+~B0.2~V. In a GOSSS spectrum no double lines are seen, likely due to an unfavorable epoch, and the resulting spectral classification is a poorer B0:~IV. Its {\it Gaia}~EDR3 distance is $4.02^{+0.39}_{-0.33}$~kpc, placing it in the background. Its red color is consistent with the measured distance. 

\paragraph{2MASS~J10460477$-$5949217.} This object in \VO{030} (Bochum~11) is classified as an O star for the first time with GIRAFFE data. It has a moderately high extinction and a spectral classification of O9.7~V(n). The spectrum could be a composite of a late-O and an early-B stars but more epochs are needed to test that hypothesis.

\paragraph{2MASS~J10444803$-$5954297.} This object is an Oe star with strong Balmer emission and no previous classification as O type. As usual with Oe stars, spectral classification is of poor quality and we can only give O7:~Ve using GIRAFFE and O8:~Ve in GOSSS~IV using GOSSS. The {\it Gaia}~EDR3 parallax indicates a background object at a distance of $4.71^{+0.79}_{-0.59}$~kpc.

\paragraph{CPD~$-$59~2618.} \citet{Alexetal16} classified this system as B2~V. A GIRAFFE spectrum indicates it is of an earlier subtype and with an anomalous composition, yielding a classification of B1:~V(n)p~He~rich. In Villafranca~III we use \lili\ data to confirm the helium enrichment and to further discover it is an SB2, classifying it as B0.7:~V(n)p~He~rich~+~B1:~V.

\paragraph{ALS~\num{15225}.} We identify this star in \VO{028} (Collinder~228) as a He-rich B star for the first time using both GIRAFFE and GOSSS spectroscopy here.

\paragraph{V662 Car.} \citet{Niemetal06a} identified this system as an  O5.5~Vz~+~O9.5~V SB2 and an eclipsing binary with a period of 1.41355~d. In GOSSS~III it was classified as O5~V(n)z + B0:~V. The GIRAFFE data shows there are two separate components in He\,{\sc ii} and both narrow, with a third component clearly separated in He\,{\sc i}, making it an SB3. However, given the multiple epochs in the GIRAFFE data, we can only classify it as O+O+B. Both O stars have narrow lines, so the GOSSS~III (n) suffix is likely due to combination of two O stars. The second O star is likely a third light not participating in the orbit and appears to be of mid-O subtype. The first O star has He\,{\sc ii}~4542 $>$ He\,{\sc ii}~4471 and should be close to O5. This system needs further high-resolution spectroscopy covering the whole classification range in a single epoch at large velocity separation.

\paragraph{ALS~\num{15203}~A,B.} In Villafranca~II this \VO{002} (Trumpler~14) object was identified as an SB3 with a classification of B0~V~+~B~+~B. In GIRAFFE we see some double lines but He\,{\sc ii} lines are very weak, invalidating the \citet{VijaDril93} classification as O7~V. As {\it Gaia}~EDR3 detects two sources of similar magnitude separated by 1\farcs2 (confirmed by HST imaging), we reanalyzed the GOSSS long slit with the best seeing and proper orientation and we were able to spatially separate the two visual components. ALS~\num{15203}~A is an SB2 with a spectral classification of B0.5~V~+~B1:~V, which corresponds to those of the secondary and tertiary in Villafranca~II, and ALS~\num{15203}~B has a classification of B0~V, which corresponds to the primary in Villafranca~II. There is a hint of emission at the bottom of H$\beta$ for ALS~\num{15203}~B but it is unclear whether it is of stellar origin or is due to an incorrect nebular subtraction. In any case, this SB3 system is now a SB2+Cas following the SBS nomenclature of \citet{Maizetal19b}

\paragraph{2MASS~J10435902$-$5933196.} We identify this star in \VO{002} (Trumpler~14) as a He-rich B star for the first time using GIRAFFE data.

\paragraph{2MASS~J10441829$-$5942296.} We identify this star in \VO{003} (Trumpler~16~W) as a He-rich B star for the first time using GIRAFFE data.

\paragraph{2MASS~J10453807$-$5944095.} This object in \VO{025} (Trumpler~16~E) is identified as an O star for the first time here using GIRAFFE data. It has an O8~Vz spectral classification and a high extinction.

\paragraph{2MASS~J10440744$-$5916399.} This is a background object caught as an SB2 but with an uncertain GIRAFFE classification of O9.7: + B0.5:. If confirmed, it would be another new O star. The derived {\it Gaia}~EDR3 distance is $d = 4.45^{+0.56}_{-0.45}$~kpc.

\paragraph{[ESK2003]~148 = [S87b]~IRS~41.} This system was first identified as an O-type candidate at the distance of Car~OB1 by \citet{Damietal17a}. The identification was based on a photometric analysis with CHORIZOS \citep{Maiz04c} and resulted in values of $\Teff = 42.4\pm4.4$~kK, 
$\EBV = 1.351\pm0.020$~mag, and $\RV = 4.92\pm 0.09$, which indicates an O star with both large color excess and anomalous extinction. It is classified as O9.2~V(n) both in GOSSS~IV and here using GIRAFFE, so the \Teff\ appears to be slightly lower than the value measured with CHORIZOS. It is highly reddened but with a position and parallax consistent with being in \VO{025} (Trumpler~16~E), likely slightly behind the rest of the cluster and immersed in the molecular cloud. 

\paragraph{2MASS~J10471498$-$5953374.} A GIRAFFE spectrum yields the spectral classification B6:~IIIe, with a double-peaked emission line in H$\alpha$ but no emission in H$\gamma$ (no other Balmer lines are covered by the GIRAFFE data). The derived {\it Gaia}~EDR3 distance is $d = 3.14^{+0.29}_{-0.25}$~kpc, placing it in the background.

\paragraph{2MASS~J10443089$-$5914461.} This object is classified as a highly extinguished supergiant with spectral type O7.5~II(f) using either GOSSS data in GOSSS~IV or GIRAFFE data here. \citet{Alexetal16} classified it as O8~V but it is clearly not a dwarf. It has a large parallax uncertainty: it could not be discarded as being in Car~OB1 but it is more likely a background object. 

\paragraph{2MASS~J10453185$-$6000293.} This highly extinguished O star in \VO{030} (Bochum~11) is identified as an O star for the first time. It receives a spectral classification of O7.5~V in GOSSS~IV using GOSSS data and a slightly later one of O8.5~V here using GIRAFFE data. The latter is rather noisy but some lines show signs of asymmetry, indicating a possible spectroscopic binary.

\paragraph{[ARV2008]~217 = [S87b]~IRS~42.} This object is one of the most interesting discoveries in this paper. Using GOSSS data in GOSSS~IV the authors give it an O3:~III: spectral classification and using GIRAFFE here we arrive at the same classification but with an (n) suffix. In both cases the O3: classification is based on the apparent absence of He\,{\sc i}~4471 but the two spectra are too noisy to provide a more accurate classification based on the N lines. Therefore, it is a new member of the limited family of Galactic O stars with spectral types earlier than O4. It was first identified as an O-type candidate at the distance of Car~OB1 by \citet{Damietal17a}. The CHORIZOS analysis there gives $\Teff = 42.0\pm4.2$~kK, $\EBV = 1.932\pm0.021$~mag, and $\RV = 4.60\pm 0.06$, that is, an early-type O star with both large color excess and anomalous extinction. That analysis is in good agreement with the spectral classification. That object position and parallax are consistent with [ARV2008]~217 being in \VO{025} (Trumpler~16~E), making it the earliest O-type star there.

\paragraph{2MASS~J10431945$-$5944488.} The {\it Gaia}~EDR3 parallax for this object yields $d = 794^{+28}_{-26}$~pc, clearly making it a foreground (and very blue) object. The existence of broad H$\gamma$ and He\,{\sc ii} lines indicates that the spectrum is dominated by an sdO. However, He\,{\sc ii}~4542 and He\,{\sc ii}~4686, observed at different epochs, have different velocities and some lines appear to originate in a later-type star. Therefore, the system is a spectroscopic binary.


\begin{figure*}
\centerline{\includegraphics[width=14cm]{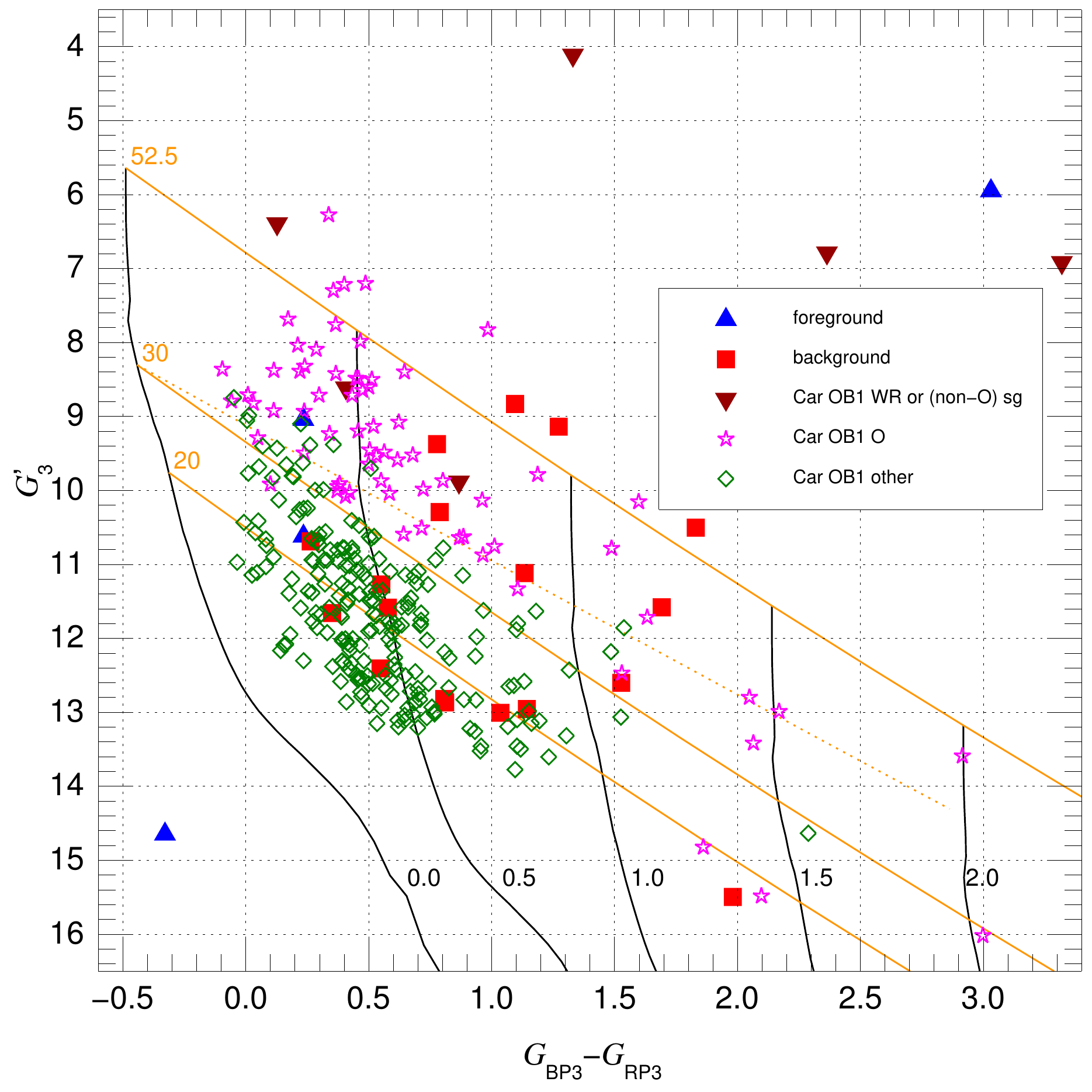}}
\caption{First panel, see next page for the second one: {\it Gaia}~EDR3 CMD for the stars with spectral types in this paper. Different symbols and colors are used to represent stars with parallaxes compatible with being (or otherwise assumed to be) in the foreground (4), in the background (18), or in Car~OB1 (294). Of the Car~OB1 stars, 6 are of Wolf-Rayet or non-O supergiant (II to Ia) spectral class, 74 are of O type, and 214 are of non-supergiant B type. Four of the \citet{Preietal21} stars are outside the frame towards the lower right due to their high extinction. Black lines show the average main sequence at a distance of 2.35~kpc with no extinction and with values of \EBV\ of 0.5, 1.0, 1.5, and 2.0 (labelled) using the extinction law of \citet{Maizetal14a} with a value of \RV\ of 4.5, which is typical of the region but with a large dispersion  \citep{MaizBarb18}. Solid orange lines show the \RV~=~4.5 extinction tracks for average MS stars of \Teff\ of 52.5~kK, 30~kK, and 20~kK (labelled), respectively. The dotted orange line shows the \RV~=~3.0 extinction track for \Teff~=~30~kK.}
\label{Gaia_CMD}
\end{figure*}

\addtocounter{figure}{-1}

\begin{figure*}
\centerline{\includegraphics[width=14cm]{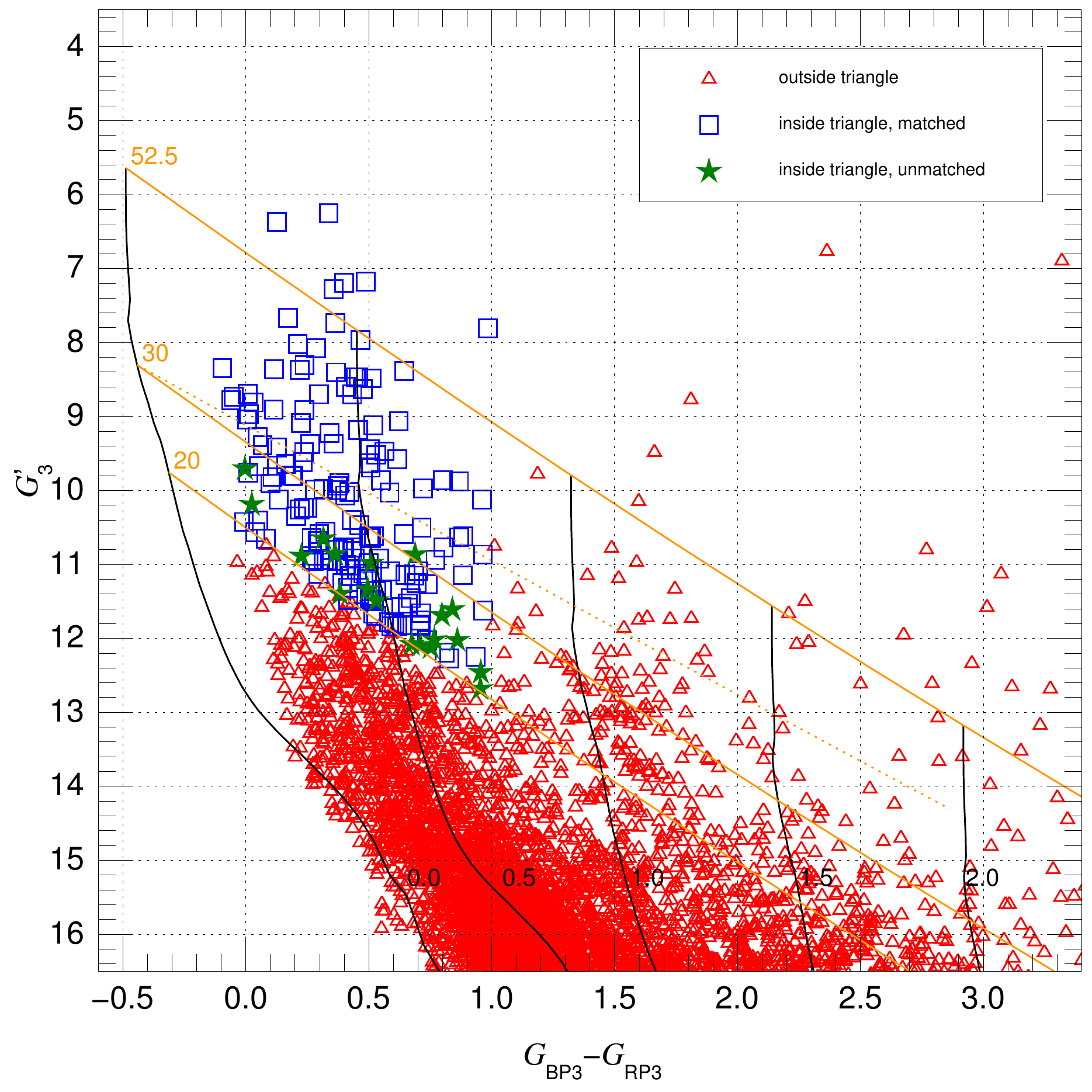}}
\caption{Second panel, see previous page for the first one: Equivalent plot but for all {\it Gaia}~EDR3 stars in the region of interest with corrected parallaxes that are compatible with the distance to Car~OB1 and positive and with catalog values of \BPRP. The plotted objects are classified according to whether they are located inside or outside of the area limited by $\BPRP = 1.0$ and the \RV~=~4.5 extinction track for average MS stars with $\Teff = 20$~kK. Stars inside that area are further divided into those matched with objects in the first panel (154) and those unmatched (19). Note that an additional three stars inside the above mentioned area in the first panel (HD~\num{93129}~Ab, CPD~$-$59~2636~A,B, and ALS~\num{19740}) plus $\eta$~Car outside that area are not shown either because they are not included in {\it Gaia}~EDR3 or have no parallaxes there. }
\end{figure*}

\paragraph{High-extinction population of \citet{Preietal21}.} That paper lists several stars that were too faint to be observed with GIRAFFE in the blue-violet region. Regarding their {\it Gaia}~EDR3 parallaxes, most of them are similar to or smaller than that of Car~OB1 but with larger uncertainties. There is only one with negative parallax, so it is likely a background object: 2MASS~J10452648$-$5946188 (=[HSB2012]~3994), which was already identified as a highly-extincted B~star by \citet{Damietal17a} using CHORIZOS.

\section{Results and discussion} \label{discussion}

\subsection{The observed CMD and completeness}	\label{complete}

We first discuss the observed {\it Gaia}~EDR3 CMD for the sample of 316 objects in this paper, which is plotted on the first panel of Fig.~\ref{Gaia_CMD}. Of those, only four are located in the foreground but two of them are in distinct regions of the CMD: HD~\num{93420}, a RSG in the upper right, and 2MASS~J10431945$-$5944488, a sdO in the lower left. The main group, the 294 objects in Car~OB1, does not follow the typical isochrone of a cluster or association because of the strong differential extinction present in the region. The majority concentrates between the extinguished isochrones that correspond to \EBV\ of 0.3 and 0.6 (assuming an \RV\ of 4.5) but some are significantly more extinguished than that, including the four \citet{Preietal21} objects outside the frame towards the lower right. The 18 background objects have, on average, a higher extinction than the Car~OB1 population, an expected effect of the extinction associated with the Carina Nebula. They appear mixed in the vertical direction with the Car~OB1 population but one should consider that if we plotted absolute magnitude on the vertical axis, they would move up. For example, five of the 18 objects are B supergiants. 

The majority of the stars lie between the \RV~=~4.5 extinction tracks for average MS stars of \Teff\ = 20~kK and 52.5~kK. A significant fraction lies below the track of \Teff\ = 20~kK, due to a combination of different effects: an average-age B2.5~V star can have a \Teff\ somewhat lower than 20~kK, ZAMS stars should be lower in the CMD than average-age ones, and the extinction tracks for $\RV > 4.5$ (which is known to be  appropriate for some stars in the Carina Nebula, see \citealt{MaizBarb18}) are steeper than the plotted ones. Above the extinction track of \Teff\ = 52.5~kK we find ten Car~OB1 stars: $\eta$~Car, the two RSGs, WR~24, four O supergiants, and HD~\num{93250}~A.B \citep[a close binary with two very early type components, see][]{lebouquin17}, that is, all of them objects that are expected to be there.

The most notorious feature of the first panel of Fig.~\ref{Gaia_CMD} is how well the Car~OB1 O and B stars are separated in the CMD, with the O stars mostly above the average-age extinction track of \Teff\ = 30~kK for \RV~=~4.5 and the B stars below it. This is an indirect confirmation of the quality of the spectral classifications. The separation is not perfect but it is not expected to be for several reasons: B giants and supergiants (plus some early B + early B binaries) are expected to be above the average-age extinction track of \Teff\ = 30~kK for \RV~=~4.5  and late O-dwarfs near the ZAMS below that track. In addition, variations in \RV\ among sightlines should produce some mixing, as low values of \RV\ can move B stars into the O-star territory and high values of \RV\ can move O stars into the B-star territory. Examples of the latter possibility are two of the stars from the \citet{Preietal21} sample, 2MASS~J10454595$-$5949075 and 2MASS~J10452013$-$5950104. If they are confirmed to be normal O dwarfs, their value of \RV\ should be high. 

When building a census, one of the most important questions that have to be addressed is how complete it is and that is especially important when the sample is built from multiple sources such as in this paper. To answer that question, we have plotted in the second panel of Fig.~\ref{Gaia_CMD} all the {\it Gaia}~EDR3 sources found within the footprint that have positive corrected parallaxes consistent with being at the distance of Car~OB1 and that have catalog values of \BPRP. The second panel shows that the first panel is just the tip of the iceberg in terms of a moderately extinguished well-populated main sequence. In addition to that main sequence, a significant population of red stars is present. By comparison with the first panel, some of those are extinguished OB stars but a comparison with other Galactic sightlines indicates that most of them must be intrinsically red stars. For example, the diagonal series of stars that follows the extinction track of \Teff\ = 30~kK around $\BPRP\sim 1.5$ is the red-clump extinction sequence, ubiquitously seen in the Galactic plane when plotting absolute magnitude in the vertical axis (as we are effectively doing here by selecting a population consistent with being at the same distance). That sequence starts around $\BPRP\sim 1.1$ for zero extinction and here we are just seeing it with an extinction distribution not too different from that of the OB stars in Car~OB1. 

Given the dominance of the late-type population for red colors (something that needs to be addressed with additional data such as NIR photometry), we do not have the means to determine how complete the sample is for high extinctions. Indeed, that is why a paper as recent as \citet{Preietal21} was able to find several new O stars in Car~OB1: thick dust clouds can easily hide OB stars if one does not have access to IR data and even in that case finding the hot needle in the cool haystack is not always straightforward. Therefore, we concentrate on the low- and moderate-extinction part of the sample, defined as those OB stars with $\BPRP<1.0$ (just to the left of the point where the first red clump stars are expected to appear). Also, as for fainter stars one expects any sample to be less complete, we restrict the completeness analysis to the region above the extinction track of \Teff\ = 20~kK in Fig.~\ref{Gaia_CMD}. In other words, we are assessing how complete the sample is regarding low/moderate extinction O and early-B stars. 

We cross-matched the two samples (the one used throughout this paper and the full {\it Gaia}~EDR3 one) inside that area and found 154 coincidences. Three objects in the main sample are not present in the {\it Gaia}~EDR3 sample either because they lack parallaxes (CPD~$-$59~2636~A,B and ALS~\num{19740}) or because they are completely absent (HD~\num{93129}~Ab), note that $\eta$~Car would be also absent if it were inside that area. {\it Gaia}~EDR3 is quite complete barring a few small-separation binaries. As for the other way around, 19 systems in the {\it Gaia}~EDR3 sample are not present in the main one (green stars in the second panel). Of those, five have large external parallax uncertainties ($>0.1$~mas), so chances are they are not real Car~OB1 members. Therefore, we estimate that our sample is around 90\% complete for low/moderate extinction O and early B systems. Furthermore, the location of the 19 green stars in the second panel of Fig.~\ref{Gaia_CMD}, all of them below the extinction track of \Teff\ = 30~kK, suggests that those missing objects are likely of early-B type. Therefore, we conclude that we are missing very few or even no low/moderate O-type systems in Car~OB1 within our footprint in our sample. As mentioned above, objects with high extinction may be another story. In any case, the 74 Car~OB1 O-type systems in this paper are the largest nearly complete sample of objects of that spectral type in any part of a Galactic OB association.

\subsection{Binary fraction}\label{bin} 

It is well known that multiplicity among massive stars is ubiquitous. Commonly, multiplicity is divided into that which is detected through spectroscopy (velocity changes and differences) and imaging (or visual multiplicity) and it is important to indicate which one is being used, as some previous studies have conflated them and caused confusion. In GOSSS~II we analyzed the population of Galactic southern stars and found out that 65-91\% of them are multiple stars of one type or another, with the values for spectroscopic and visual multiples being 50-60\% and 30-76\%, respectively. One consequence of those numbers is that a significant fraction (at least 15\%) are at the same time spectroscopic and visual multiples, and most of those involved three stars, as in 2014 the number of pairs detected simultaneously with spectroscopy and imaging was quite low. An analysis of known multiple O stars in the northern hemisphere \citep{Maizetal19a} confirmed the trend towards systems of three or more stars and revealed that simple binaries are a minority once spectroscopic and visual multiples are included. For example, hierarchical triples composed of a short-period (less than 1 month) system orbited by a companion in a long-period (years or more) orbit are quite common. 

\begin{figure*}[t!]
\centering
\includegraphics[width=15.0cm, trim={0 0cm 0 0cm},clip]{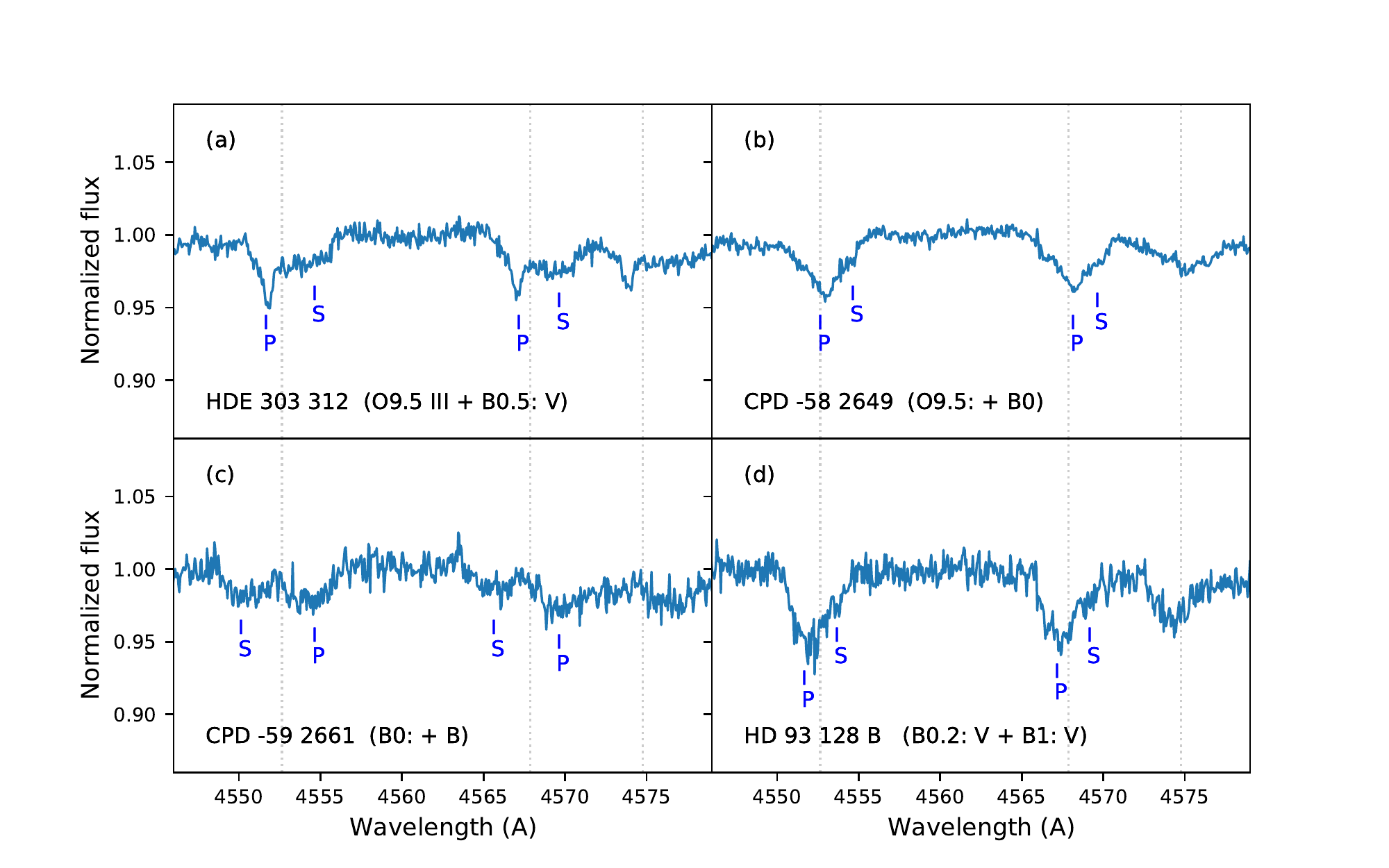}
\caption{Example of new spectroscopic binary systems reported in this work: Si\,{\sc iii} line profiles from GES spectra shown at their original resolution. For reference, P and S letters indicate the position of the primary and secondary component, respectively. See figures on Appendix~\ref{ap_list} for full spectra details.}
\label{sb2_new}
\end{figure*}

In the census presented here we find 20 spectroscopic binary systems containing at least one O-type star (listed in Table~\ref{obinaries}), one of them located in the background. There are six new systems reported for the first time in this work, either from GES and/or GOSSS~IV observations.  Excluding the background system, the total number of O stars in the 19 systems of Car~OB1 is 30. This represents a fraction of 0.35 (30 out of a total of 85 O-stars, see Table~\ref{table_binclus} where binary statistics and fractions for the spectroscopic systems containing at least one O-type star in the different Villafranca groups are summarized).
This number is still far from those reported in GOSSS~II and also from the 0.44 fraction of O-type stars in binaries quoted by \cite{sana11} or even the somewhat more than 0.50 indicated by \cite{sana17}. This indicates that there is a significant number of binaries still to be identified in the region. We highlight that the multiplicity statistics reported in this work are, however, incomplete. We only report double-line spectroscopic binaries. Visual binaries are not considered and only a small fraction of the sample has significant multi-epoch coverage for single-line spectroscopic binary detection.
In spite of this,  we significantly increase the fractions quoted by \cite{sana11} in Trumpler 14 and Trumpler 16, for which these authors quoted fractions of zero and 0.48, respectively (see Table~\ref{table_binclus}).
\begin{table*}
\caption{Binary statistics and fractions for the spectroscopic systems containing at least one O-type star present in the census.}
\label{table_binclus}
\centering
\begin{tabular}{lcccr}
\hline\hline \\[-1.5ex]
Group & Single & Binary & O stars & Fraction \\
      & O-stars & O-systems & in binaries &            \\
\hline \\[-1.5ex]

\hline \\[-1.5ex]
O-002 (Trumpler 14)   & 10   & 3  & 5 & 0.33 \\
O-003 (Trumpler 16 W)   & 2   & 2  & 3 & 0.60  \\
O-025 (Trumpler 16 E)  & 8  & 6  & 10 & 0.55  \\
O-027 (Trumpler 15) & 3   & -  & - & 0    \\
O-028 (Collinder 228) & 16   & 3   & 4 & 0.15   \\
O-029 (Collinder 232) & 2   & -   & - & 0    \\
O-030 (Bochum 11) & 4   & 1   & 2 & 0.33  \\
Car OB1 & 10   & 4   & 6 & 0.37  \\
\hline \\[-1.5ex]
Whole sample & 55  & 19  & 30  & 0.35     \\
\hline\\[-1.5ex]

\multicolumn{5}{l}{\footnotesize{Notes. Background and foreground members are excluded from the statistics. We}}\\ 
\multicolumn{5}{l}{\footnotesize{separate numbers considering the different Villafranca groups of Carina. Car OB1}}\\ \multicolumn{5}{l}{\footnotesize{group refers to the  stars just falling in the gaps between defined Villafranca groups}}\\
\multicolumn{5}{l}{\footnotesize{(see Appendix~\ref{ap_list}) for further details.}}
\end{tabular}
\end{table*}

In addition, we found 18 spectroscopic binary systems formed by early B-type stars, one of them a background system and 13 new binary detections from GES, GOSSS and \lili~ observations (see Table~\ref{bbinaries}). In Fig.~\ref{sb2_new} we show an example of the new spectroscopic binary detections from GES spectra at their original resolution. The Si~III triplet at $\uplambda$4552-68-75 $\AA$ is shown for binary late-O (a and b panels) and early-B (c and d panels) systems. See figures in Appendix~\ref{ap_list} for full spectra details.

\begin{figure}[t!]
\centering
\includegraphics[width=8cm]{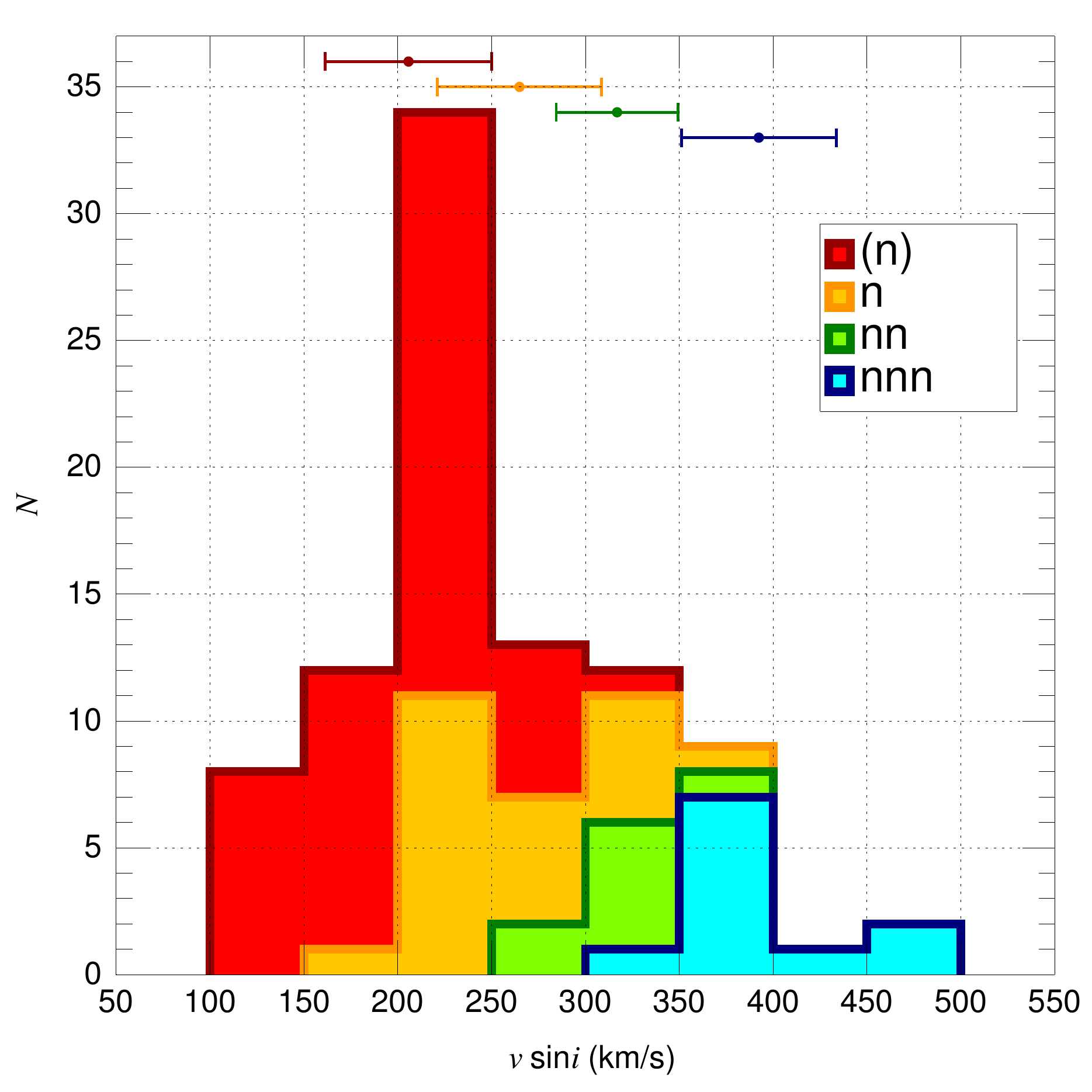}
\caption{ $v\sin i$ histogram of those OB stars with any rotation index in their spectral classification. Broadening is denoted by (n), n, nn and nnn indexes, progressing from somewhat to more broadened lines. Points and horizontal lines on the top of the figure indicate the mean $v\sin i$ value for each rotating group and the corresponding dispersion, respectively.}
\label{rot_index}
\end{figure}

\subsection{The spectroscopic n-qualifier as an indicator for rotation}\label{rot}

As stated in Sect.~\ref{classif}, the MGB tool has been used for the spectral classification of GES data. This tool allows the user to obtain not only the spectral subtype and luminosity classification, but also spectral peculiarities and the rotation index. Broadening is denoted by (n), n, nn and nnn indexes, progressing from somewhat to more and even more broadened lines.  Therefore, this qualifier has been traditionally interpreted as a sign for high rotational velocity. As a consistency check, we have determined the projected rotational velocity of stars with this qualifier, in order to know whether there is a 1:1 relationship between both.

To that aim we used \texttt{iacob-broad},  a user-friendly tool for the line-broadening characterization of OB stars \citep{ssimon07,ssimon14a}. It is based on a combined Fourier Transform (FT) and the Goodness-of-fit (GOF) method that allows us to  determine easily the stellar projected rotational velocity ($v\sin i$) and the amount of extra broadening ($v_{\rm mac}$) from a specific diagnostic line. The FT technique is based on the identification of the first zero in the Fourier transform of a given line profile \citep{gray08,ssimon07}. The GOF technique is based on a comparison between the observed line profile and a synthetic one that is convolved with different values of $v\sin i$ and $v_{\rm mac}$ to obtain the best-fit by means of a $\chi^{2}$ optimization. 
The main advantage of this methodology is that we obtain two independent measurements of the $v\sin i$ (resulting from either the FT or the GOF analysis) whose comparison is used as a consistency check and to better understand problematic cases. Since  metallic lines do not suffer from strong Stark broadening or nebular contamination, they are best suited for obtaining accurate $v\sin i$ values. GIRAFFE set-ups cover the Si\,{\sc iii}~$\uplambda$4552 diagnostic line, while UVES/FEROS/HARPS set-ups also cover the O\,{\sc iii}~$\uplambda$5592 diagnostic line. In case none of them are present or are too weak, we then use the nebular free or weakly contaminated He\,{\sc i} lines (\ion{He}{i}~$\uplambda$4387, $\uplambda$4471, $\uplambda$4713).

\begin{figure*}[t!]
\centering
\includegraphics[width=18cm, trim={1.6cm 0 0 0},clip]{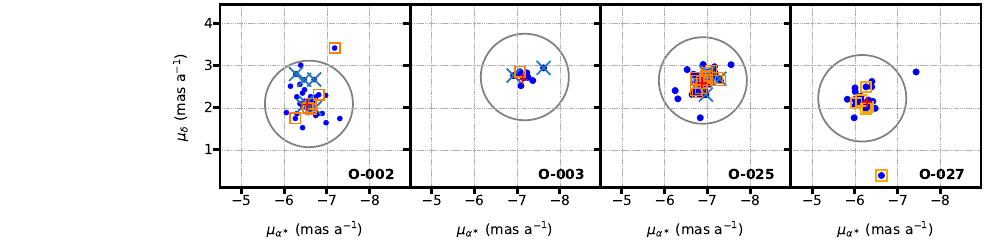}
\includegraphics[width=13.5cm, trim={1.2cm 0 0 0},clip]{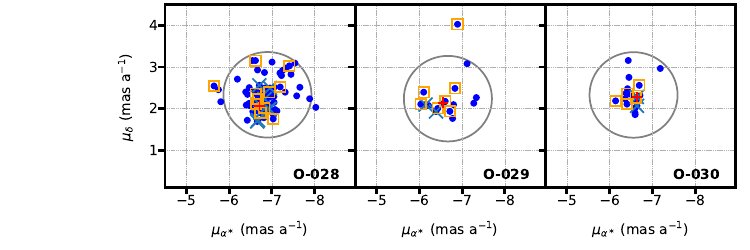}
\caption{Proper motion distribution from \textit{Gaia}~EDR3 astrometry for all stars of our census in each assigned Villafranca group. Orange squares indicate those OB stars analyzed in this work that are rotating at $v\sin i$ $\geq$ 200 km s$^{-1}$. Blue crosses represent identified binary systems. Circles represent group proper motion constraints, whose centers $\mu_{\alpha *,g}$~ and ~$\mu_{\delta,g}$ are those shown in Table~\ref{pm_groups} in the central columns. For comparison, red plus symbols indicate group centers from Villafranca II and III works. Note that stars labelled as Car~OB1 members are not included in the panels.}
\label{pm_fast}
\end{figure*}

Fig.~\ref{rot_index} presents the $v$ sin $i$ histogram of those OB stars included in our census with any broadening index in their spectral classification (see Table~\ref{table_spclas}). Mean  $v$ sin $i$ values for (n), n, nn and nnn are 206, 265, 317 and 392 km s$^{-1}$, respectively, which confirms the trend that the higher the rotation index the higher the projected rotational velocity. However, we find a significant overlap in the ranges of projected rotational velocities. For example, the (n)- and n-type stars peak at the same bin: between 200 and 250 \kms; and the fastest n-star rotates 88 \kms faster than the slowest nn-star (but the distribution of (n)-stars is slanted towards the left from that point while that of n-stars is slanted towards the right). There are two likely explanations for the overlap. On the one hand, a non-negligible fraction of these stars could end up being actually spectroscopic binaries since such broadened lines may prevent us from detecting binary line profiles when multi-epoch observations are not available. On the other hand, the sample is dominated by B1-B2.5 dwarfs, which have few intrinsically deep and narrow lines in the analyzed wavelength range (which is only a part of the standard blue-violet classification range), making the n-type indexes more unreliable than for e.g. O stars or B supergiants.

\subsection{Runaway candidates}

\begin{table*}
\caption{Group proper motions in $\alpha$* and $\delta$ derived in this work and those from the Villafranca II and III works, all based on \textit{Gaia} EDR3 astrometry.}
\label{pm_groups}
\centering
\begin{tabular}{lccccc}
\hline\hline \\[-1.5ex]
& \multicolumn{3}{c}{\rule[2pt]{2.0cm}{0.4pt} This work \rule[2pt]{2.0cm}{0.4pt}} & \multicolumn{2}{c}{\rule[2pt]{1.2cm}{0.4pt} Villafranca II-III} \rule[2pt]{1.2cm}{0.4pt} \\
\hline \\[-1.5ex]
Group & $N$ & $\mu_{\alpha *,g}$ & $\mu_{\delta,g}$   & $\mu_{\alpha *,g}$ & $\mu_{\delta,g}$\\
\hline \\[-1.5ex]
O-002 (Trumpler 14)     &  32  & -6.580 $\pm$ 0.240  & 2.089 $\pm$ 0.246  & -6.534 $\pm$ 0.023  & 2.076 $\pm$  0.023\\
O-003 (Trumpler 16 W)   & 8   & -7.179 $\pm$ 0.102  & 2.730 $\pm$ 0.103 & -7.128 $\pm$ 0.024  & 2.670 $\pm$  0.024\\
O-025 (Trumpler 16 E)   &  47  & -6.894 $\pm$ 0.214  & 2.647 $\pm$ 0.177 & -6.877 $\pm$ 0.023  & 2.596 $\pm$  0.023\\
O-027 (Trumpler 15)     &  17  & -6.172 $\pm$ 0.164  & 2.224 $\pm$ 0.175 & -6.282 $\pm$  0.023  & 2.131 $\pm$  0.023\\
O-028 (Collinder 228)   & 53   & -6.896 $\pm$ 0.334  & 2.332 $\pm$ 0.396 & -6.713 $\pm$  0.021  & 2.070 $\pm$  0.021\\
O-029 (Collinder 232)   & 12   & -6.667 $\pm$ 0.414  & 2.238 $\pm$ 0.294 & -6.552 $\pm$  0.023  & 2.142 $\pm$  0.023\\
O-030 (Bochum 11)       &  19  & -6.559 $\pm$ 0.204  & 2.328 $\pm$ 0.307 & -6.635 $\pm$  0.021  & 2.279 $\pm$  0.021\\
\hline\\[-1.5ex]

\multicolumn{6}{l}{\footnotesize{Note. Group uncertainties reported in this work refer to the standard deviation of the selected OB stars while} }\\
\multicolumn{6}{l}{\footnotesize{those in Villafranca II-III correspond to the standard deviation of the mean (with the angular covariance term}}\\
\multicolumn{6}{l}{\footnotesize{included) of all the stars identified as group members,  a much larger number than the one used in this work.} }\\ 
\end{tabular}
\end{table*}

Benefiting from the high-precision astrometry that \textit{Gaia}~EDR3 provides in Carina, we have investigated the proper motions of the stars in our census to identify bona-fide runaways as a first step for future studies. Following the ideas of \cite{demink13} (who propose that fast rotating stars are the product of post-interacting binaries and therefore could also have been ejected from binary systems in which the mass donor exploded as supernova) we are interested in exploring whether there is a connection between O and B-type stars with the spectroscopic n-qualifier\footnote{as a proxy for fast rotation, although we emphasize that, even if there is a good correlation with the average $v\sin i$, not all stars with this qualifier have a high projected rotational velocity, as shown in Sect.~\ref{rot}.} and the runaway status.

The proper motion distribution for each Villafranca group is shown in Fig.~\ref{pm_fast}. We define group centers through an iterative process assuming the average values of each group members, but excluding detected binaries, objects with a RUWE (renormalised unit weight error) $>$ 1.4 and those stars that do not comply with the proper motion constraint described below. For reference, group proper motions in $\alpha *$ and $\delta$ derived in this work and those from the Villafranca II and III works are shown in Table~\ref{pm_groups}. As in Villafranca II, we find the proper motion of O-025 not identical to that of O-003, indicating that both groups in Trumpler 16 are well separated. We iteratively filter stars with proper motions larger than the mean values for each group by more than three sigma. To this aim, we calculated for each group $\sigma_{g}= \sqrt{\sigma^{2}_{\mu_{\alpha *}}+\sigma^{2}_{\mu_{\delta}}}$, deriving a final mean $\sigma_{g}$ of 0.342 mas a$^{-1}$ and thus a three sigma value of 1.03 mas a$^{-1}$.
We find four stars with the n-qualifier in their spectral classification that do not meet the imposed constraint (those stars falling outside the circles in Fig.~\ref{pm_fast}). Three of them (2MASS~J10440866-5933488 in O-002, CPD~-59~2541 in O-028 and 2MASS~J10451588-5929563 in O-029) can be considered firm runaway candidates. We note, however, that the large RUWE value for CPD~-58~2657 (in the O-027 group) indicates inaccurate astrometric measurements (see Table~\ref{runaways}). The rest of stars with the n-qualifier are homogeneously distributed around the core motion of each group. Interestingly, the two extreme very fast B rotators of our sample, ALS~\num{15248} and 2MASS~J10433865-5934444 rotating both at $v\sin i$ $>$ 450 km s$^{-1}$, do not show peculiar proper motions, and so are consistent with the main values of each group.
We also find four further runaway candidates that are not included in the group with the n-qualifier but show peculiar proper motions. Two of them are RSG stars. We note that two stars identified in Villafranca I as possible runaway stars ejected from Trumpler 14 (HDE~303\,313 and ALS~16\,078) spatially fall in the gaps of the redefined Villafranca groups. Therefore, they have been labelled as just Car OB1 members and are not discussed here\footnote{a direct comparison between the runaway candidates identified in both works must be done with caution since different methods have been used. Note that in Villafranca works, stars are selected as candidate runaway/walkaway objects when their proper motion points in the opposite direction to that of the center of the group (within some margins, see Villafranca I-II for further details).}.

Thus we have four out of 90 stars with the n-qualifier identified as candidate runaway objects and another four out of 168 without the n-qualifier, which means fractions of 4.4$\%$ and 2.4$\%$, respectively \footnote{Note that detected spectroscopic binary systems, with or without the n-qualifier,  have been excluded from the statistics.}.
This points to a connection between runaways and fast rotators, as pointed out by other works \citep[see f.e.][and references therein]{demink13, holgado22} particularly if we consider that the viewing angle may be affecting the projected rotational velocities, resulting in less broadened lines.
However, given the limitations of our work, further research on this topic (in particular, a distribution of projected rotational velocities and a more detailed study of the runaway condition) is needed in order to obtain a firm conclusion.

Finally, we remark that the distribution of binary systems in the proper motion diagram (crosses in Fig~\ref{pm_fast}), contrary to what might be expected, is homogeneously distributed. A similar pattern was found in the Cygnus OB2 association \citep{berlanas20}, implying that these systems may still keep their original velocities.

\section{Conclusions}\label{conclusion}

We present a new census of massive stars in the central part of Carina, Car OB1, based on high-quality spectroscopic data provided by GES, GOSSS, \lili ~and additional sources from the literature.  It contains a total of 316 massive stars. We separated stars with distances compatible with Car OB1 (assigning group membership) from those in the foreground and background, finding four systems in the foreground and 18 in the background. Of the 294 stellar systems in Car OB1, 74 are of O type, 214 are of non-supergiant B type and six are of WR or non-O supergiant (II to Ia) spectral class. 
We estimate that our sample is around 90\% complete for low/moderate extinction O and early B systems, missing very few or even no O stars within our footprint. The 74 Car OB1 O-type systems quoted in this paper are the largest nearly complete sample of objects of that spectral type in any part of a Galactic OB association.

Among the stellar census, we identified  20 spectroscopic binary systems that contain at least one O-type star. Six of them are new identifications and one is located in the background. The observed binary fraction of O stars found in the Car~OB1 region is 0.35, although this number only refers to double-lined spectroscopic binaries and represents, therefore, a lower limit. Visual binaries are not considered and only a small fraction of the sample has significant multi-epoch coverage for single-lined spectroscopic binary detection. Thus, this number should be considered as a lower limit. In addition, we found another 18 spectroscopic binary systems with a B-star primary, one of them being a background system and 13 of them new binary detections from GES, GOSSS and \lili~ observations.

We explore the correlation between the spectroscopic rotation index, n, and the actual projected rotational velocities of the stars. We find a good correlation of the average $v\sin i$  values with the qualitative classification of each group ((n), n, nn, nnn). However, there is a significant overlap in their $v\sin i$ ranges. We note that it is possible that a non-negligible fraction of these stars are actually spectroscopic binaries contaminating the fast rotator sample.

Finally, we investigated the proper motion distribution for the sample of those O and B-type stars with a spectroscopic n-qualifier. Our results indicate a connection between runaways and fast rotators. 
Furthermore, the distribution of binary systems in the proper motion diagram is homogeneously distributed, implying that these systems may still keep their original velocities.

\begin{acknowledgements}

This paper is based mainly on data products from spectroscopic observations made with ESO Telescopes at the Paranal Observatory under programme ID 188.B-3002. These data products have been processed by the Cambridge Astronomy Survey Unit (CASU) at the Institute of Astronomy, University of Cambridge, and by the FLAMES/UVES reduction team at INAF/Osservatorio Astrofisico di Arcetri. These data have been obtained from the \textit{Gaia}-ESO Survey Data Archive, prepared and hosted by the Wide Field Astronomy Unit, Institute for Astronomy, University of Edinburgh, which is funded by the UK Science and Technology Facilities Council.
This work was partly supported by the European Union FP7 programme through ERC grant number 320360 and by the Leverhulme Trust through grant RPG-2012-541. We acknowledge the support from INAF and Ministero dell' Istruzione, dell' Universit\`a' e della Ricerca (MIUR) in the form of the grant 'Premiale VLT 2012'. The results presented here benefit from discussions held during the \textit{Gaia}-ESO workshops and conferences supported by the ESF (European Science Foundation) through the GREAT Research Network Programme. Additional spectra were obtained using the     \href{http://www.lco.cl/irenee-du-pont-telescope/}{2.5~m du Pont Telescope} at the Observatorio de Las Campanas (LCO) and the     \href{https://www.eso.org/public/teles-instr/lasilla/mpg22/}{2.2~m MPG/ESO Telescope} at the Observatorio de La Silla (LSO).

This work has made use of data from the European Space Agency (ESA) mission 
\href{https://www.cosmos.esa.int/gaia}{\it Gaia}, processed by the {\it Gaia} Data Processing and Analysis  Consortium (\href{https://www.cosmos.esa.int/web/gaia/dpac/consortium}{DPAC}). Funding for the DPAC has been provided by national institutions, in particular the institutions participating in the {\it Gaia} Multilateral Agreement. 

This research is partially funded by the Spanish Government Ministerio de Ciencia e Innovaci\'on  and Agencia Estatal de Investigaci\'on (MCIN/AEI/\num{10.13039}/\num{501100011033}/FEDER, UE) through grants PGC2018-\num{093741}-B-C21/C22, PGC2018-\num{095049}-B-C21/C22 and PID2021-\num{122397}NB-C21/C22.
SRB also acknowledges funding by MCIN under the Juan de la Cierva - Formación grant (contract FJC 2020-\num{045785}-I) and NextGeneration EU/PRTR and MIU (UNI/551/2021) through grant Margarita Salas-ULL. AH also acknowledges support by the Severo Ochoa Program through CEX2019-000920-S. EJA also acknowledges financial support from the State Agency for Research of the Spanish MCIU through the “Center of Excellence Severo Ochoa” award to the Instituto de Astrofísica de Andalucía (SEV-2017-0709). MB is supported through the Lise Meitner grant from the Max Planck Society. We acknowledge support by the Collaborative Research centre SFB 881 (projects A5, A10), Heidelberg University, of the Deutsche Forschungsgemeinschaft (DFG, German Research Foundation).  This project has received funding from the European Research Council (ERC) under the European Union’s Horizon 2020 research and innovation programme (Grant agreement No. 949173).

\end{acknowledgements}

%
%

\def\bibname{References}

\bibliographystyle{aa}
\bibliography{biblio.bib}

\begin{appendix}

\section{Tables and spectrograms} \label{ap_list}

In this Appendix we present the tables with the information for the stars in the field of the Carina Nebula analyzed in this paper and the figures with the spectrograms that have not appeared in previous papers.

Table~\ref{table_main} lists the basic information for the stars: name, coordinates, identifications, \GGc\ magnitude, {\it Gaia}~EDR3 parallax and group membership. Regarding the latter, the following algorithm is used:

\begin{itemize}
 \item All stars are initially labelled as Car~OB1.
 \item A search is done to see if the star is located inside the region of the sky defined by the center and radius of one of the groups defined
       in Villafranca~II~or~III (\VO{002}~=~Trumpler~14, \VO{003}~=~Trumpler~16~W, \VO{025}~=~Trumpler~16~E, \VO{027}~=~Trumpler~15, 
       \VO{028}~=~Collinder~228, \VO{029}~=~Collinder~232, and \VO{030}~=~Bochum~11). If found, then the membership is changed to that group.
       Note that, as mentioned in the Villafranca papers, the traditional division into such groups is to some point arbitrary: \VO{002},
       \VO{025}, and \VO{027} are likely real clusters while the rest of the groups are just subassociations of the larger Car~OB1. As the 
       apertures in the Villafranca papers are circular, some stars just fall in the gaps and remain labelled as Car~OB1.
 \item Stars without a parallax or with corrected parallax uncertainties greater than 0.1~mas are given in bold face, to indicate that the
       {\it Gaia}~EDR3 information is not sufficient to determine their distances. Note that many of those objects are known to be located in Car~OB1 for other reasons (e.g. $\eta$~Car or HD~\num{93129}~Ab).
 \item Stars whose parallax is larger than 0.44~mas by more than three sigmas are labelled as foreground.
 \item Stars whose parallax is smaller than 0.41~mas by more than three sigmas or is negative are labelled as background. The limits here and in the previous steps are determined from Villafranca~II~and~III.
\end{itemize}

Table~\ref{table_spclas} lists the spectral types for the stars in this paper. Three types of spectral types are given: those derived from \textit{Gaia}-ESO spectra (all new), those derived from GOSSS spectra (most from previous papers and from GOSSS~IV but some new, marked as TW), and those derived from \lili\ and STIS spectra as well as from the literature (all previously published).

Tables~\ref{non-members} and~\ref{six-members} list those stars of the census identified in the foreground or in the background, and those of WR or non-O supergiant (II to Ia) spectral class, respectively. 

Tables~\ref{obinaries} and  ~\ref{bbinaries} list spectroscopic binary systems identified in our census containing at least one O-type and those formed by early B-type stars, respectively.  

Table~\ref{runaways} lists runaway candidates identified in this work.

Figure~\ref{UVES_spectra} shows the UVES spectra,
Fig.~\ref{GIRAFFE_spectra} the GIRAFFE spectra, and Fig.~\ref{GOSSS_spectra} the GOSSS spectra.

\begin{landscape}
\input{table_A11}
\end{landscape}

\begin{landscape}
\addtocounter{table}{-1}
\input{table_A12}
\end{landscape}

\begin{landscape}
\addtocounter{table}{-1}
\input{table_A13}
\end{landscape}

\begin{landscape}
\addtocounter{table}{-1}
\input{table_A14}
\end{landscape}

\begin{landscape}
\addtocounter{table}{-1}
\input{table_A15}
\end{landscape}

\begin{landscape}
\addtocounter{table}{-1}
\input{table_A16}
\end{landscape}

\begin{landscape}
\input{table_A21}
\end{landscape}

\begin{landscape}
\addtocounter{table}{-1}
\input{table_A22}
\end{landscape}

\begin{landscape}
\addtocounter{table}{-1}
\input{table_A23}
\end{landscape}

\begin{landscape}
\addtocounter{table}{-1}
\input{table_A24}
\end{landscape}

\begin{landscape}
\addtocounter{table}{-1}
\input{table_A25}
\end{landscape}

\begin{landscape}
\addtocounter{table}{-1}
\input{table_A26}
\end{landscape}

\input{non_members}

\input{six_members}

\input{obinaries}

\input{bbinaries}

\input{runaways}

\begin{figure*}
\centerline{
\includegraphics[width=\linewidth]{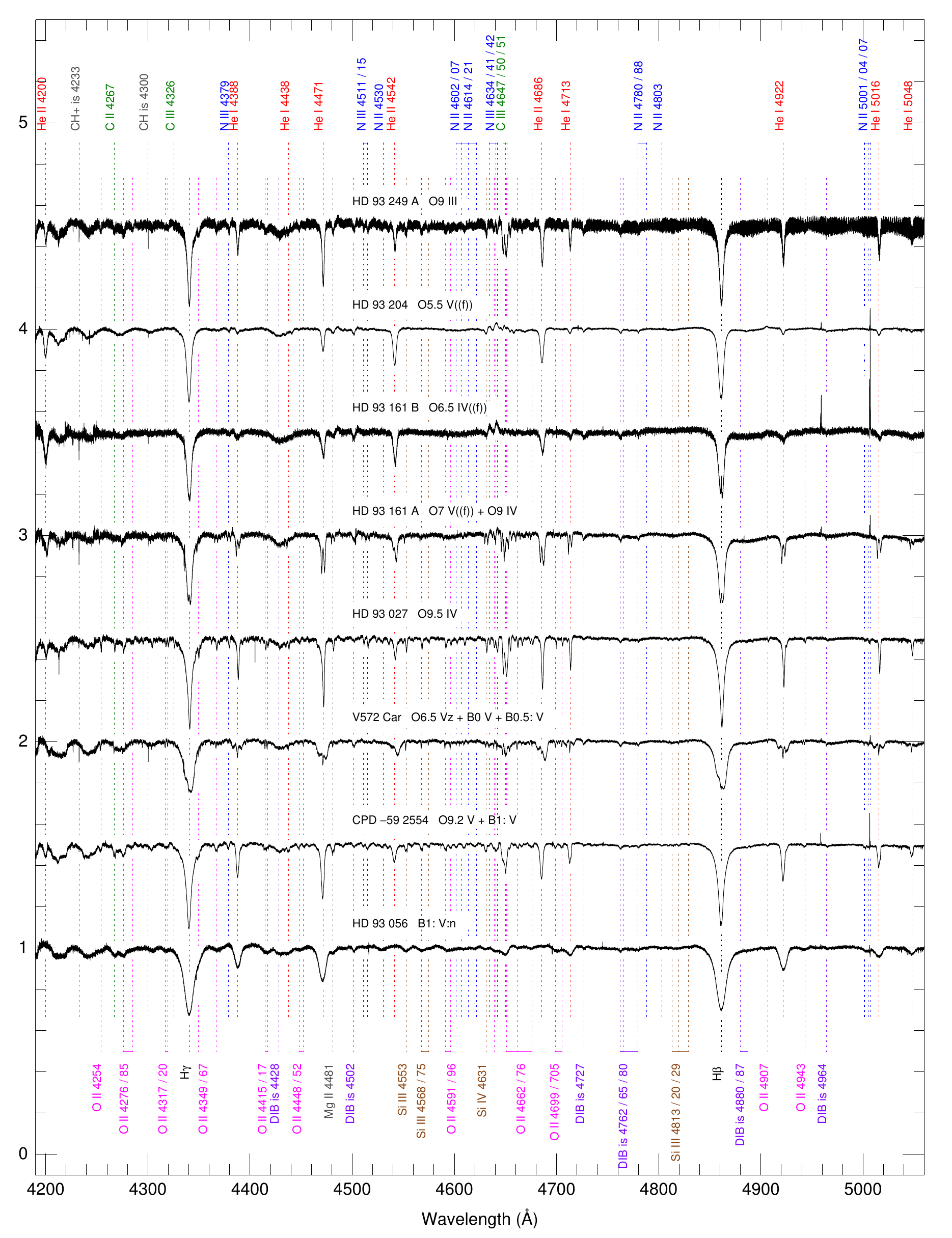}
}
\caption{UVES spectra shown at their original resolution.}
\label{UVES_spectra}
\end{figure*}

\addtocounter{figure}{-1}
\begin{figure*}
\centerline{
\includegraphics[width=\linewidth]{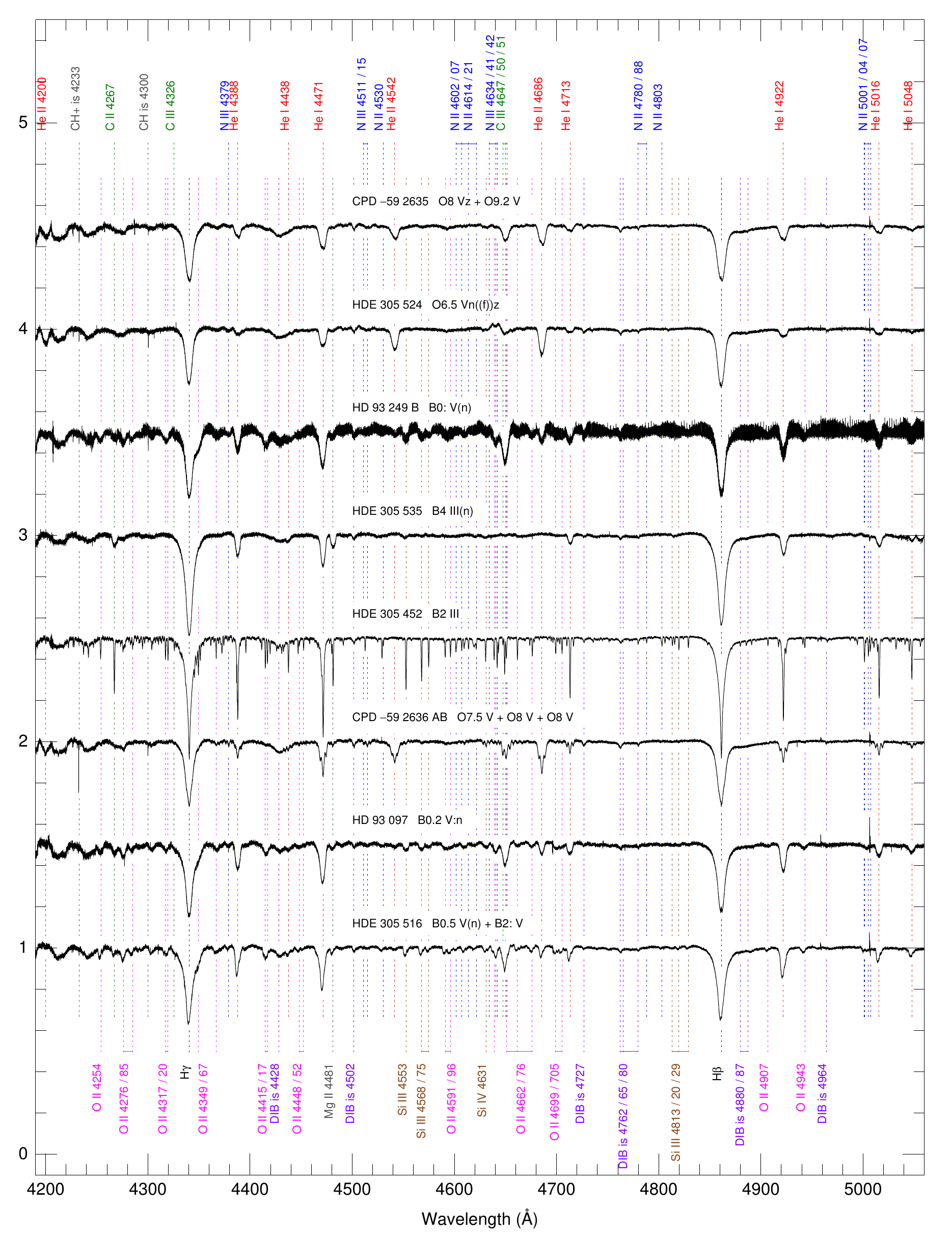}
}
\caption{(Continued).}
\end{figure*}

\begin{figure*}
\centerline{
\includegraphics[width=\linewidth]{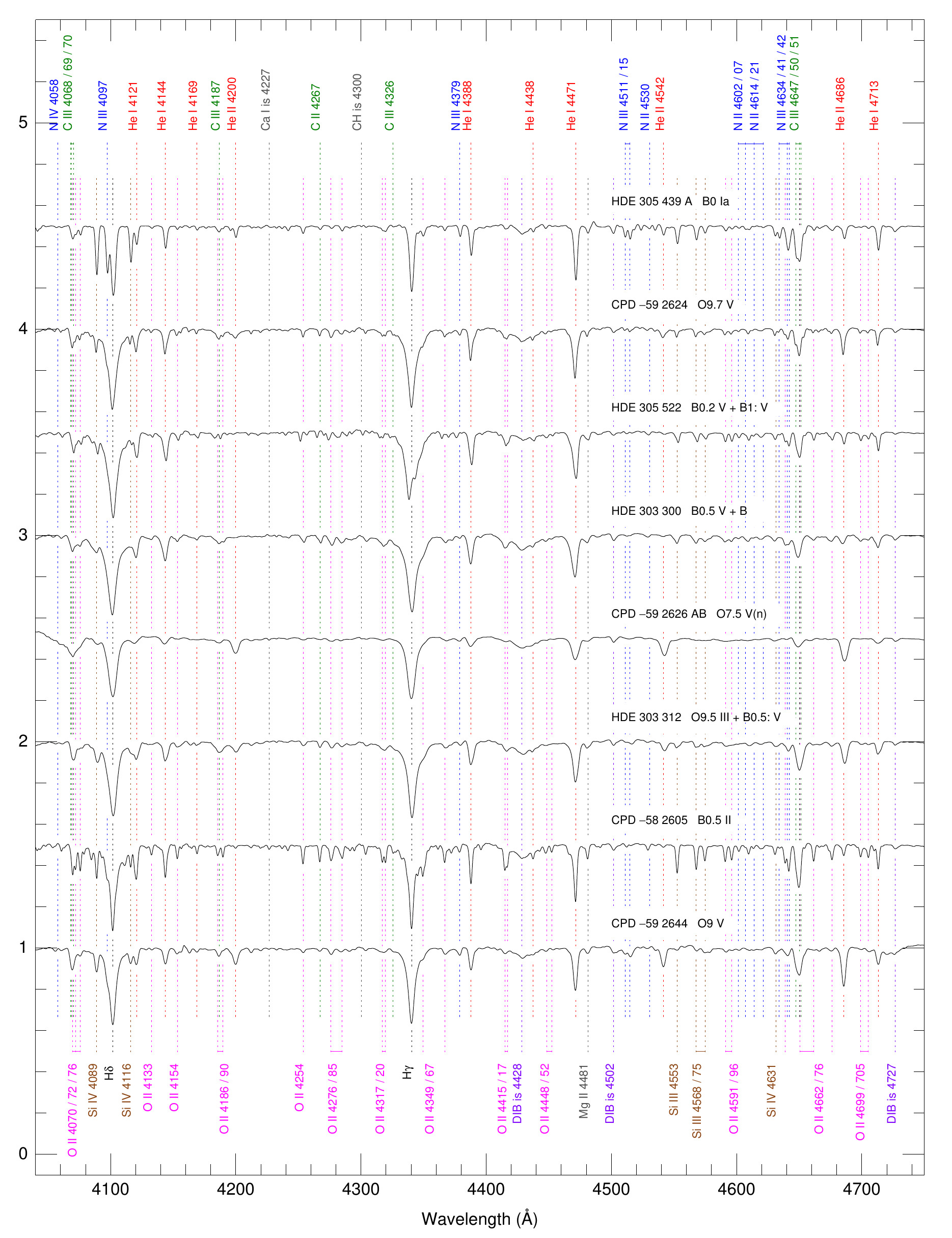}
}
\caption{GIRAFFE spectra shown at a resolution $R = 2500$.}
\label{GIRAFFE_spectra}
\end{figure*}

\addtocounter{figure}{-1}
\begin{figure*}
\centerline{
\includegraphics[width=\linewidth]{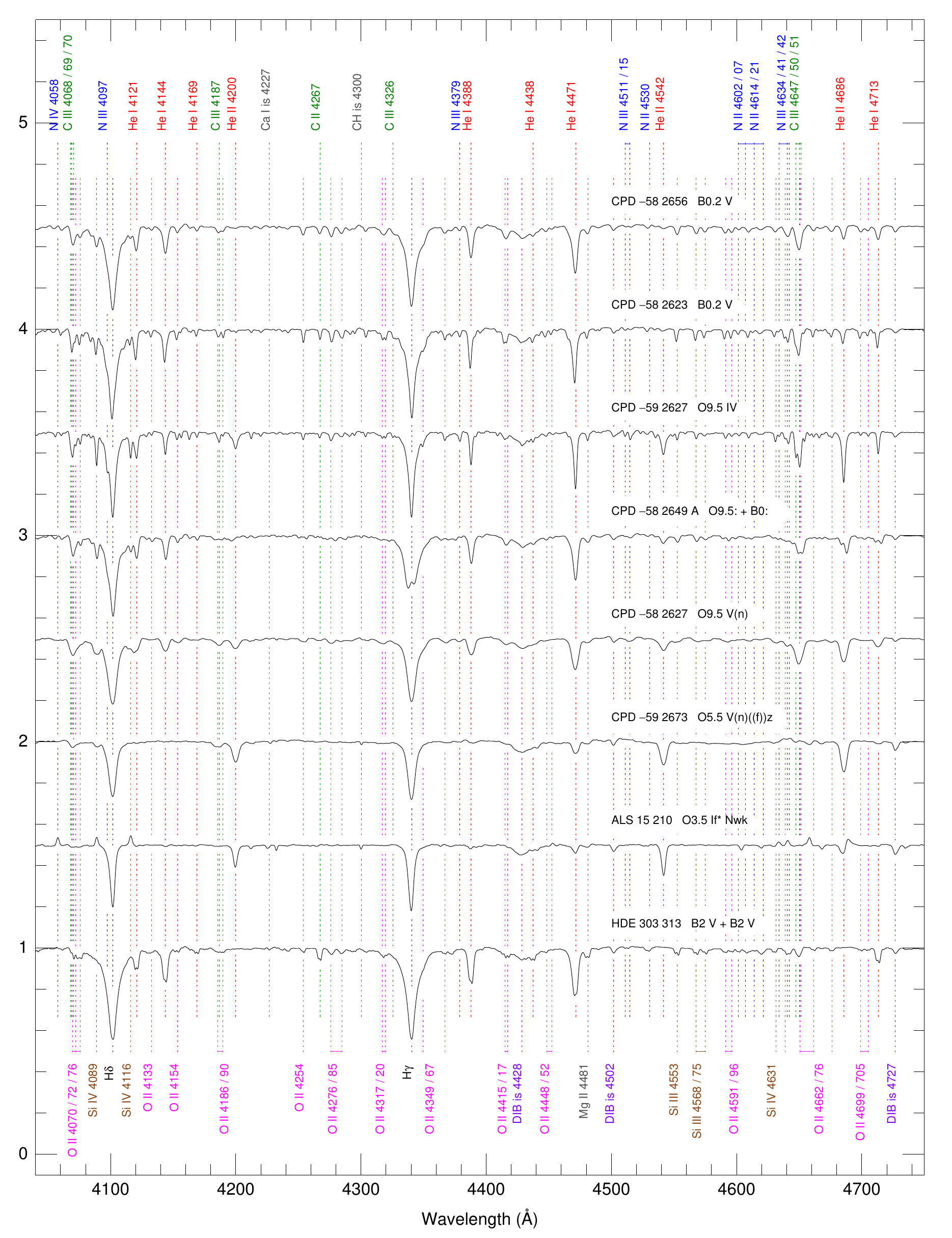}
}
\caption{(Continued).}
\end{figure*}

\addtocounter{figure}{-1}
\begin{figure*}
\centerline{
\includegraphics[width=\linewidth]{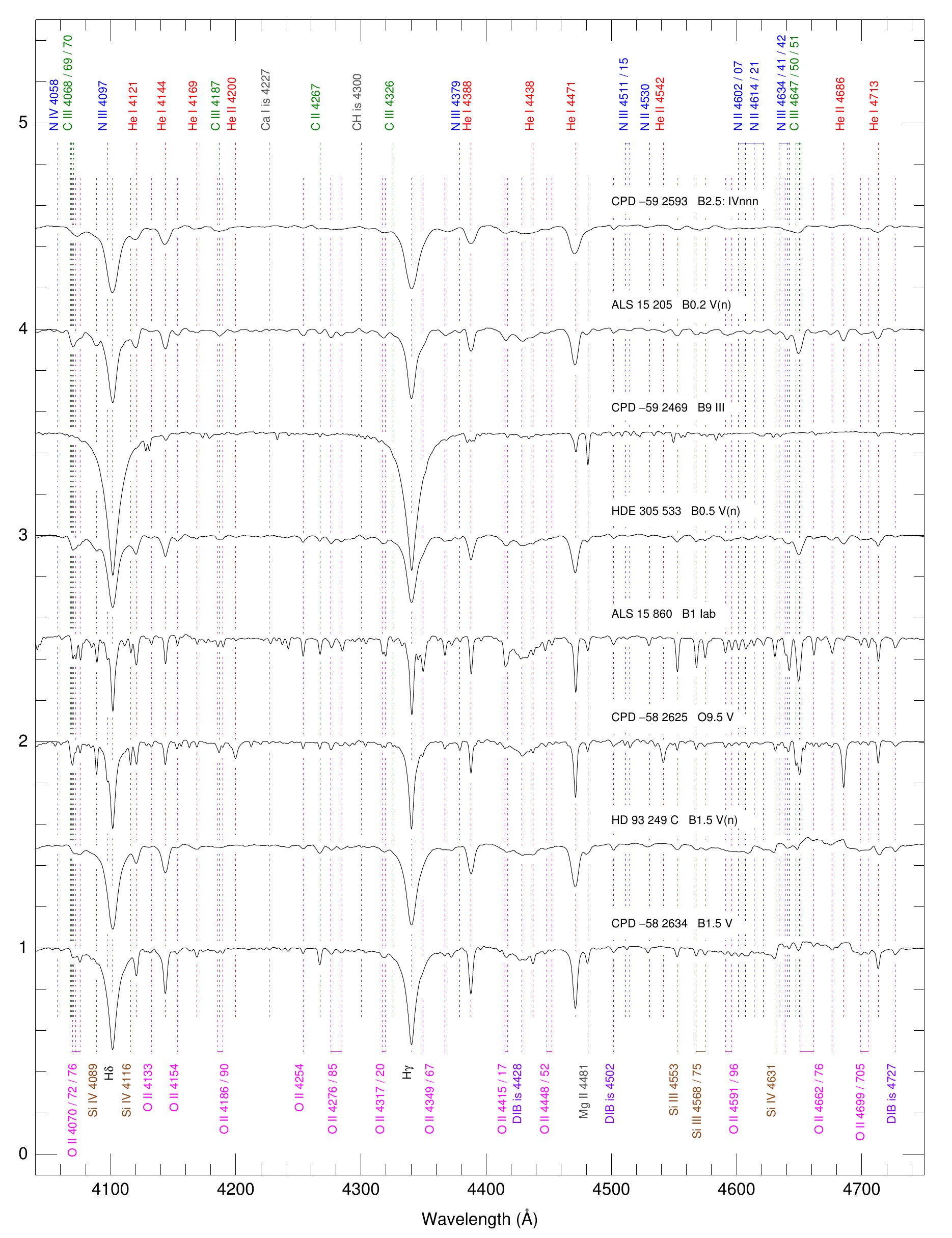}
}
\caption{(Continued).}
\end{figure*}

\addtocounter{figure}{-1}
\begin{figure*}
\centerline{
\includegraphics[width=\linewidth]{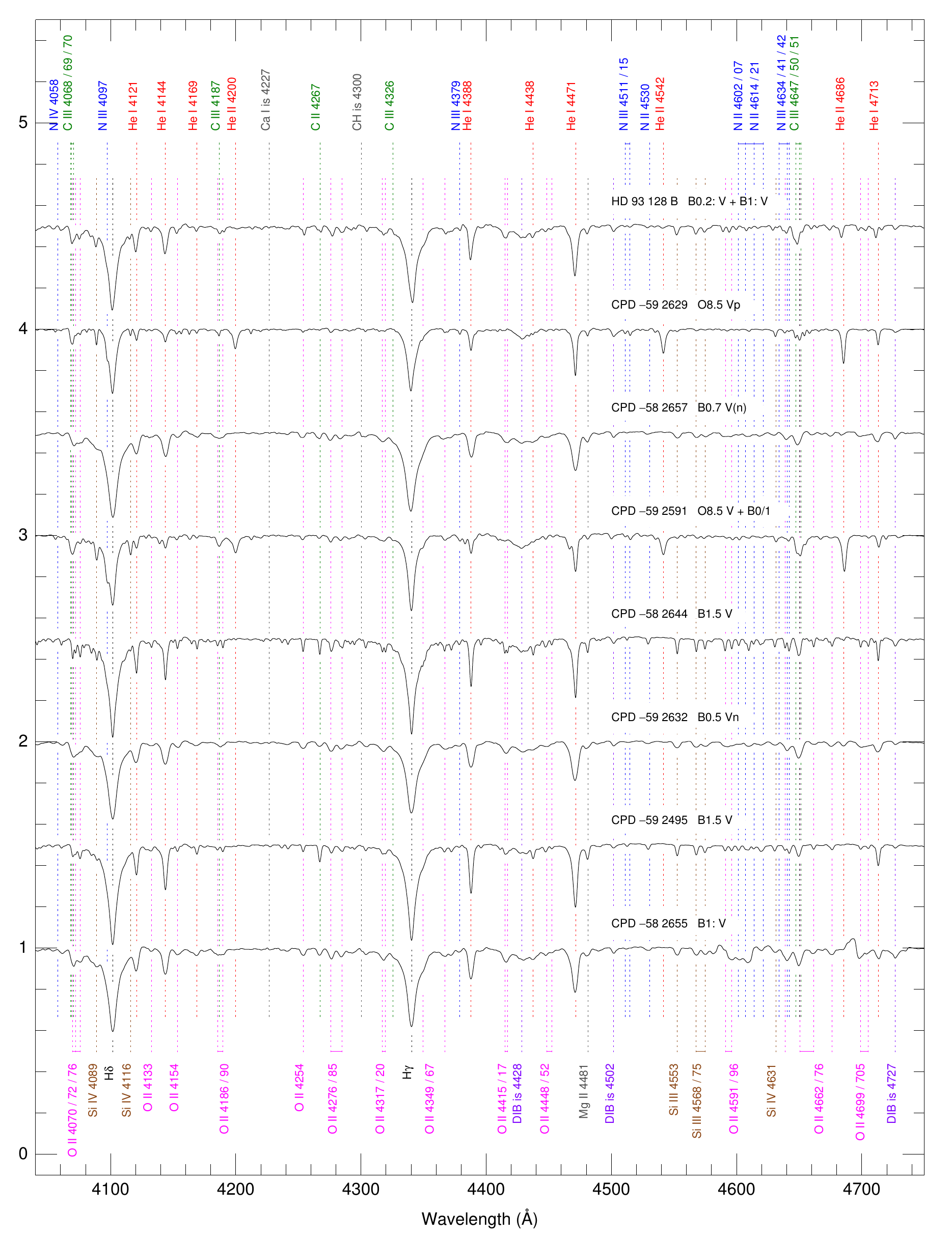}
}
\caption{(Continued).}
\end{figure*}

\addtocounter{figure}{-1}
\begin{figure*}
\centerline{
\includegraphics[width=\linewidth]{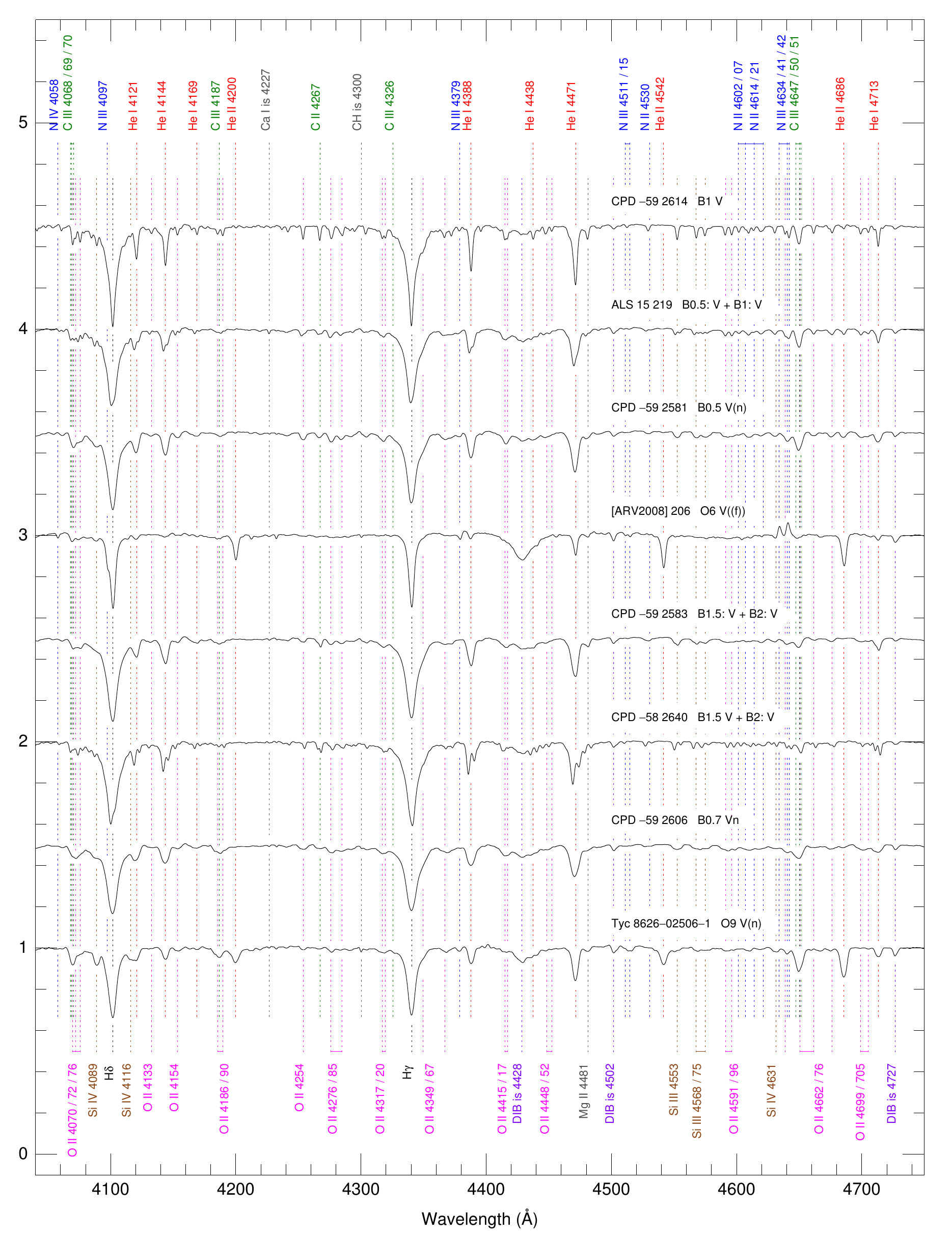}
}
\caption{(Continued).}
\end{figure*}

\addtocounter{figure}{-1}
\begin{figure*}
\centerline{
\includegraphics[width=\linewidth]{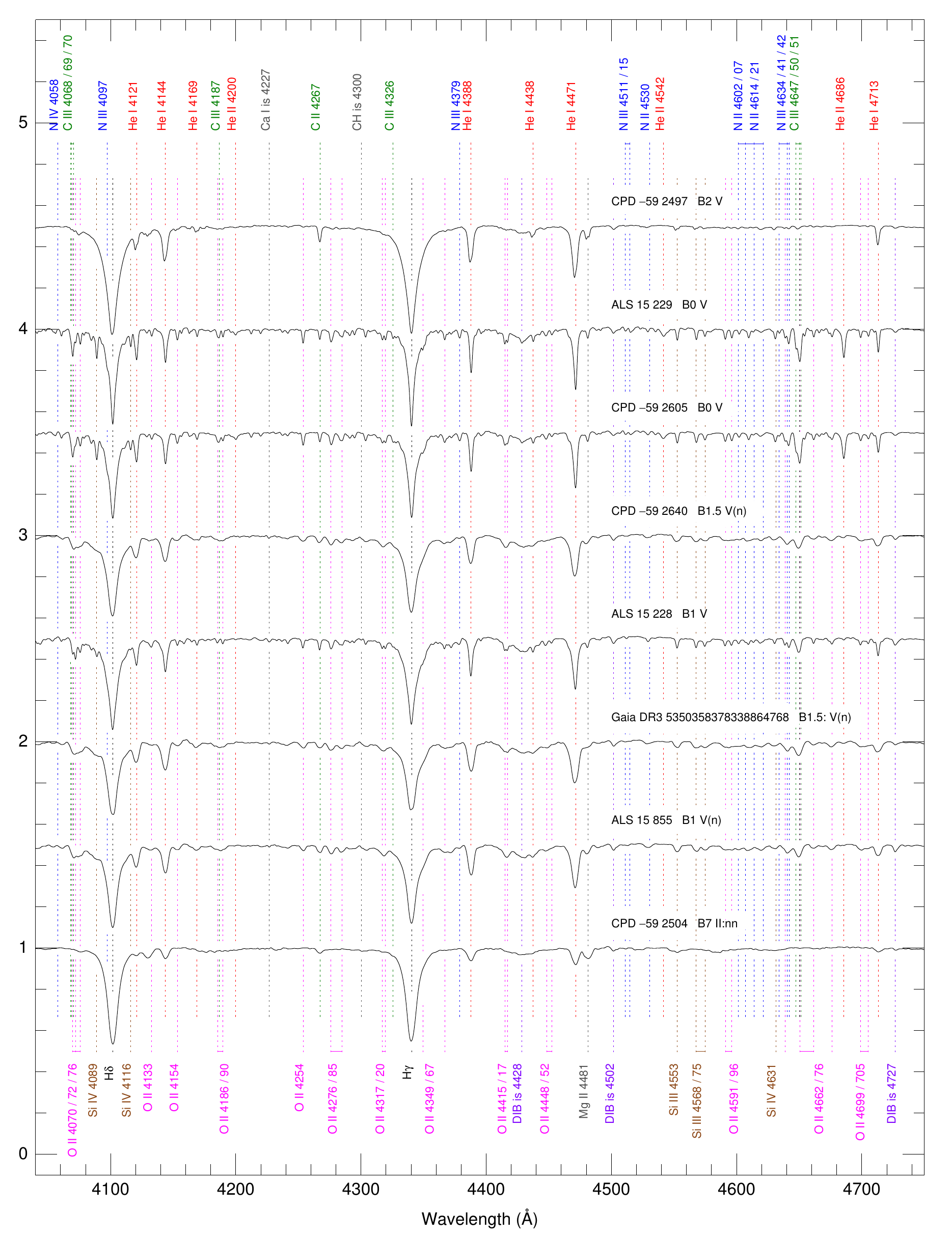}
}
\caption{(Continued).}
\end{figure*}

\addtocounter{figure}{-1}
\begin{figure*}
\centerline{
\includegraphics[width=\linewidth]{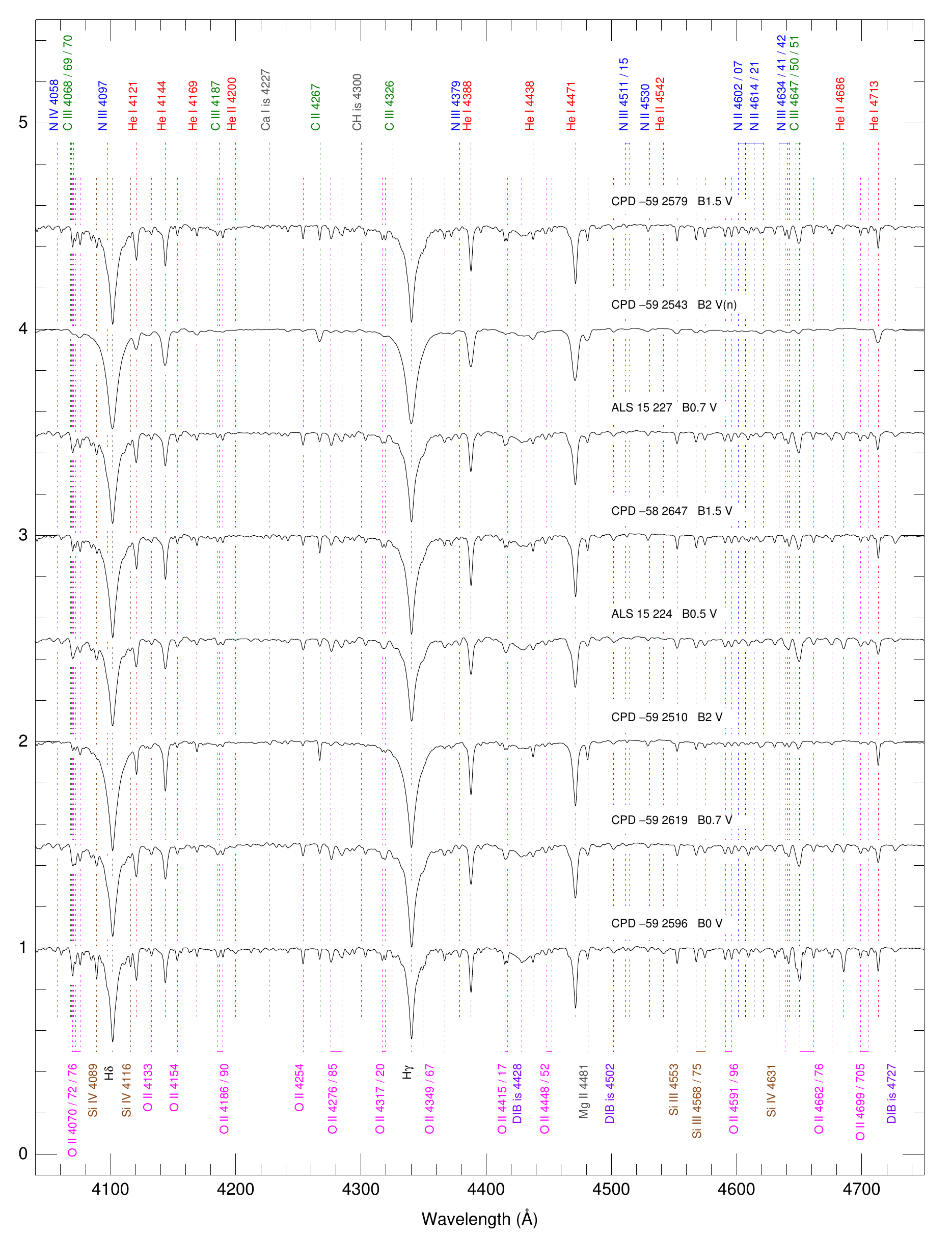}
}
\caption{(Continued).}
\end{figure*}

\addtocounter{figure}{-1}
\begin{figure*}
\centerline{
\includegraphics[width=\linewidth]{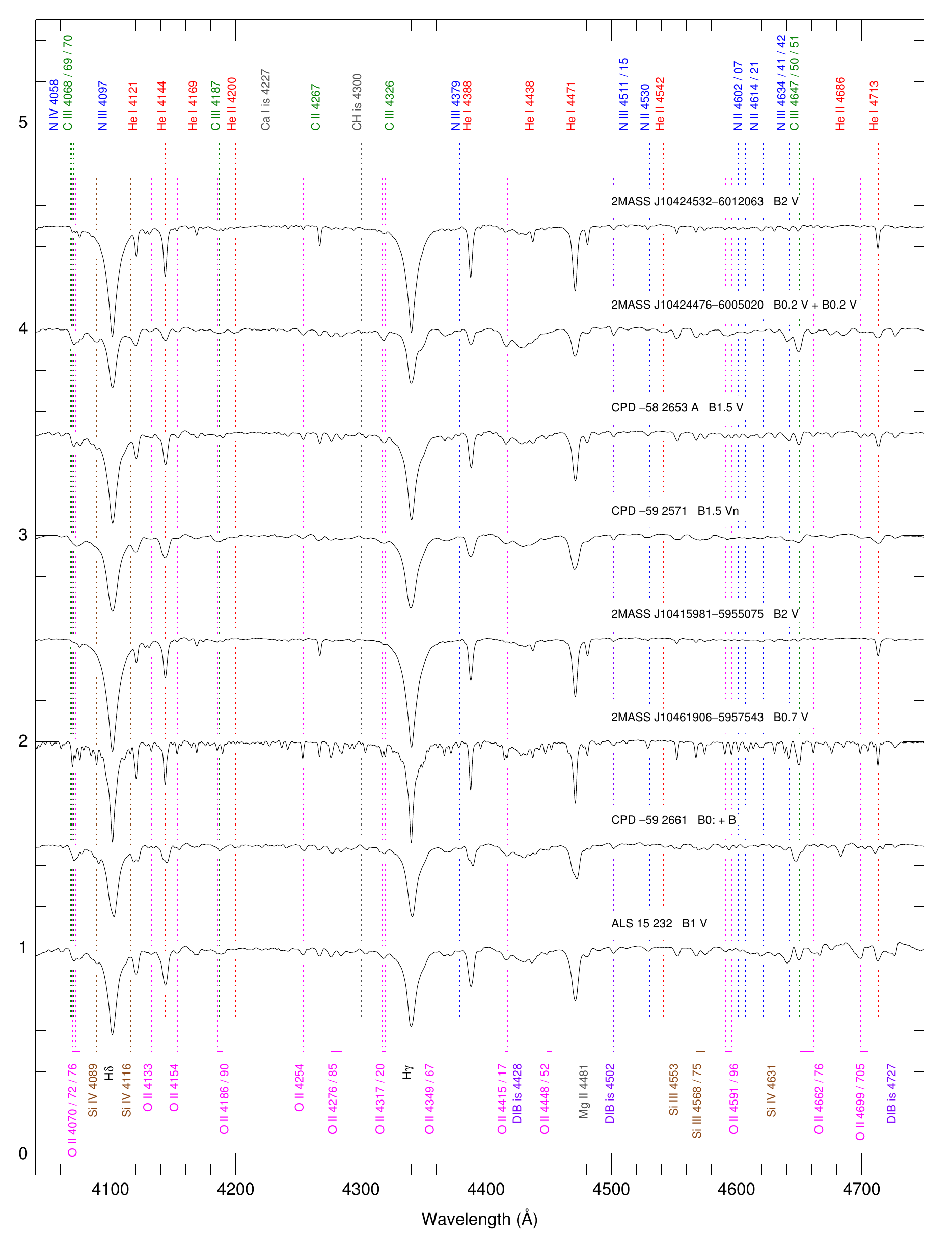}
}
\caption{(Continued).}
\end{figure*}

\addtocounter{figure}{-1}
\begin{figure*}
\centerline{
\includegraphics[width=\linewidth]{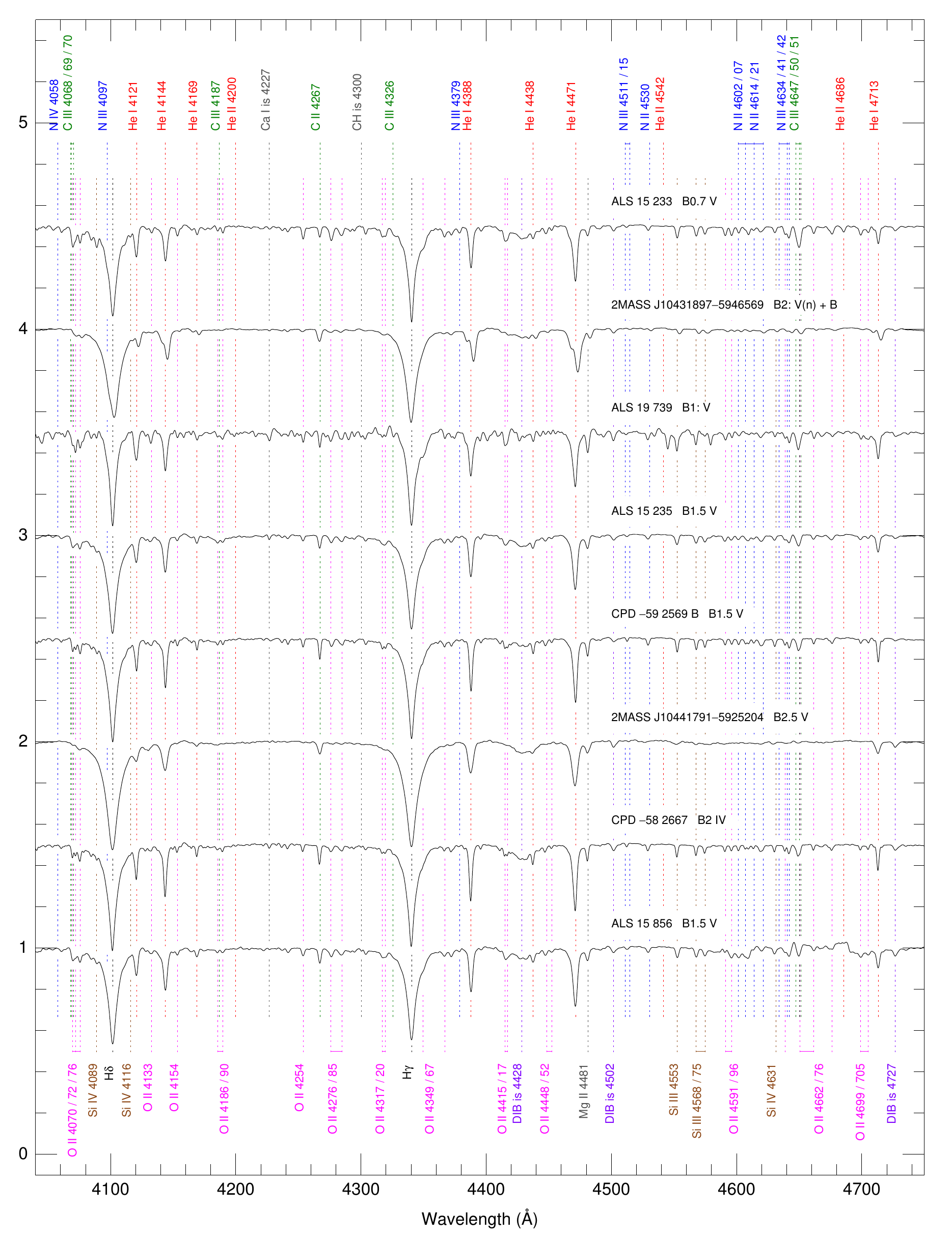}
}
\caption{(Continued).}
\end{figure*}

\addtocounter{figure}{-1}
\begin{figure*}
\centerline{
\includegraphics[width=\linewidth]{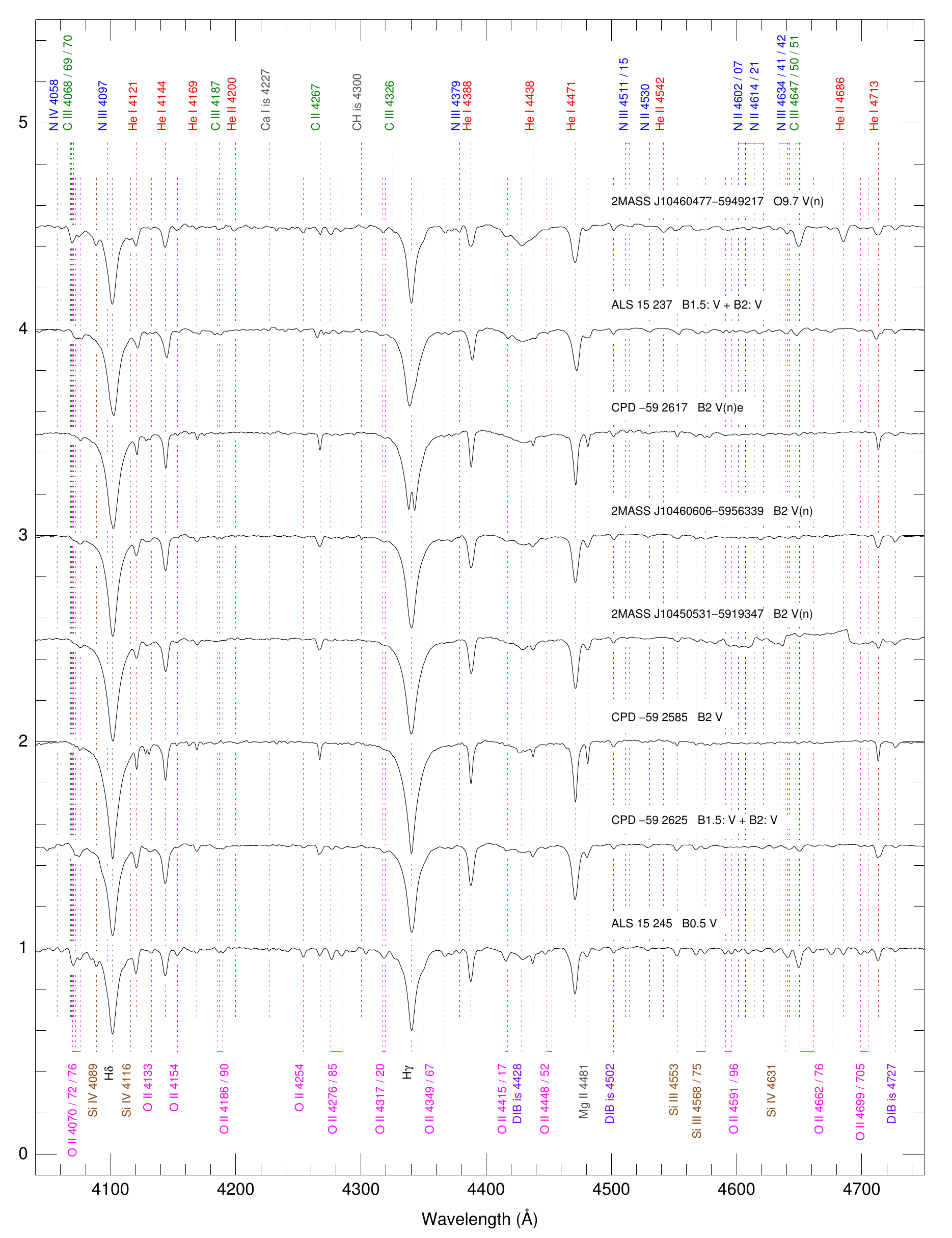}
}
\caption{(Continued).}
\end{figure*}

\addtocounter{figure}{-1}
\begin{figure*}
\centerline{
\includegraphics[width=\linewidth]{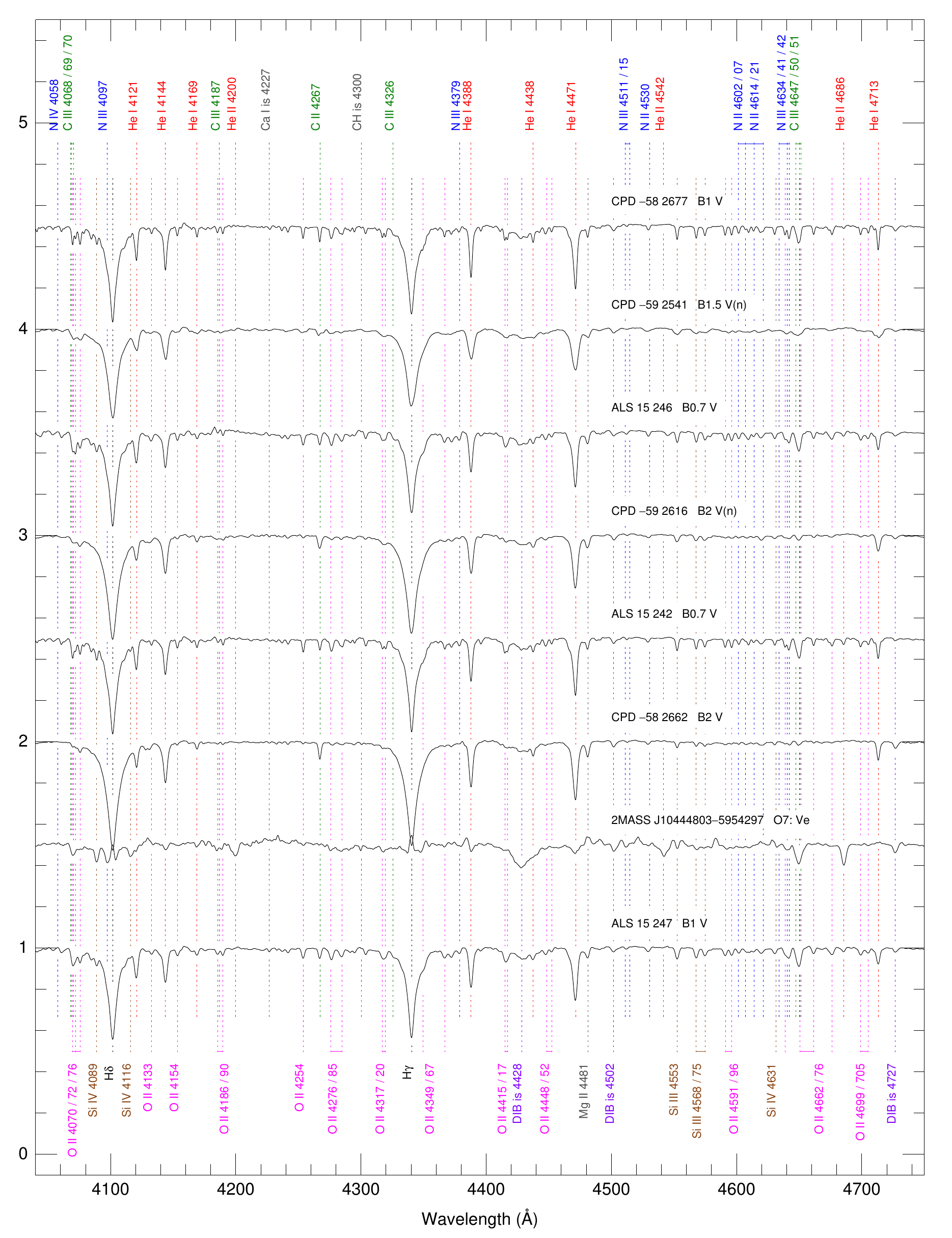}
}
\caption{(Continued).}
\end{figure*}

\addtocounter{figure}{-1}
\begin{figure*}
\centerline{
\includegraphics[width=\linewidth]{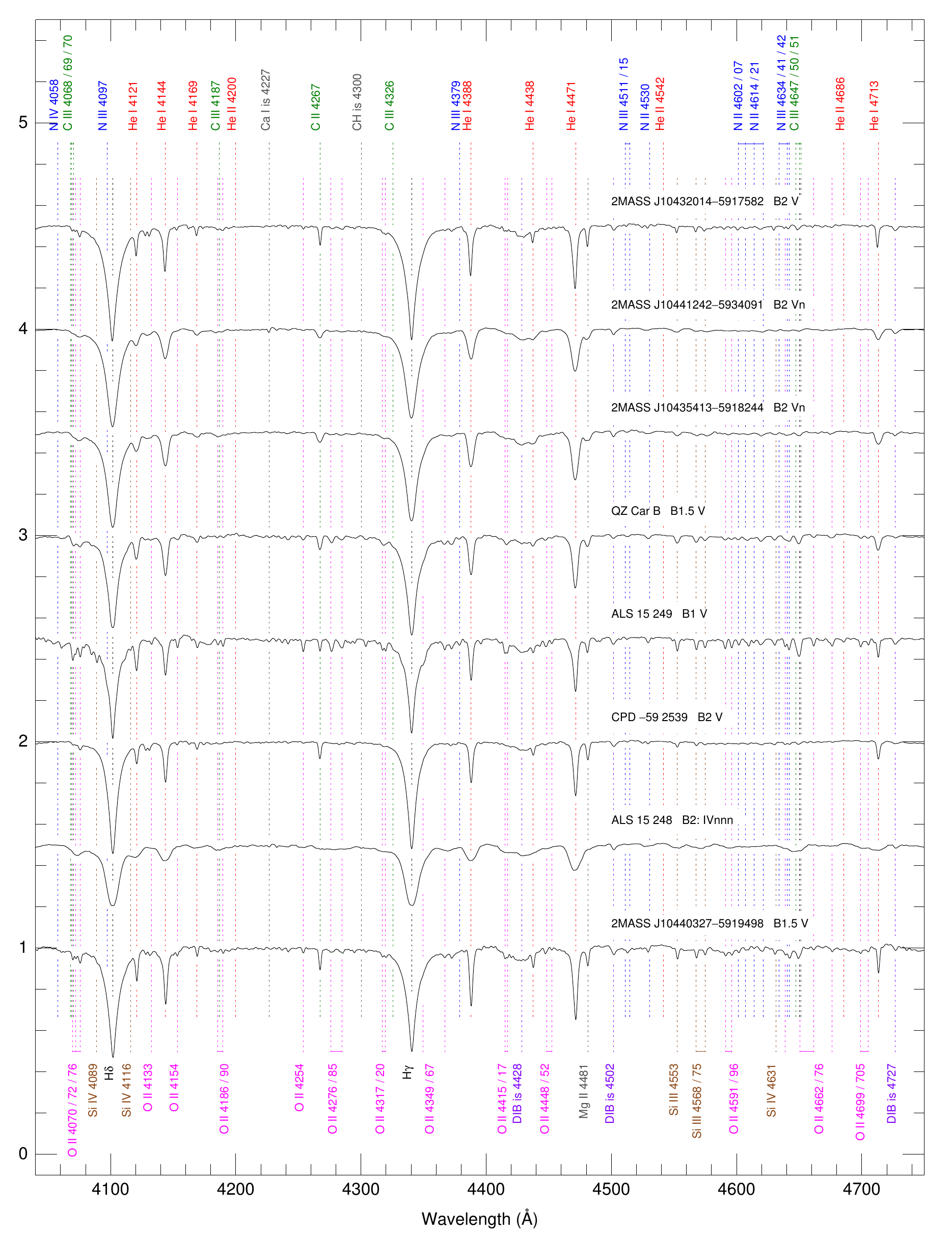}
}
\caption{(Continued).}
\end{figure*}

\addtocounter{figure}{-1}
\begin{figure*}
\centerline{
\includegraphics[width=\linewidth]{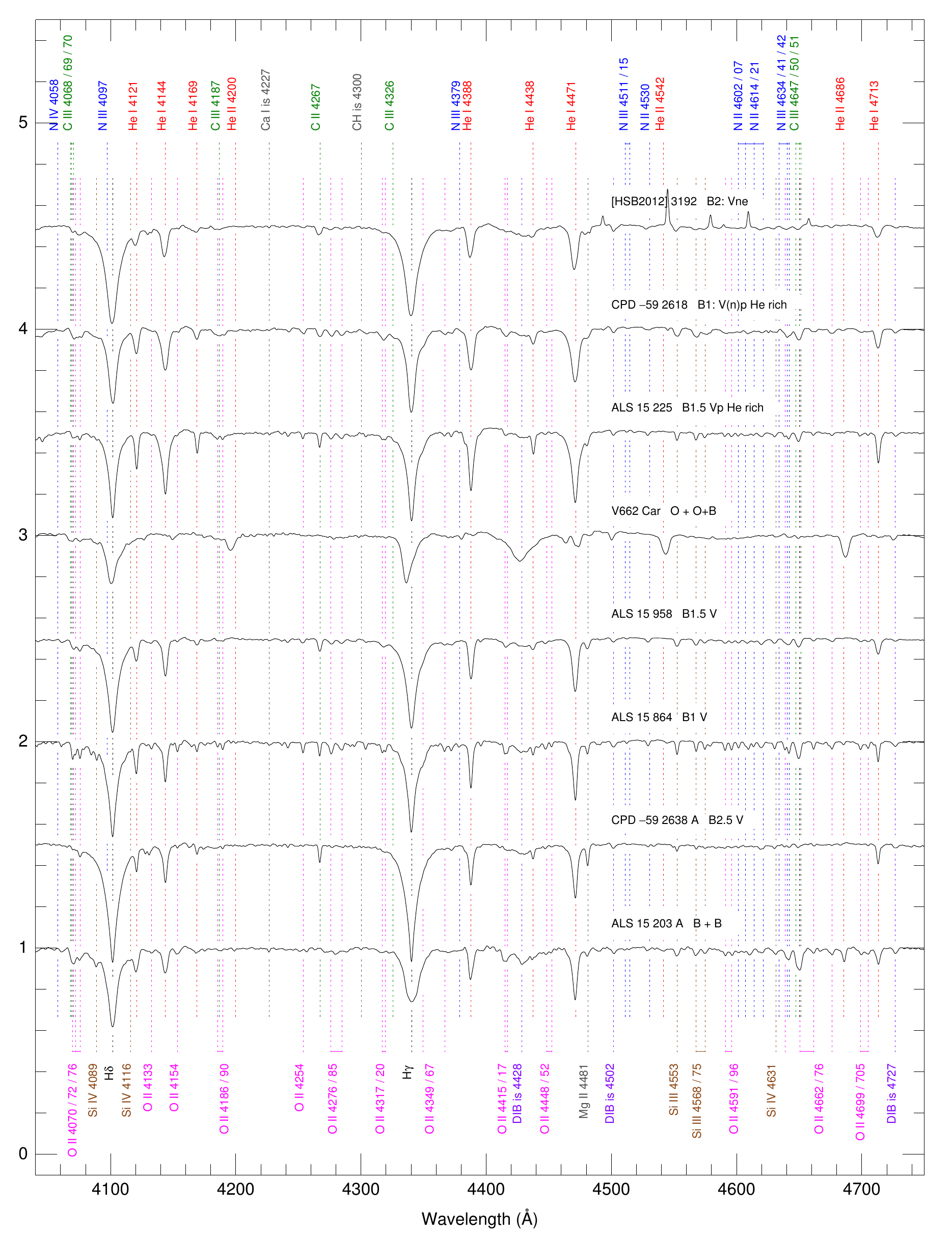}
}
\caption{(Continued).}
\end{figure*}

\addtocounter{figure}{-1}
\begin{figure*}
\centerline{
\includegraphics[width=\linewidth]{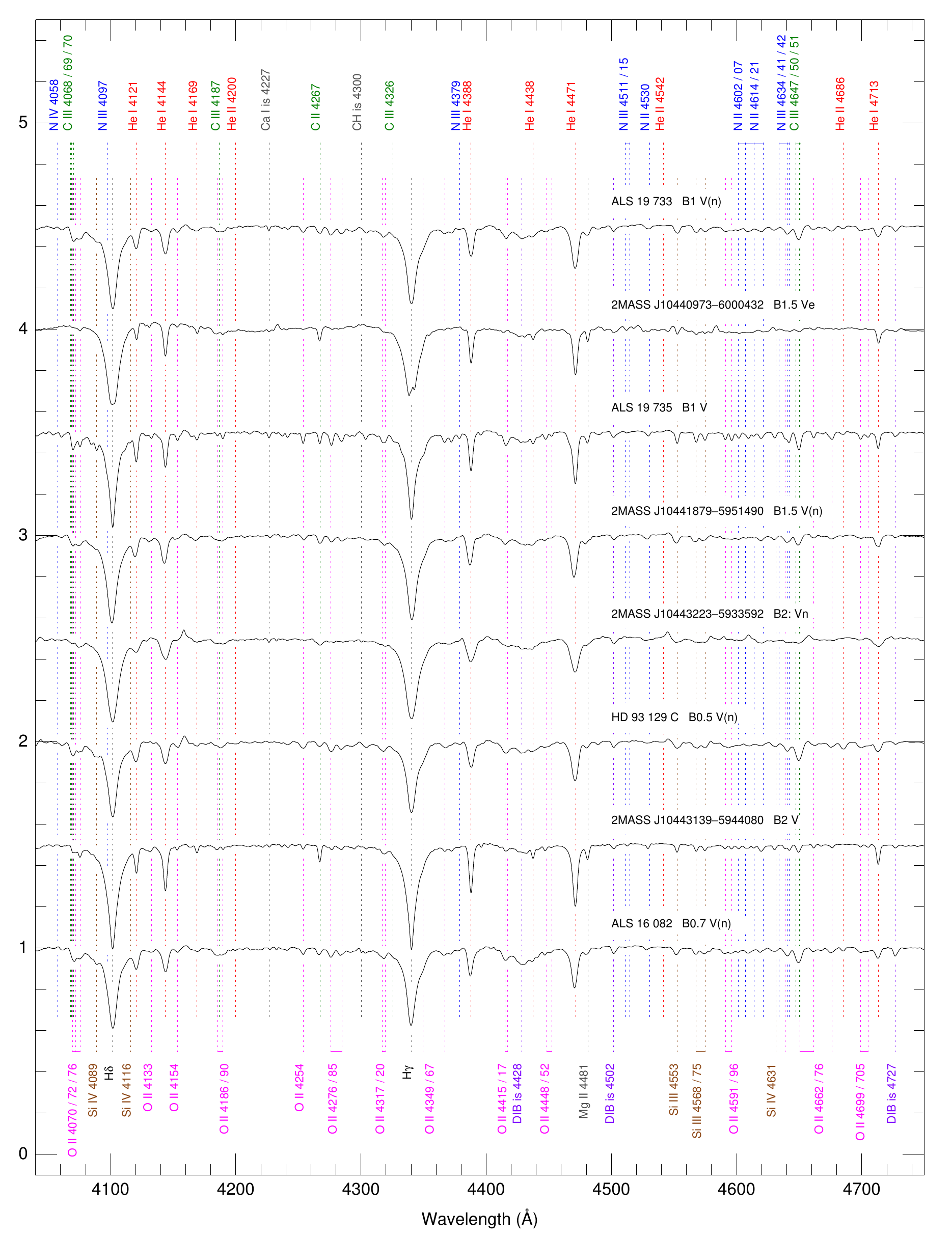}
}
\caption{(Continued).}
\end{figure*}

\addtocounter{figure}{-1}
\begin{figure*}
\centerline{
\includegraphics[width=\linewidth]{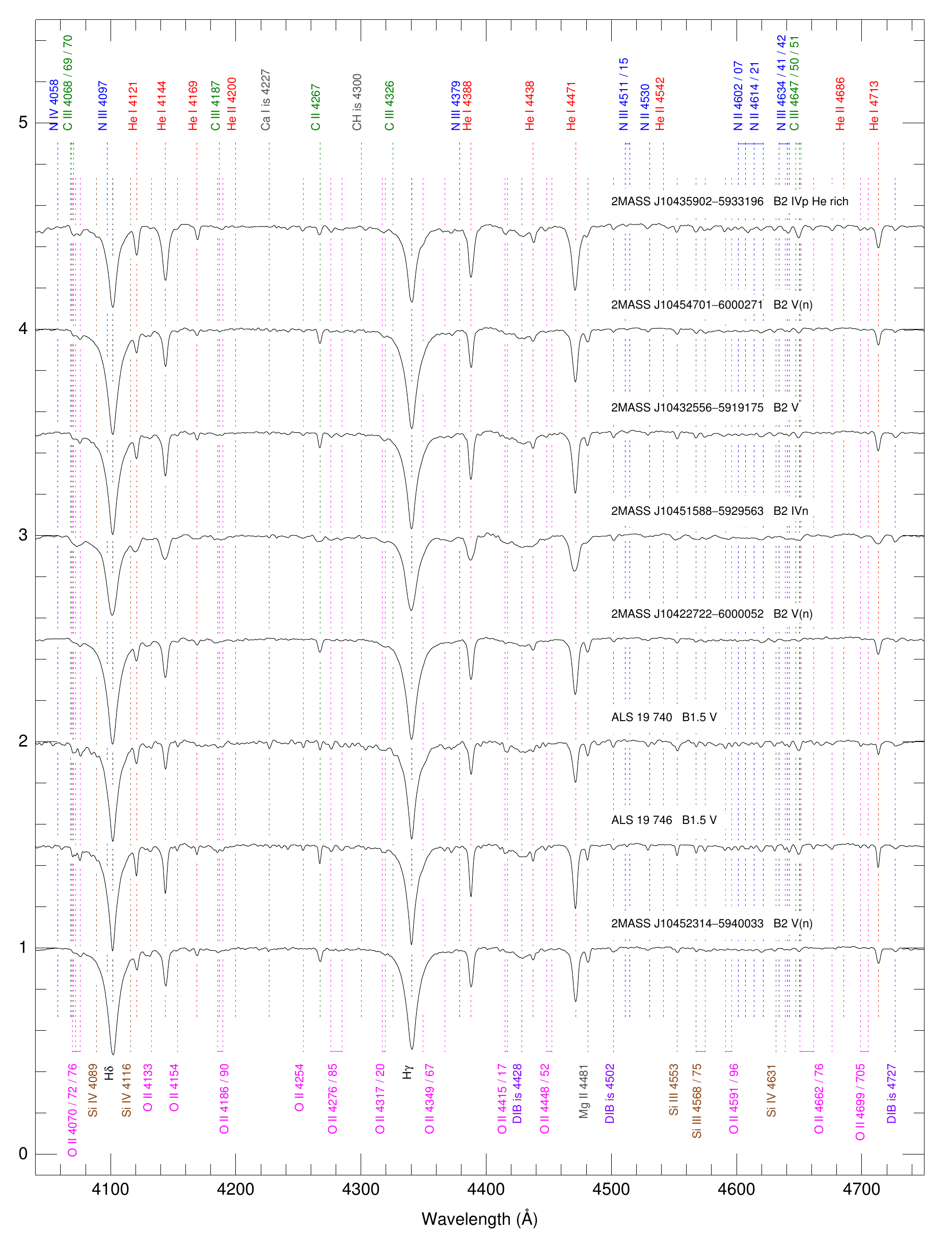}
}
\caption{(Continued).}
\end{figure*}

\addtocounter{figure}{-1}
\begin{figure*}
\centerline{
\includegraphics[width=\linewidth]{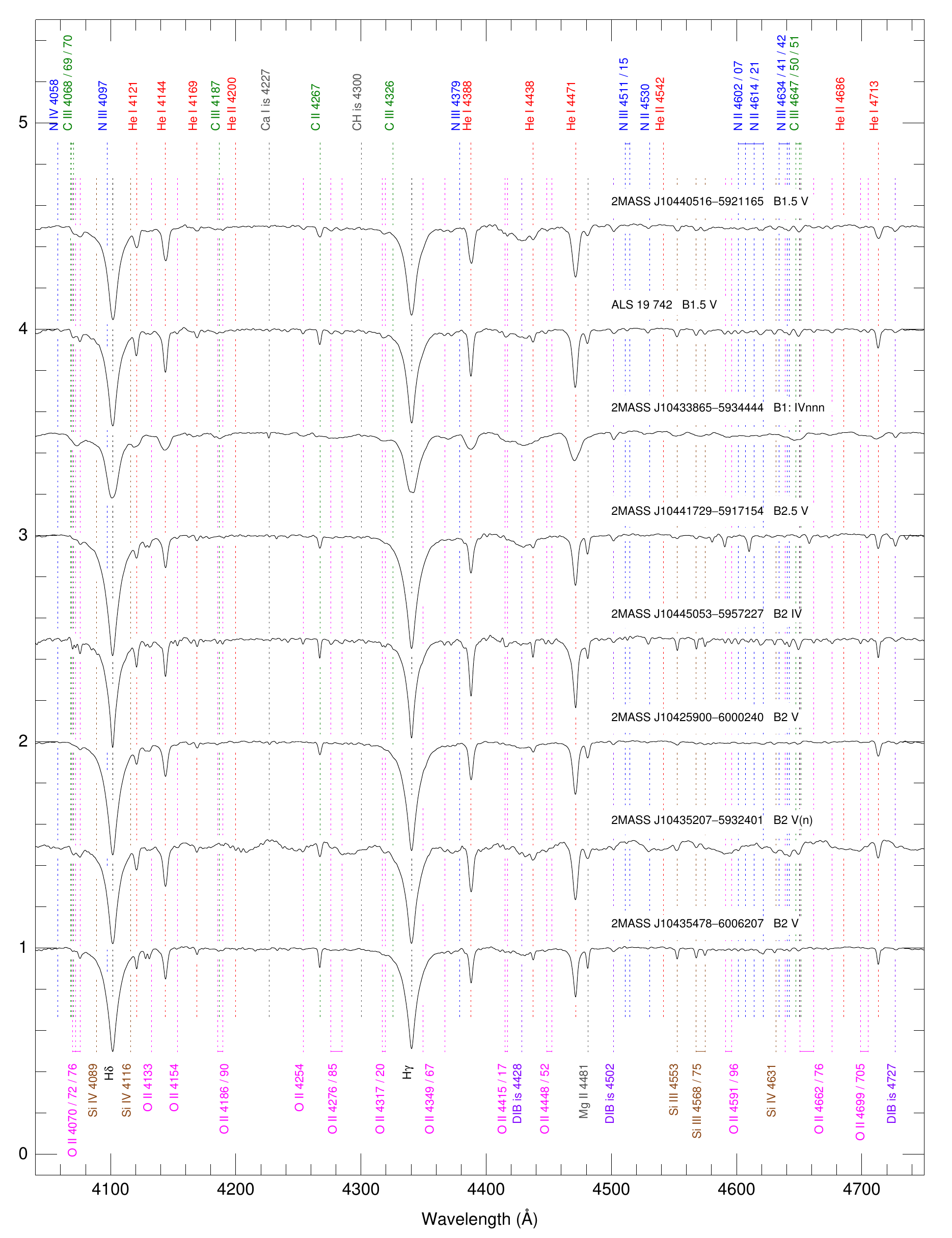}
}
\caption{(Continued).}
\end{figure*}

\addtocounter{figure}{-1}
\begin{figure*}
\centerline{
\includegraphics[width=\linewidth]{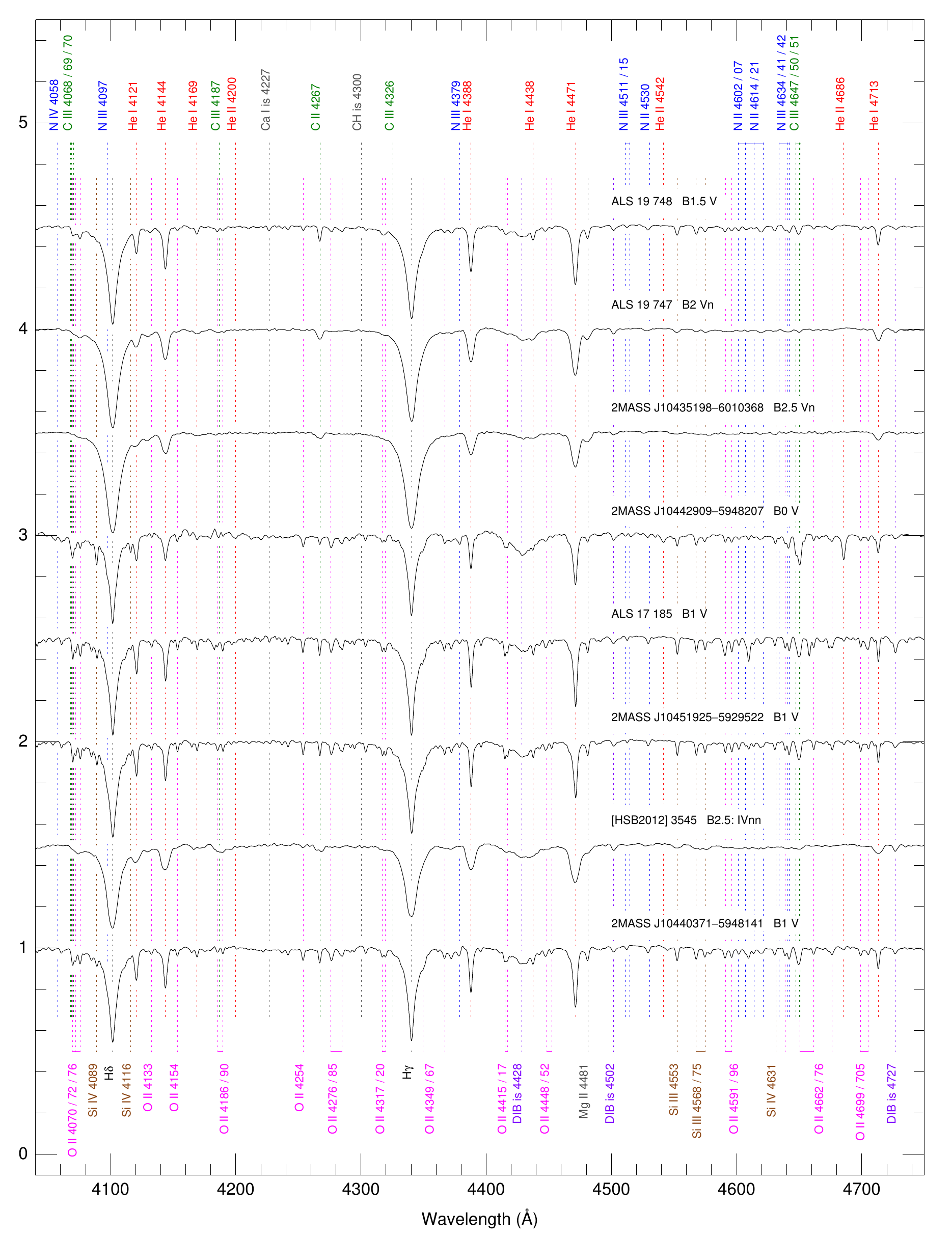}
}
\caption{(Continued).}
\end{figure*}

\addtocounter{figure}{-1}
\begin{figure*}
\centerline{
\includegraphics[width=\linewidth]{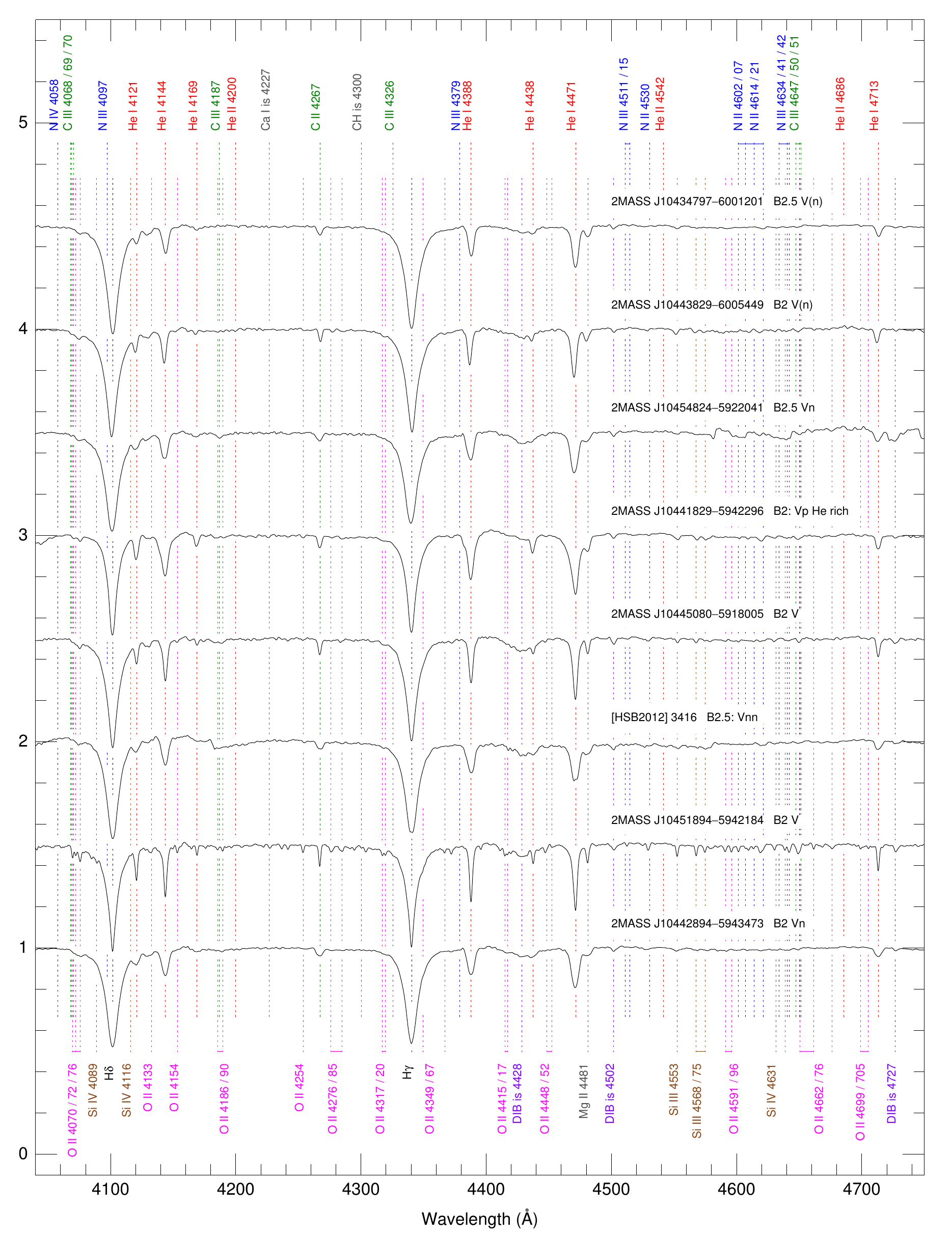}
}
\caption{(Continued).}
\end{figure*}

\addtocounter{figure}{-1}
\begin{figure*}
\centerline{
\includegraphics[width=\linewidth]{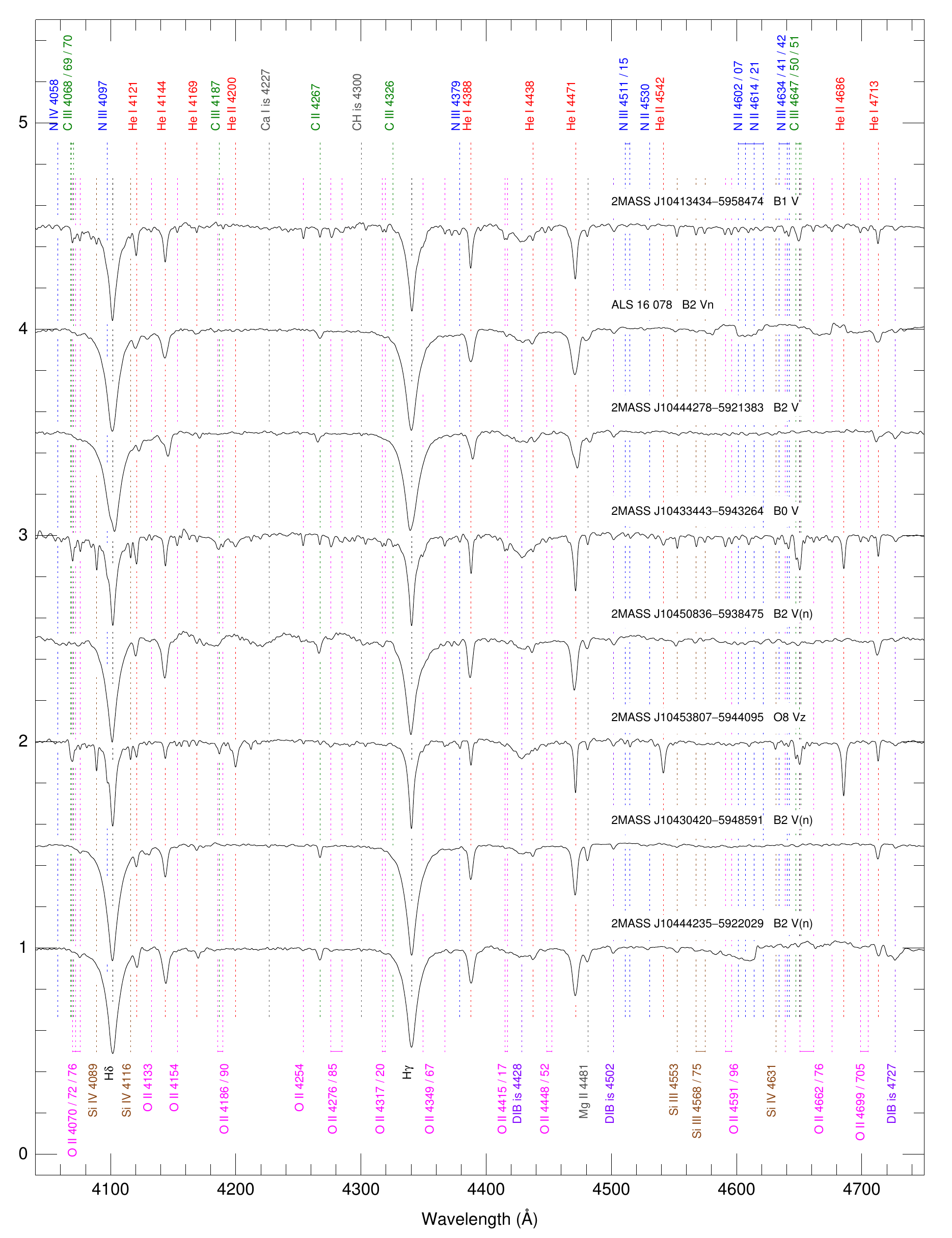}
}
\caption{(Continued).}
\end{figure*}

\addtocounter{figure}{-1}
\begin{figure*}
\centerline{
\includegraphics[width=\linewidth]{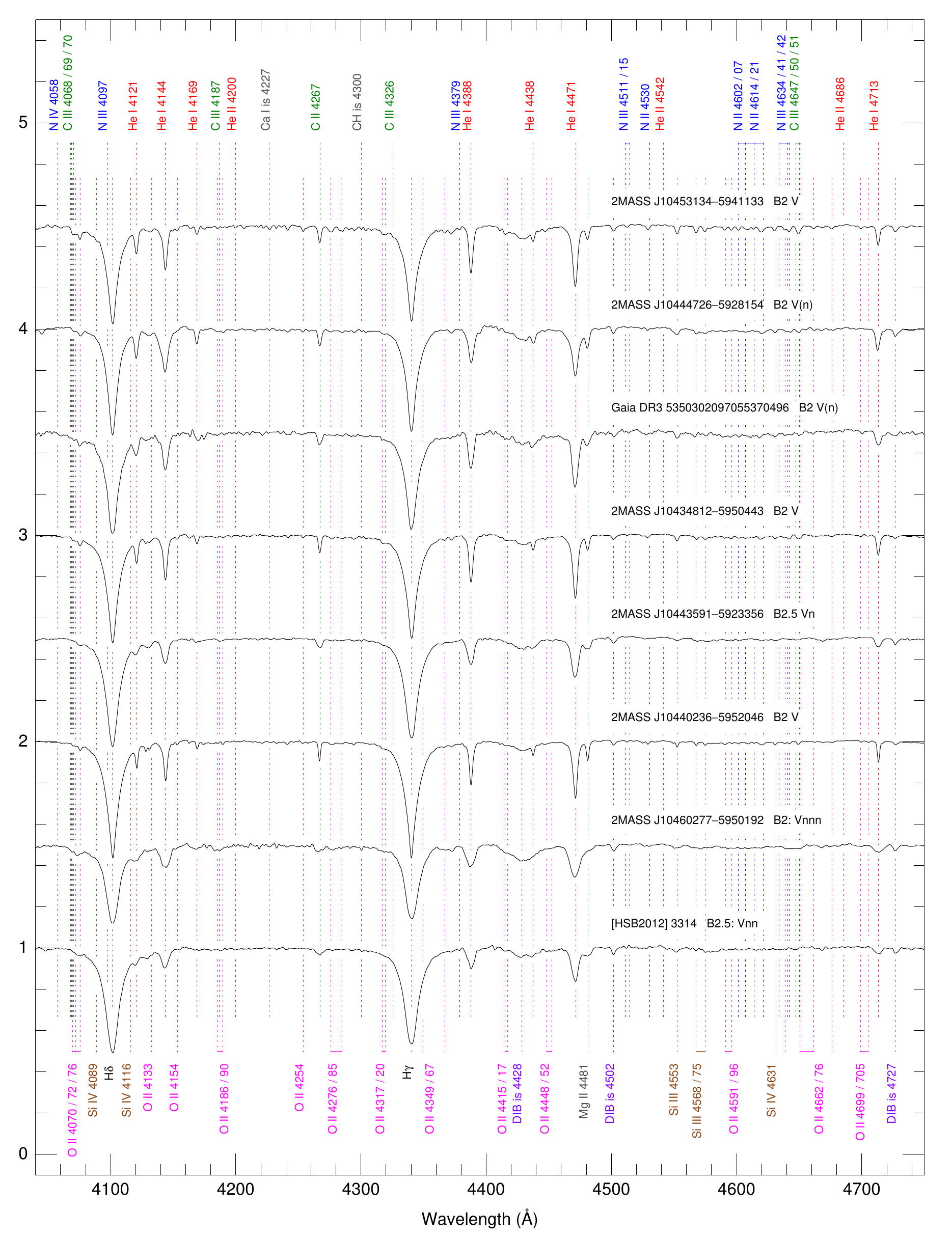}
}
\caption{(Continued).}
\end{figure*}

\addtocounter{figure}{-1}
\begin{figure*}
\centerline{
\includegraphics[width=\linewidth]{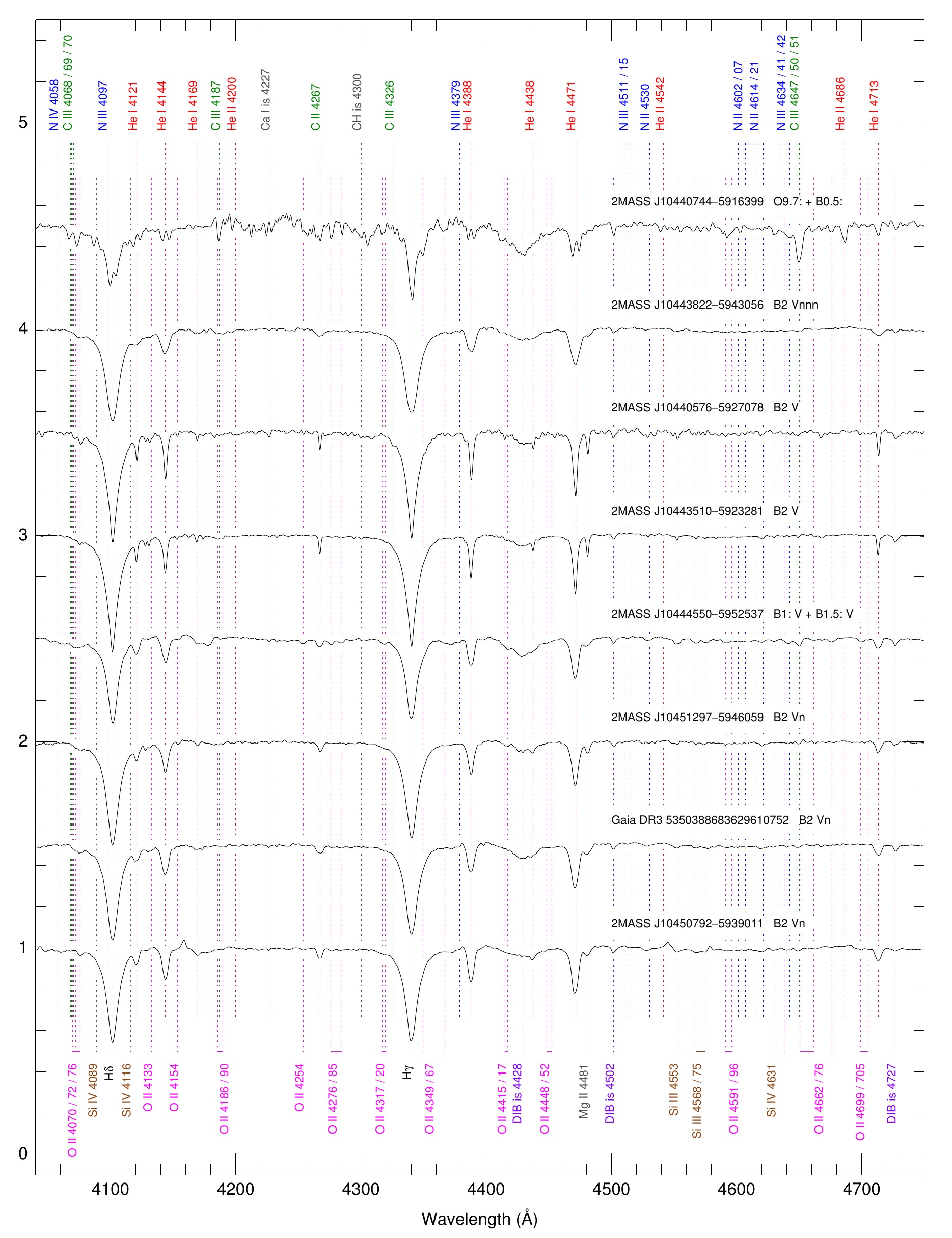}
}
\caption{(Continued).}
\end{figure*}

\addtocounter{figure}{-1}
\begin{figure*}
\centerline{
\includegraphics[width=\linewidth]{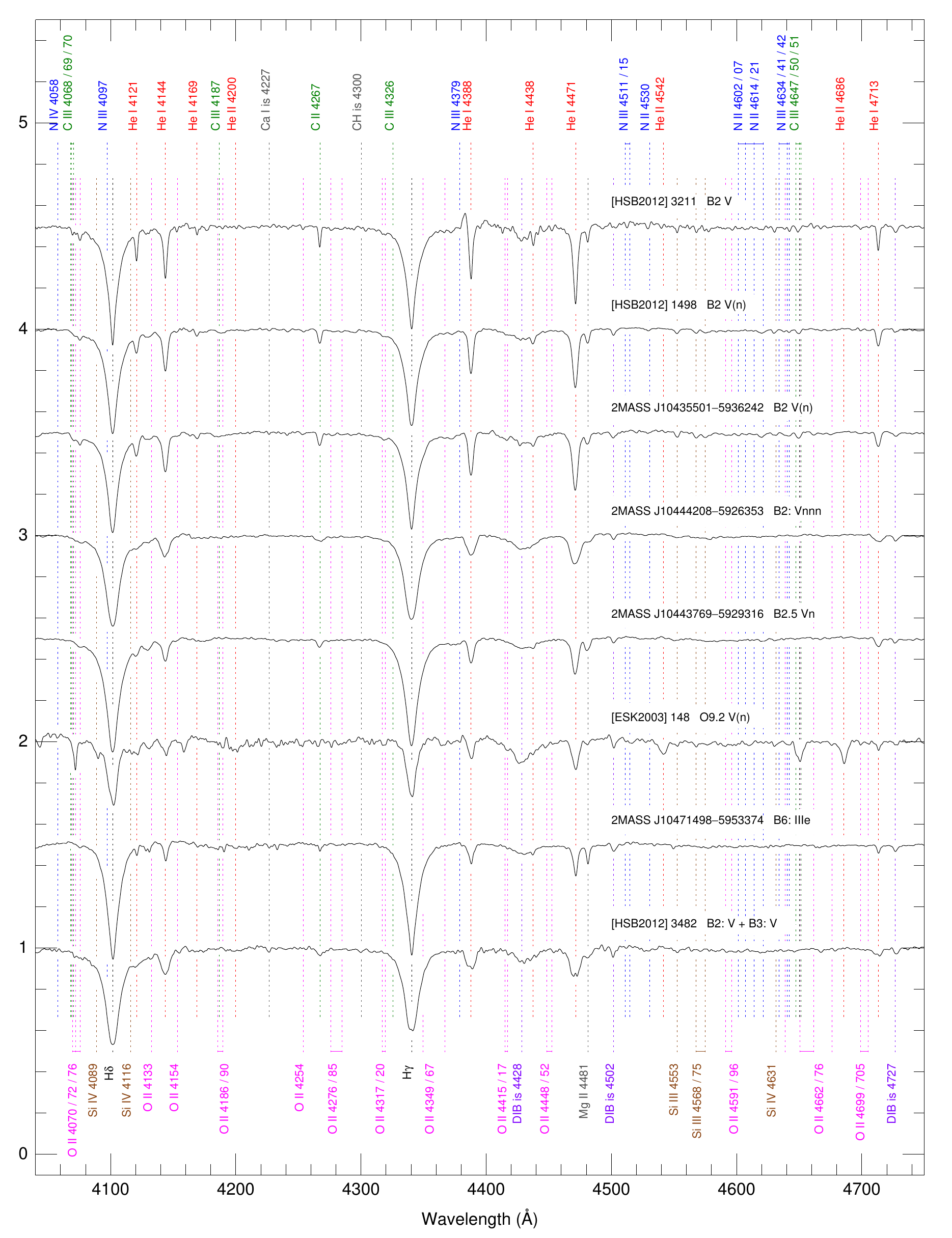}
}
\caption{(Continued).}
\end{figure*}

\addtocounter{figure}{-1}
\begin{figure*}
\centerline{
\includegraphics[width=\linewidth]{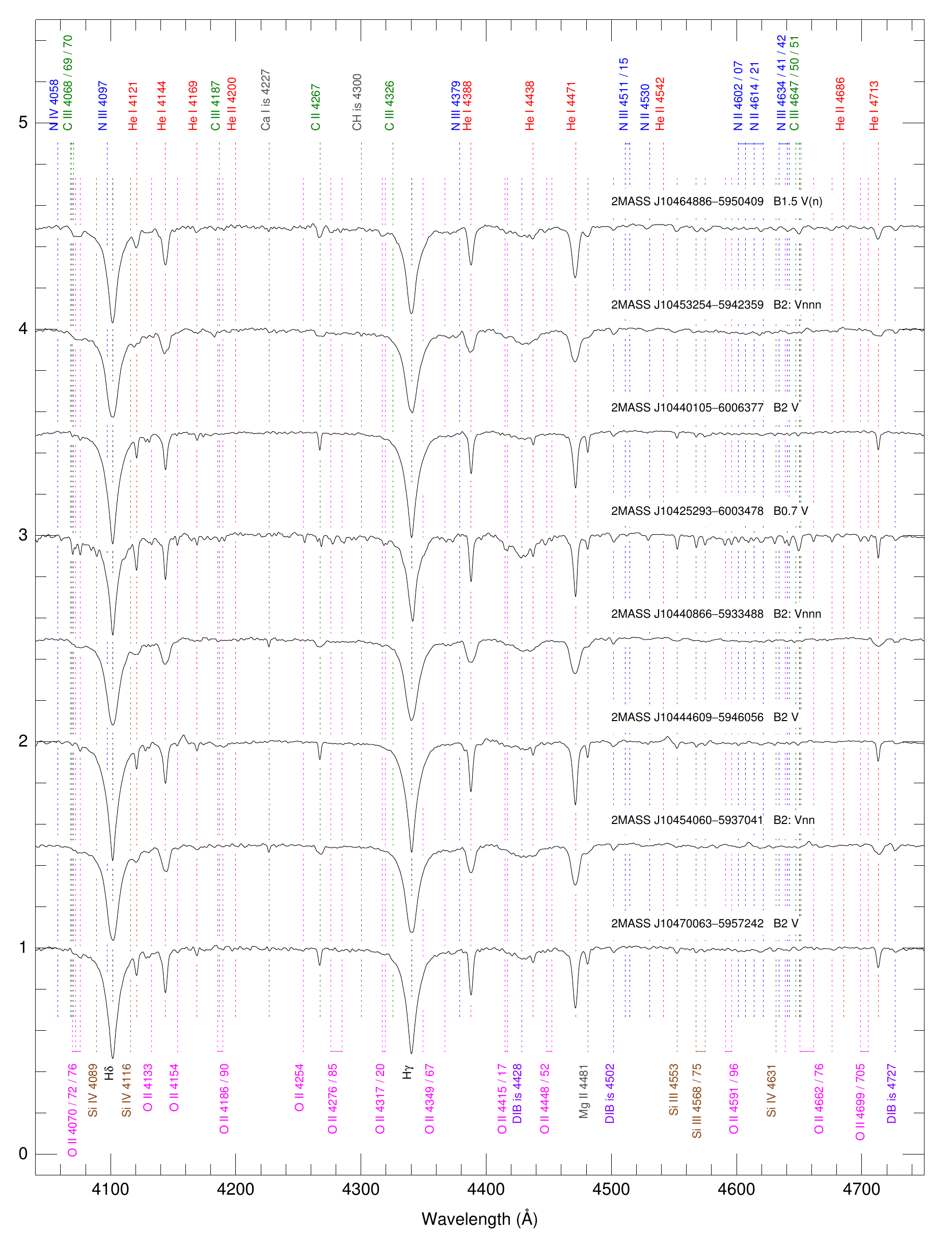}
}
\caption{(Continued).}
\end{figure*}

\addtocounter{figure}{-1}
\begin{figure*}
\centerline{
\includegraphics[width=\linewidth]{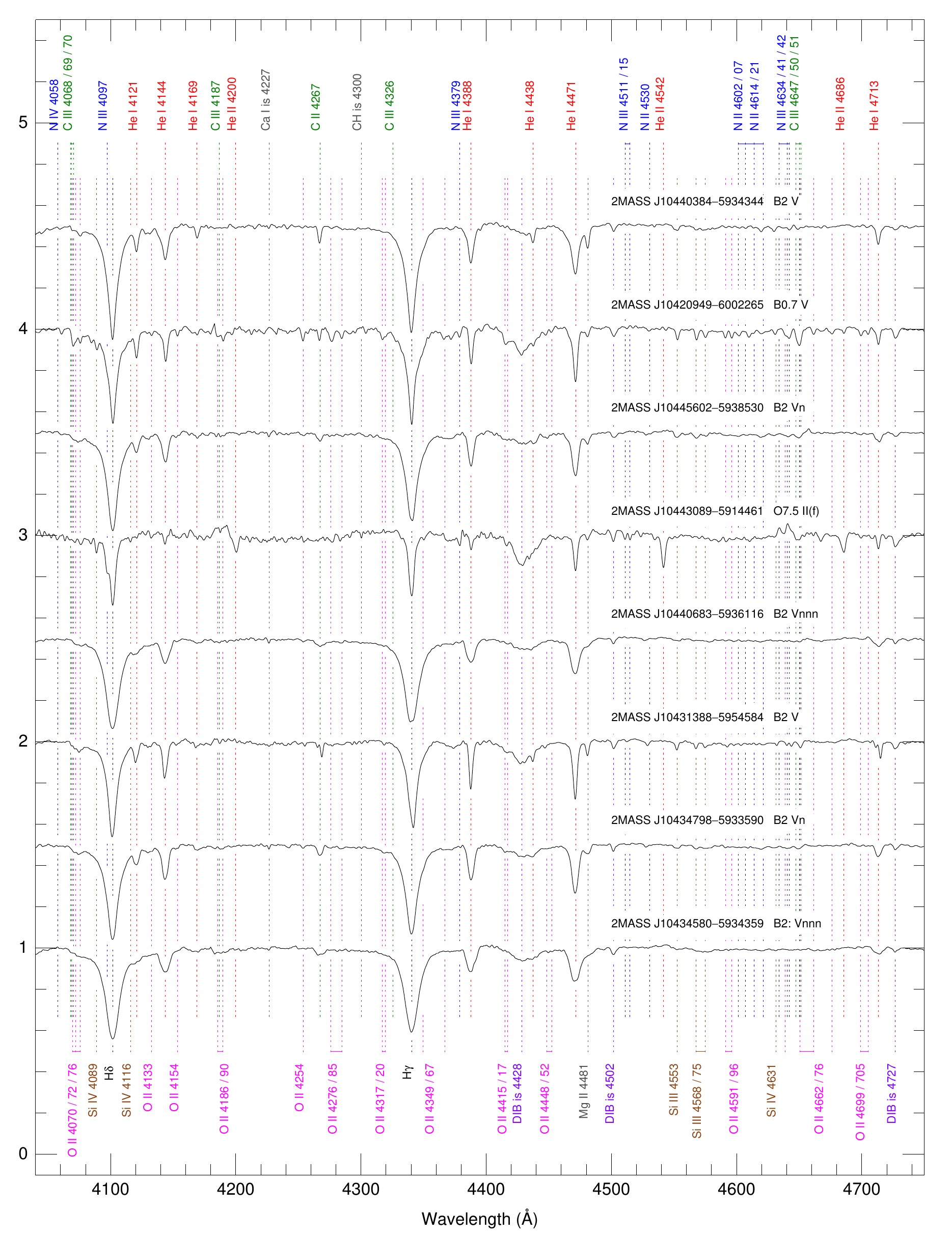}
}
\caption{(Continued).}
\end{figure*}

\addtocounter{figure}{-1}
\begin{figure*}
\centerline{
\includegraphics[width=\linewidth]{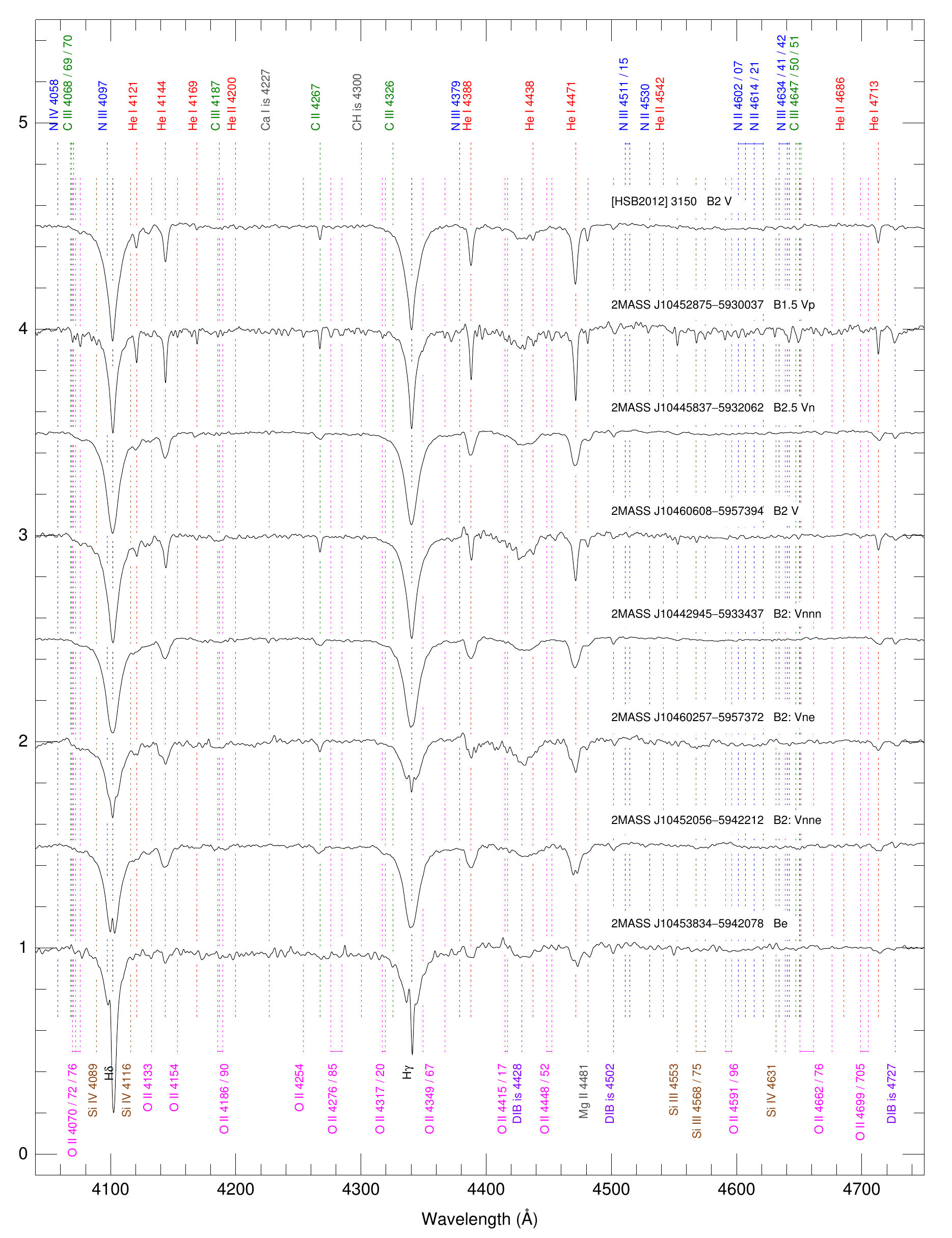}
}
\caption{(Continued).}
\end{figure*}

\addtocounter{figure}{-1}
\begin{figure*}
\centerline{
\includegraphics[width=\linewidth]{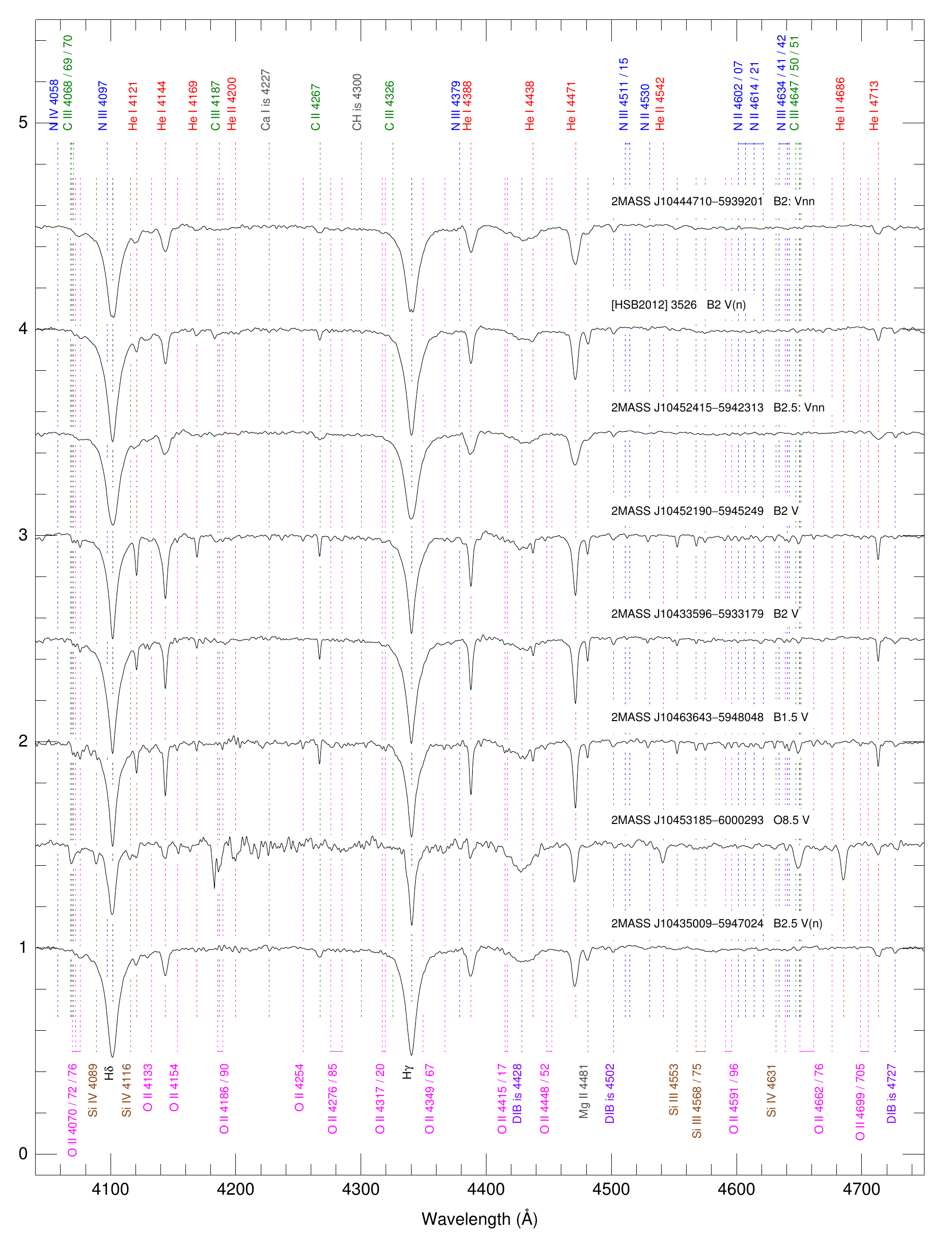}
}
\caption{(Continued).}
\end{figure*}

\addtocounter{figure}{-1}
\begin{figure*}
\centerline{
\includegraphics[width=\linewidth]{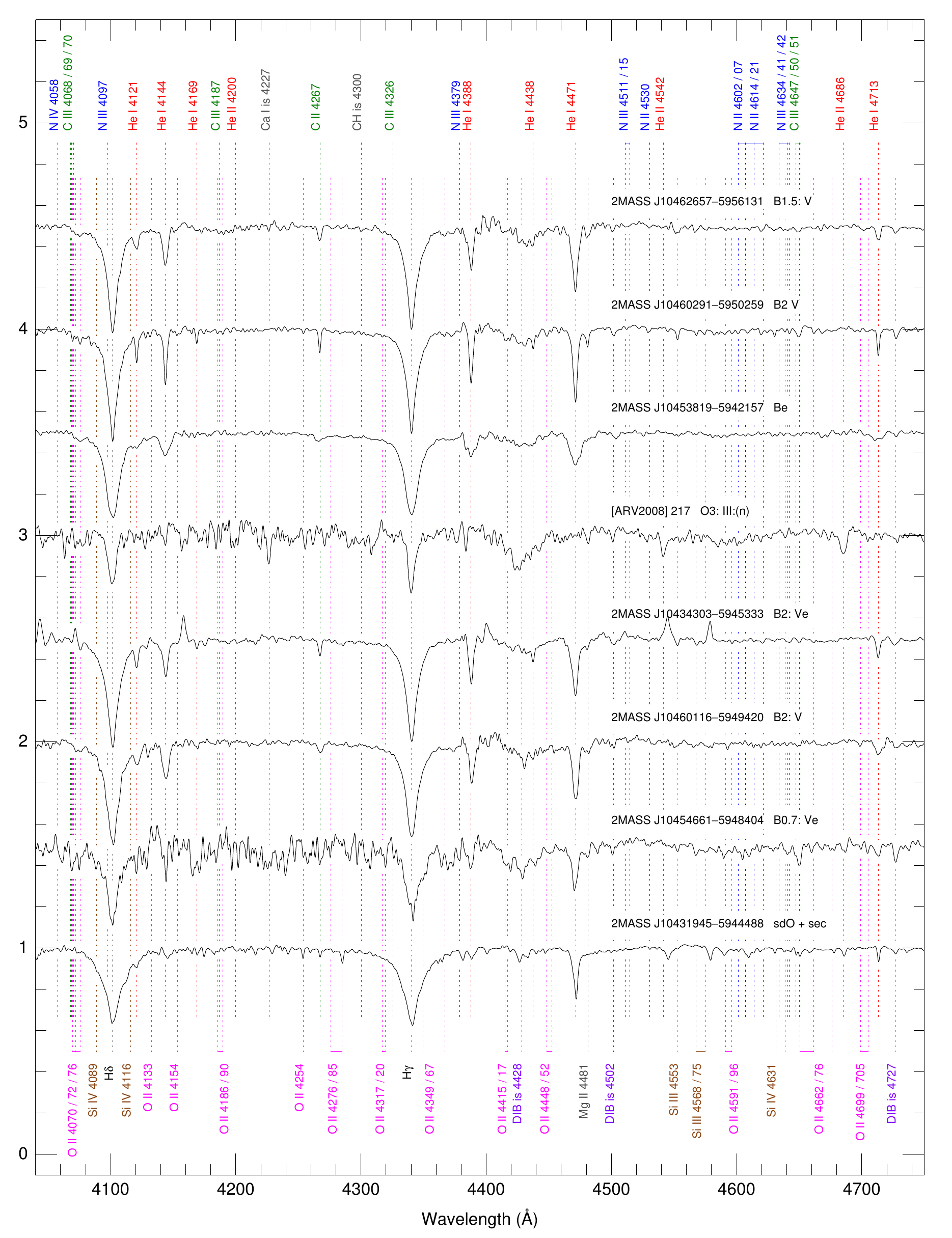}
}
\caption{(Continued).}
\end{figure*}

\begin{figure*}
\centerline{
\includegraphics[width=\linewidth]{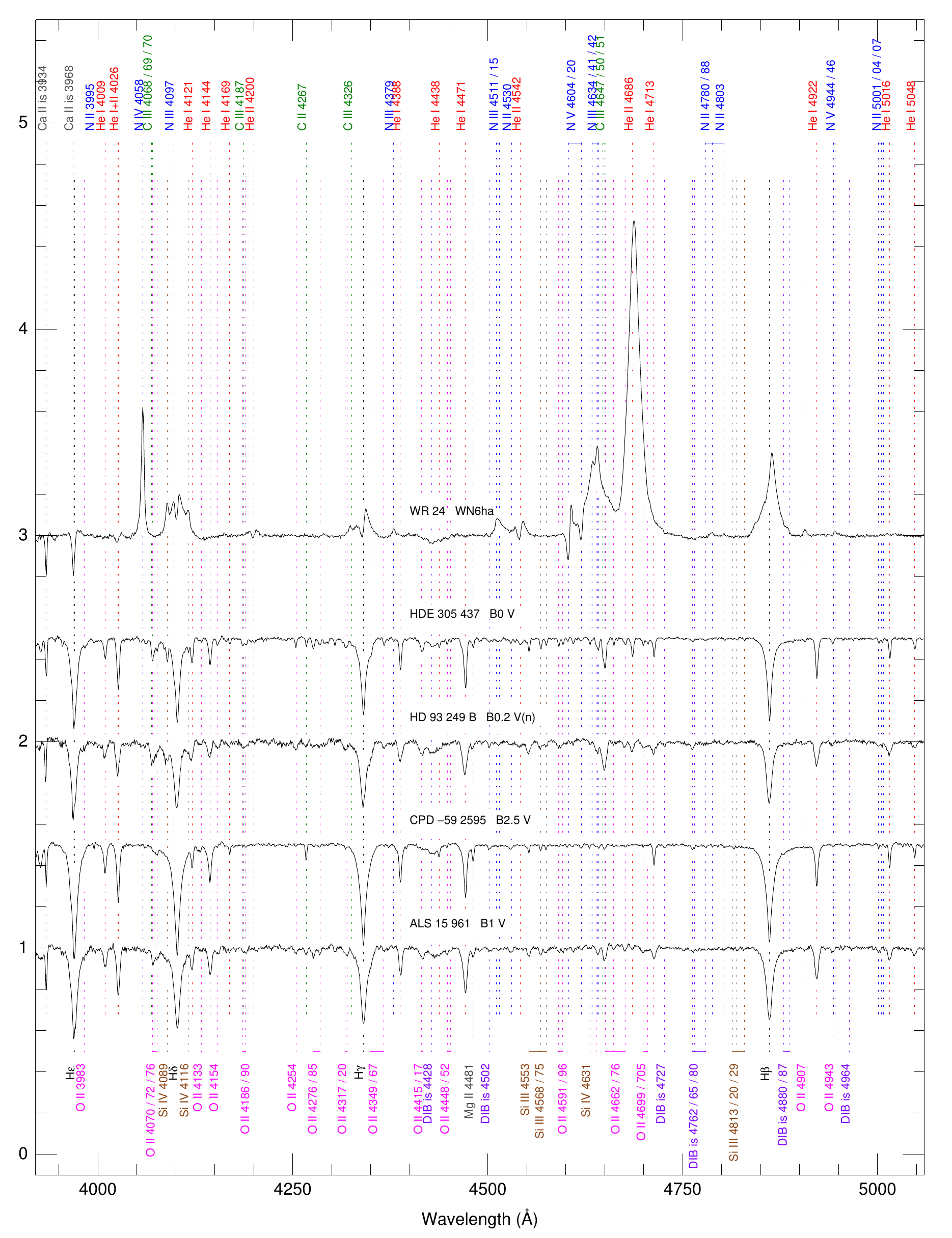}
}
\caption{GOSSS spectra shown at a resolution $R = 2500$.}
\label{GOSSS_spectra}
\end{figure*}

\addtocounter{figure}{-1}
\begin{figure*}
\centerline{
\includegraphics[width=\linewidth]{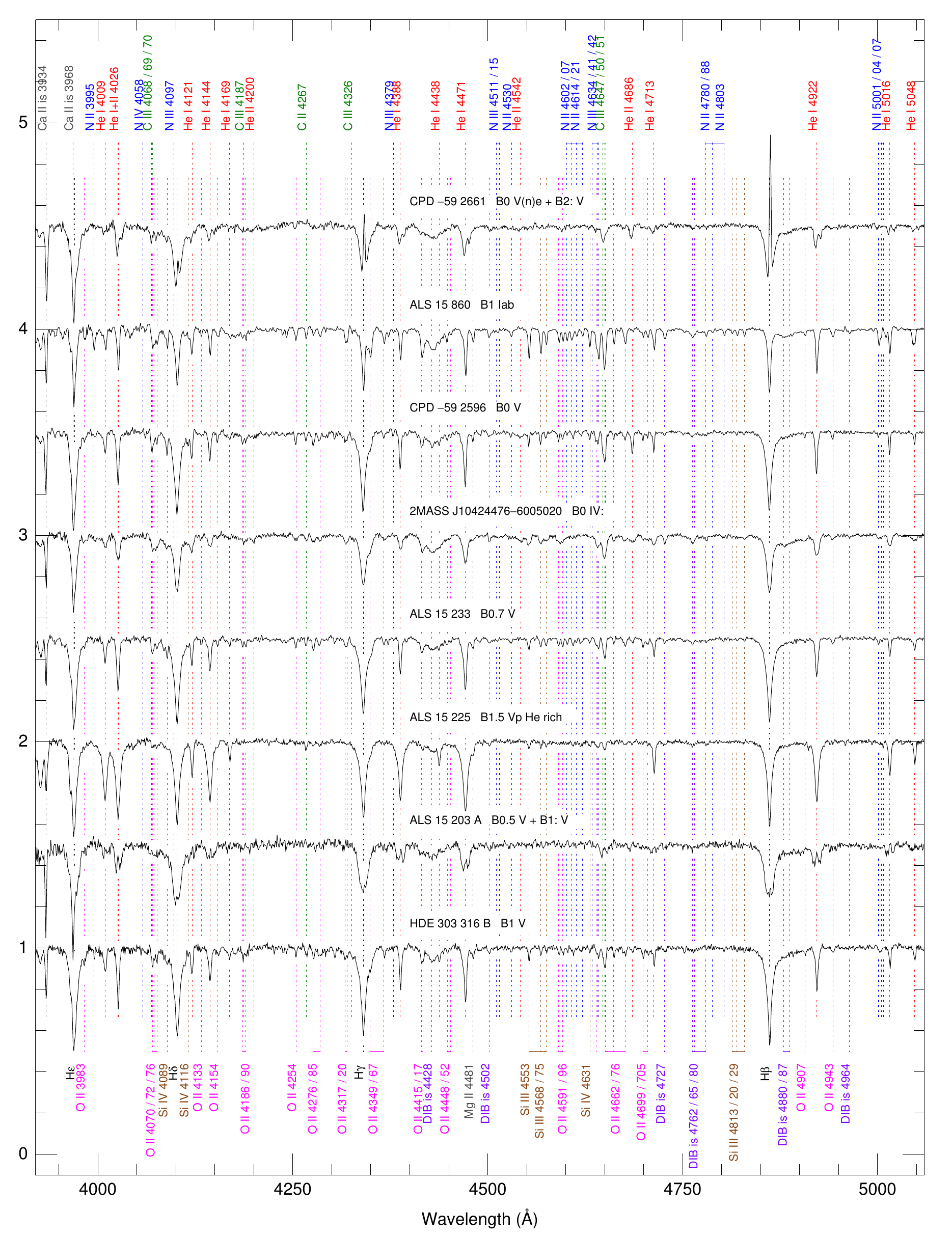}
}
\caption{(Continued).}
\end{figure*}

\addtocounter{figure}{-1}
\begin{figure*}
\centerline{
\includegraphics[width=\linewidth]{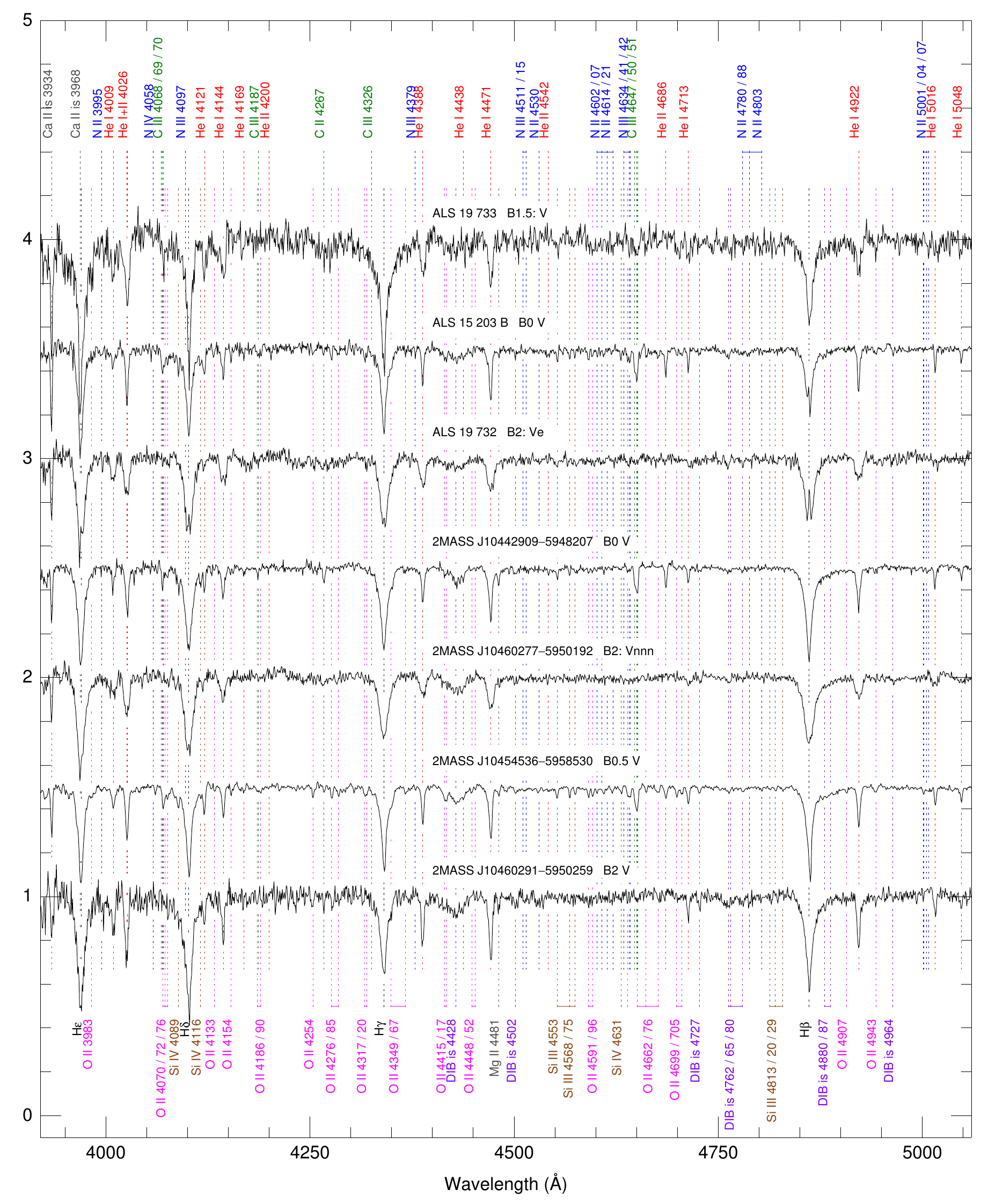}
}
\caption{(Continued).}
\end{figure*}

\end{appendix}

\end{document}

%% file: table_A11.tex
\begin{table}
\caption{Stars in the field of the Carina Nebula analyzed in this paper sorted by \GGc.}
\label{table_main}
\centerline{
\tiny
\addtolength{\tabcolsep}{-1mm}
\begin{tabular}{lccccclrrrc}
\hline
\hline
Name                                &     RA       &      dec       &    GOS/GBS/GWS ID    &         GES ID      &          Gaia ID          & \mci{ALS ID}&      \mci{\GGc}       &      \mci{\BPRP}      & \mci{\pic (mas)}&    group     \\
\hline
    $\eta$~Car                      & 10:45:03.546 & $-$59:41:03.95 & 287.60$-$00.63$\_$01 &                 --- & \num{5350358584482202880} &        1868 &  4.1133               &  1.3306               & \mci{---}       &   \bf{O-025} \\
    HD~93\,420                      & 10:45:50.661 & $-$59:29:19.33 & 287.59$-$00.41$\_$01 &                 --- & \num{5350343775447416704} &         --- &  5.9483               &  3.0309               & 0.658$\pm$0.071 &   foreground \\
    QZ~Car~Aa,Ac                    & 10:44:22.910 & $-$59:59:35.95 & 287.67$-$00.94$\_$01 &                 --- & \num{5350346970905044480} &  1839~Aa,Ac &  6.2727               &  0.3369               & 0.771$\pm$0.436 &   \bf{O-028} \\
    WR~24                           & 10:43:52.258 & $-$60:07:04.02 & 287.67$-$01.08$\_$01 &                 --- & \num{5254268071479968512} &        1817 &  6.3929               &  0.1271               & 0.408$\pm$0.062 &        O-028 \\
    HD~93\,281                      & 10:44:57.341 & $-$59:56:06.38 & 287.70$-$00.86$\_$03 &                 --- & \num{5350302097082780416} &         --- &  6.7871               &  2.3637               & 0.429$\pm$0.049 &        O-028 \\
    HDE~303\,310                    & 10:44:47.146 & $-$59:24:48.16 & 287.44$-$00.41$\_$01 &                 --- & \num{5350389095946525568} &         --- &  6.9176               &  3.3193               & 0.470$\pm$0.069 &        O-027 \\
    HD~93\,129~Aa                   & 10:43:57.462 & $-$59:32:51.27 & 287.41$-$00.57$\_$01 &                 --- & \num{5350363910256783488} &      1820~A &  7.2004               &  0.4867               & 0.407$\pm$0.044 &        O-002 \\
\end{tabular}
\addtolength{\tabcolsep}{1mm}
}
\end{table}

%% file: table_A12.tex
\begin{table}
\caption{(Continued).}
\centerline{
\tiny
\addtolength{\tabcolsep}{-1mm}
\begin{tabular}{lccccclrrrc}
\hline
\hline
Name                                &     RA       &      dec       &    GOS/GBS/GWS ID    &         GES ID      &          Gaia ID          & \mci{ALS ID}&      \mci{\GGc}       &      \mci{\BPRP}      & \mci{\pic (mas)}&    group     \\
\hline
    CPD~$-$59~2636~A,B              & 10:45:12.870 & $-$59:44:19.24 & 287.64$-$00.67$\_$01 & u10451288$-$5944192 & \num{5350357862947802880} & 15\,194~A,B &  9.5201               &  0.6791               & \mci{---}       &   \bf{O-025} \\
    HDE~303\,316~A                  & 10:43:11.178 & $-$59:44:21.02 & 287.41$-$00.79$\_$01 &                 --- & \num{5350355251602278528} &        1800 &  9.5280               &  0.5305               & 0.439$\pm$0.025 &        O-028 \\
    HDE~305\,518                    & 10:43:44.006 & $-$59:48:17.96 & 287.51$-$00.81$\_$01 &                 --- & \num{5350353567958779392} &        1810 &  9.5862               &  0.6167               & 0.503$\pm$0.033 &        O-028 \\
    HDE~305\,534                    & 10:44:47.517 & $-$59:57:58.88 & 287.70$-$00.89$\_$01 &                 --- & \num{5350301994003543936} &        1862 &  9.6264               &  0.2318               & 0.446$\pm$0.028 &        O-028 \\
    CPD~$-$59~2624                  & 10:45:05.828 & $-$59:43:07.57 & 287.62$-$00.66$\_$01 & g10450584$-$5943077 & \num{5350357966022158720} &     15\,197 &  9.6423               &  0.4992               & 0.445$\pm$0.149 &   \bf{O-025} \\
    HDE~305\,522                    & 10:44:14.961 & $-$60:00:05.71 & 287.66$-$00.96$\_$01 & g10441496$-$6000057 & \num{5350347142704520576} &        1836 &  9.6518               &  0.1651               & 0.414$\pm$0.030 &        O-028 \\
    HDE~305\,543                    & 10:43:10.072 & $-$60:02:11.78 & 287.55$-$01.05$\_$01 &                 --- & \num{5254273672117339520} &        1799 &  9.6785               &  0.0534               & 0.410$\pm$0.028 &        O-028 \\
\hline
\end{tabular}
\addtolength{\tabcolsep}{1mm}
}
\end{table}

%% file: table_A13.tex
\begin{table}
\caption{(Continued).}
\centerline{
\tiny
\addtolength{\tabcolsep}{-1mm}
\begin{tabular}{lccccclrrrc}
\hline
\hline
Name                                &     RA       &      dec       &    GOS/GBS/GWS ID    &         GES ID      &          Gaia ID          & \mci{ALS ID}&      \mci{\GGc}       &      \mci{\BPRP}      & \mci{\pic (mas)}&    group     \\
\hline
    CPD~$-$59~2632                  & 10:45:11.197 & $-$59:41:11.29 & 287.61$-$00.62$\_$01 & g10451120$-$5941113 & \num{5350358481418096896} &     15\,209 & 10.7331               &  0.2975               & 0.416$\pm$0.024 &        O-025 \\
    CPD~$-$59~2495                  & 10:42:36.164 & $-$59:59:26.20 & 287.47$-$01.04$\_$01 & g10423616$-$5959262 & \num{5254273500318621184} &     15\,956 & 10.7438               &  0.0837               & 0.426$\pm$0.026 &        O-028 \\
    ALS~15\,204                     & 10:43:41.237 & $-$59:35:48.18 & 287.40$-$00.63$\_$02 &                 --- & \num{5350363326141056768} &     15\,204 & 10.7551               &  1.0114               & 0.434$\pm$0.024 &        O-002 \\
    CPD~$-$58~2655                  & 10:44:42.119 & $-$59:22:30.53 & 287.41$-$00.38$\_$02 & g10444212$-$5922305 & \num{5350389508263432320} &         --- & 10.7725               &  0.3943               & 0.434$\pm$0.022 &        O-027 \\
    CPD~$-$59~2614                  & 10:44:57.331 & $-$60:00:46.79 & 287.74$-$00.93$\_$01 & g10445734$-$6000467 & \num{5350300207297141120} &        1866 & 10.7763               &  0.4319               & 0.471$\pm$0.085 &        O-028 \\
    ALS~15\,219                     & 10:43:59.863 & $-$59:35:24.40 & 287.43$-$00.61$\_$01 & g10435986$-$5935244 & \num{5350362638946370688} &     15\,219 & 10.7772               &  0.8030               & 0.426$\pm$0.026 &        O-002 \\
    CPD~$-$59~2581                  & 10:44:30.480 & $-$59:41:40.41 & 287.54$-$00.67$\_$01 & g10443049$-$5941406 & \num{5350358893734785664} &     15\,215 & 10.7796               &  0.4012               & 0.462$\pm$0.021 &      Car OB1 \\
\hline
\end{tabular}
\addtolength{\tabcolsep}{1mm}
}
\end{table}

%% file: table_A14.tex
\begin{table}
\caption{(Continued).}
\centerline{
\tiny
\addtolength{\tabcolsep}{-1mm}
\begin{tabular}{lccccclrrrc}
\hline
\hline
Name                                &     RA       &      dec       &    GOS/GBS/GWS ID    &         GES ID      &          Gaia ID          & \mci{ALS ID}&      \mci{\GGc}       &      \mci{\BPRP}      & \mci{\pic (mas)}&    group     \\
\hline
    CPD~$-$59~2625                  & 10:45:05.877 & $-$59:44:18.88 & 287.63$-$00.68$\_$01 & g10450588$-$5944189 & \num{5350357755541882880} &     15\,238 & 11.4258               &  0.3799               & 0.407$\pm$0.033 &        O-025 \\
    ALS~15\,245                     & 10:45:14.985 & $-$59:43:23.28 & 287.64$-$00.65$\_$02 & g10451499$-$5943233 & \num{5350358275259591296} &     15\,245 & 11.4535               &  0.6717               & 0.432$\pm$0.028 &        O-025 \\
    CPD~$-$58~2677                  & 10:45:22.137 & $-$59:37:38.53 & 287.60$-$00.56$\_$01 & g10452214$-$5937385 & \num{5350382086559684480} &     15\,244 & 11.4716               &  0.5113               & 0.453$\pm$0.035 &        O-025 \\
    CPD~$-$59~2541                  & 10:43:48.824 & $-$60:00:36.71 & 287.61$-$00.99$\_$01 & g10434882$-$6000366 & \num{5254271056429426688} &     15\,962 & 11.4791               &  0.4154               & 0.317$\pm$0.079 &        O-028 \\
    ALS~15\,246                     & 10:45:09.744 & $-$59:42:57.21 & 287.62$-$00.65$\_$01 & g10450974$-$5942572 & \num{5350357897302465536} &     15\,246 & 11.4902               &  0.5053               & 0.409$\pm$0.033 &        O-025 \\
    CPD~$-$59~2616                  & 10:45:00.234 & $-$59:43:34.53 & 287.61$-$00.67$\_$01 & g10450023$-$5943345 & \num{5350357931646616448} &     19\,745 & 11.4969               &  0.5213               & 0.454$\pm$0.030 &        O-025 \\
    ALS~15\,242                     & 10:44:37.185 & $-$59:40:01.49 & 287.54$-$00.64$\_$01 & g10443719$-$5940015 & \num{5350359065533555712} &     15\,242 & 11.5250               &  0.6599               & 0.439$\pm$0.028 &        O-025 \\
\hline
\end{tabular}
\addtolength{\tabcolsep}{1mm}
}
\end{table}

%% file: table_A15.tex
\begin{table}
\caption{(Continued).}
\centerline{
\tiny
\addtolength{\tabcolsep}{-1mm}
\begin{tabular}{lccccclrrrc}
\hline
\hline
Name                                &     RA       &      dec       &    GOS/GBS/GWS ID    &         GES ID      &          Gaia ID          & \mci{ALS ID}&      \mci{\GGc}       &      \mci{\BPRP}      & \mci{\pic (mas)}&    group     \\
\hline
    2MASS~J10435478$-$6006207       & 10:43:54.787 & $-$60:06:20.77 & 287.67$-$01.07$\_$01 & g10435479$-$6006208 & \num{5254269475881471104} &         --- & 12.0899               &  0.1552               & 0.453$\pm$0.033 &        O-028 \\
    ALS~19\,748                     & 10:45:19.427 & $-$59:39:37.44 & 287.62$-$00.59$\_$01 & g10451943$-$5939374 & \num{5350381914760925056} &     19\,748 & 12.0908               &  0.4645               & 0.423$\pm$0.028 &        O-025 \\
    ALS~19\,732                     & 10:43:48.812 & $-$59:33:35.17 & 287.40$-$00.59$\_$02 &                 --- & \num{5350363841537241088} &     19\,732 & 12.0957               &  0.5980               & 0.435$\pm$0.023 &        O-002 \\
    ALS~19\,747                     & 10:45:09.651 & $-$59:40:08.46 & 287.60$-$00.61$\_$01 & g10450968$-$5940088 & \num{5350358515777872512} &     19\,747 & 12.1571               &  0.4232               & 0.484$\pm$0.033 &        O-025 \\
    2MASS~J10435198$-$6010368       & 10:43:51.984 & $-$60:10:36.81 & 287.70$-$01.13$\_$01 & g10435198$-$6010368 & \num{5254267448656913152} &         --- & 12.1638               &  0.1415               & 0.391$\pm$0.030 &        O-028 \\
    2MASS~J10442909$-$5948207       & 10:44:29.098 & $-$59:48:20.75 & 287.59$-$00.77$\_$01 & g10442910$-$5948207 & \num{5350356656026369280} &         --- & 12.1802               &  1.4850               & 0.582$\pm$0.134 &   \bf{O-028} \\
    ALS~17\,185                     & 10:43:43.556 & $-$59:34:03.47 & 287.39$-$00.61$\_$01 & g10434356$-$5934035 & \num{5350363459259473664} &     17\,185 & 12.1839               &  0.8092               & 0.424$\pm$0.023 &        O-002 \\
\hline
\end{tabular}
\addtolength{\tabcolsep}{1mm}
}
\end{table}

%% file: table_A16.tex
\begin{table}
\caption{(Continued).}
\centerline{
\tiny
\addtolength{\tabcolsep}{-1mm}
\begin{tabular}{lccccclrrrc}
\hline
\hline
Name                                &     RA       &      dec       &    GOS/GBS/GWS ID    &         GES ID      &          Gaia ID          & \mci{ALS ID}&      \mci{\GGc}       &      \mci{\BPRP}      & \mci{\pic (mas)}&    group     \\
\hline
    2MASS~J10425293$-$6003478       & 10:42:52.934 & $-$60:03:47.83 & 287.53$-$01.09$\_$01 & g10425293$-$6003478 & \num{5254272297727705728} &         --- & 12.8649               &  0.8109               & 0.239$\pm$0.025 &   background \\
    2MASS~J10440866$-$5933488       & 10:44:08.669 & $-$59:33:48.81 & 287.44$-$00.58$\_$01 & g10440867$-$5933488 & \num{5350363085623074432} &         --- & 12.8705               &  0.6870               & 0.398$\pm$0.028 &        O-002 \\
    2MASS~J10444609$-$5946056       & 10:44:46.092 & $-$59:46:05.69 & 287.60$-$00.72$\_$01 & g10444609$-$5946057 & \num{5350356385474179840} &         --- & 12.9082               &  0.5019               & 0.416$\pm$0.035 &      Car OB1 \\
    2MASS~J10454060$-$5937041       & 10:45:40.609 & $-$59:37:04.20 & 287.63$-$00.53$\_$01 & g10454061$-$5937042 & \num{5350335597829590400} &         --- & 12.9289               &  0.7568               & 0.455$\pm$0.028 &      Car OB1 \\
    2MASS~J10470063$-$5957242       & 10:47:00.639 & $-$59:57:24.23 & 287.94$-$00.76$\_$01 & g10470064$-$5957242 & \num{5350293232270641152} &         --- & 12.9382               &  0.6593               & 0.396$\pm$0.026 &        O-030 \\
    2MASS~J10440384$-$5934344       & 10:44:03.843 & $-$59:34:34.44 & 287.43$-$00.59$\_$01 & g10440384$-$5934344 & \num{5350363051263281920} &         --- & 12.9416               &  0.5507               & 0.442$\pm$0.029 &        O-002 \\
    2MASS~J10420949$-$6002265       & 10:42:09.498 & $-$60:02:26.60 & 287.44$-$01.12$\_$01 & g10420950$-$6002267 & \num{5254284598515540608} &         --- & 12.9549               &  1.1432               & 0.237$\pm$0.028 &   background \\
\hline
\end{tabular}
\addtolength{\tabcolsep}{1mm}
}
\end{table}

%% file: table_A21.tex
\begin{table}
\caption{Spectral classifications for the stars in the field of the Carina Nebula analyzed in this paper.}
\label{table_spclas}
\centerline{
\tiny
\addtolength{\tabcolsep}{-1mm}
\begin{tabular}{lllllllllllllll}
\hline
\hline
Name                                &\multicolumn{4}{c}{\rule[2pt]{0.5cm}{0.4pt} Gaia-ESO \rule[2pt]{0.5cm}{0.4pt}}&\multicolumn{5}{c}{\rule[2pt]{1.0cm}{0.4pt} GOSSS \rule[2pt]{1.0cm}{0.4pt}}&\multicolumn{5}{c}{\lili\ + STIS + literature}\\
                                    &\mci{ST}& \mci{LC}&\mci{qual.}&   \mci{sec.}   &\mci{ST}& \mci{LC}&\mci{qual.}&   \mci{sec.}   &\mci{ref.}&\mci{ST}& \mci{LC}&\mci{qual.}&   \mci{sec.}   &\mci{ref.}\\
\hline
    $\eta$~Car                      & ---    & ---     & ---       & ---            & LBV    & ---     & ---       & ---            & M22      & F:     & I       & ---       & ---            & W77      \\
    HD~93\,420                      & ---    & ---     & ---       & ---            & ---    & ---     & ---       & ---            & ---      & M4     & Ib      & ---       & ---            & M23b     \\
    QZ~Car~Aa,Ac                    & ---    & ---     & ---       & ---            & O9.7   & Ib      & n         & ---            & S14      & O9.7   & Ib      & ---       & O9 II:         & M23a     \\
    WR~24                           & ---    & ---     & ---       & ---            & WN6    & ---     & ha        & ---            & TW       & WN6    & ---     & ha-w...   & ---            & H06      \\
    HD~93\,281                      & ---    & ---     & ---       & ---            & ---    & ---     & ---       & ---            & ---      & M1.5   & Iab     & ---       & ---            & M23b     \\
    HDE~303\,310                    & ---    & ---     & ---       & ---            & ---    & ---     & ---       & ---            & ---      & M3     & Iab     & ---       & ---            & M23b     \\
\hline
\end{tabular}
\addtolength{\tabcolsep}{1mm}
}
\end{table}

%% file: table_A22.tex
\begin{table}
\caption{(Continued).}
\centerline{
\tiny
\addtolength{\tabcolsep}{-1mm}
\begin{tabular}{lllllllllllllll}
\hline
\hline
Name                                &\multicolumn{4}{c}{\rule[2pt]{0.5cm}{0.4pt} Gaia-ESO \rule[2pt]{0.5cm}{0.4pt}}&\multicolumn{5}{c}{\rule[2pt]{1.0cm}{0.4pt} GOSSS \rule[2pt]{1.0cm}{0.4pt}}&\multicolumn{5}{c}{\lili\ + STIS + literature}\\
                                    &\mci{ST}& \mci{LC}&\mci{qual.}&   \mci{sec.}   &\mci{ST}& \mci{LC}&\mci{qual.}&   \mci{sec.}   &\mci{ref.}&\mci{ST}& \mci{LC}&\mci{qual.}&   \mci{sec.}   &\mci{ref.}\\
\hline
    CPD~$-$59~2636~A,B              & O7.5   & V       & ---       & O8 V + O8 V    & O8     & V       & ---       & O8 V           & S14      & O7     & V       & ---       & O8 V + O9 V    & A02      \\
    HDE~303\,316~A                  & ---    & ---     & ---       & ---            & O7     & V       & ((f))z    & ---            & S14      & O6     & V       & ---       & ---            & F81      \\
    HDE~305\,518                    & ---    & ---     & ---       & ---            & O9.7   & III     & ---       & ---            & S14      & O9.5   & V       & ---       & ---            & L82      \\
    HDE~305\,534                    & ---    & ---     & ---       & ---            & ---    & ---     & ---       & ---            & ---      & B0     & V       & ---       & B1: V          & M23b     \\
    CPD~$-$59~2624                  & O9.7   & V       & ---       & ---            & O9.7   & IV      & ---       & ---            & S14      & O9.5   & V       & ---       & ---            & A16      \\
    HDE~305\,522                    & B0.2   & V       & ---       & B1: V          & ---    & ---     & ---       & ---            & ---      & B0     & V       & ---       & B1 V           & A16      \\
\hline
\end{tabular}
\addtolength{\tabcolsep}{1mm}
}
\end{table}

%% file: table_A23.tex
\begin{table}
\caption{(Continued).}
\centerline{
\tiny
\addtolength{\tabcolsep}{-1mm}
\begin{tabular}{lllllllllllllll}
\hline
\hline
Name                                &\multicolumn{4}{c}{\rule[2pt]{0.5cm}{0.4pt} Gaia-ESO \rule[2pt]{0.5cm}{0.4pt}}&\multicolumn{5}{c}{\rule[2pt]{1.0cm}{0.4pt} GOSSS \rule[2pt]{1.0cm}{0.4pt}}&\multicolumn{5}{c}{\lili\ + STIS + literature}\\
                                    &\mci{ST}& \mci{LC}&\mci{qual.}&   \mci{sec.}   &\mci{ST}& \mci{LC}&\mci{qual.}&   \mci{sec.}   &\mci{ref.}&\mci{ST}& \mci{LC}&\mci{qual.}&   \mci{sec.}   &\mci{ref.}\\
\hline
    CPD~$-$59~2632                  & B0.5   & V       & n         & ---            & ---    & ---     & ---       & ---            & ---      & B1     & V       & ---       & ---            & A16      \\
    CPD~$-$59~2495                  & B1.5   & V       & ---       & ---            & ---    & ---     & ---       & ---            & ---      & B0.5   & V       & ---       & ---            & A16      \\
    ALS~15\,204                     & ---    & ---     & ---       & ---            & O7.5   & V       & z         & O9: V          & M16      & ---    & ---     & ---       & ---            & ---      \\
    CPD~$-$58~2655                  & B1:    & V       & ---       & ---            & ---    & ---     & ---       & ---            & ---      & B1     & V       & ---       & ---            & F80      \\
    CPD~$-$59~2614                  & B1     & V       & ---       & ---            & ---    & ---     & ---       & ---            & ---      & B1     & III     & ---       & ---            & A16      \\
    ALS~15\,219                     & B0.5:  & V       & ---       & B1: V          & B0.7   & V       & ---       & B1 V           & M22      & B1     & V       & ---       & ---            & M93      \\
\hline
\end{tabular}
\addtolength{\tabcolsep}{1mm}
}
\end{table}

%% file: table_A24.tex
\begin{table}
\caption{(Continued).}
\centerline{
\tiny
\addtolength{\tabcolsep}{-1mm}
\begin{tabular}{lllllllllllllll}
\hline
\hline
Name                                &\multicolumn{4}{c}{\rule[2pt]{0.5cm}{0.4pt} Gaia-ESO \rule[2pt]{0.5cm}{0.4pt}}&\multicolumn{5}{c}{\rule[2pt]{1.0cm}{0.4pt} GOSSS \rule[2pt]{1.0cm}{0.4pt}}&\multicolumn{5}{c}{\lili\ + STIS + literature}\\
                                    &\mci{ST}& \mci{LC}&\mci{qual.}&   \mci{sec.}   &\mci{ST}& \mci{LC}&\mci{qual.}&   \mci{sec.}   &\mci{ref.}&\mci{ST}& \mci{LC}&\mci{qual.}&   \mci{sec.}   &\mci{ref.}\\
\hline
    CPD~$-$59~2625                  & B1.5:  & V       & ---       & B2: V          & ---    & ---     & ---       & ---            & ---      & B2     & V       & ---       & ---            & A16      \\
    ALS~15\,245                     & B0.5   & V       & ---       & ---            & ---    & ---     & ---       & ---            & ---      & ---    & ---     & ---       & ---            & ---      \\
    CPD~$-$58~2677                  & B1     & V       & ---       & ---            & ---    & ---     & ---       & ---            & ---      & B2     & III     & ---       & ---            & A16      \\
    CPD~$-$59~2541                  & B1.5   & V       & (n)       & ---            & ---    & ---     & ---       & ---            & ---      & ---    & ---     & ---       & ---            & ---      \\
    ALS~15\,246                     & B0.7   & V       & ---       & ---            & ---    & ---     & ---       & ---            & ---      & B1     & V       & ---       & ---            & A16      \\
    CPD~$-$59~2616                  & B2     & V       & (n)       & ---            & B2     & V       & (n)       & ---            & M22      & B2     & V       & ---       & ---            & A16      \\
\hline
\end{tabular}
\addtolength{\tabcolsep}{1mm}
}
\end{table}

%% file: table_A25.tex
\begin{table}
\caption{(Continued).}
\centerline{
\tiny
\addtolength{\tabcolsep}{-1mm}
\begin{tabular}{lllllllllllllll}
\hline
\hline
Name                                &\multicolumn{4}{c}{\rule[2pt]{0.5cm}{0.4pt} Gaia-ESO \rule[2pt]{0.5cm}{0.4pt}}&\multicolumn{5}{c}{\rule[2pt]{1.0cm}{0.4pt} GOSSS \rule[2pt]{1.0cm}{0.4pt}}&\multicolumn{5}{c}{\lili\ + STIS + literature}\\
                                    &\mci{ST}& \mci{LC}&\mci{qual.}&   \mci{sec.}   &\mci{ST}& \mci{LC}&\mci{qual.}&   \mci{sec.}   &\mci{ref.}&\mci{ST}& \mci{LC}&\mci{qual.}&   \mci{sec.}   &\mci{ref.}\\
\hline
    2MASS~J10435478$-$6006207       & B2     & V       & ---       & ---            & ---    & ---     & ---       & ---            & ---      & ---    & ---     & ---       & ---            & ---      \\
    ALS~19\,748                     & B1.5   & V       & ---       & ---            & ---    & ---     & ---       & ---            & ---      & B1.5   & V       & ---       & ---            & M93      \\
    ALS~19\,732                     & ---    & ---     & ---       & ---            & B2:    & V       & e         & ---            & TW       & B2     & V       & ---       & ---            & A16      \\
    ALS~19\,747                     & B2     & V       & n         & ---            & ---    & ---     & ---       & ---            & ---      & B2     & V       & ---       & ---            & M93      \\
    2MASS~J10435198$-$6010368       & B2.5   & V       & n         & ---            & ---    & ---     & ---       & ---            & ---      & ---    & ---     & ---       & ---            & ---      \\
    2MASS~J10442909$-$5948207       & B0     & V       & ---       & ---            & B0     & V       & ---       & ---            & TW       & ---    & ---     & ---       & ---            & ---      \\
\hline
\end{tabular}
\addtolength{\tabcolsep}{1mm}
}
\end{table}

%% file: table_A26.tex
\begin{table}
\caption{(Continued).}
\centerline{
\tiny
\addtolength{\tabcolsep}{-1mm}
\begin{tabular}{lllllllllllllll}
\hline
\hline
Name                                &\multicolumn{4}{c}{\rule[2pt]{0.5cm}{0.4pt} Gaia-ESO \rule[2pt]{0.5cm}{0.4pt}}&\multicolumn{5}{c}{\rule[2pt]{1.0cm}{0.4pt} GOSSS \rule[2pt]{1.0cm}{0.4pt}}&\multicolumn{5}{c}{\lili\ + STIS + literature}\\
                                    &\mci{ST}& \mci{LC}&\mci{qual.}&   \mci{sec.}   &\mci{ST}& \mci{LC}&\mci{qual.}&   \mci{sec.}   &\mci{ref.}&\mci{ST}& \mci{LC}&\mci{qual.}&   \mci{sec.}   &\mci{ref.}\\
\hline
    2MASS~J10425293$-$6003478       & B0.7   & V       & ---       & ---            & ---    & ---     & ---       & ---            & ---      & ---    & ---     & ---       & ---            & ---      \\
    2MASS~J10440866$-$5933488       & B2:    & V       & nnn       & ---            & ---    & ---     & ---       & ---            & ---      & ---    & ---     & ---       & ---            & ---      \\
    2MASS~J10444609$-$5946056       & B2     & V       & ---       & ---            & ---    & ---     & ---       & ---            & ---      & ---    & ---     & ---       & ---            & ---      \\
    2MASS~J10454060$-$5937041       & B2:    & V       & nn        & ---            & ---    & ---     & ---       & ---            & ---      & B      & ---     & ---       & ---            & D17      \\
    2MASS~J10470063$-$5957242       & B2     & V       & ---       & ---            & ---    & ---     & ---       & ---            & ---      & ---    & ---     & ---       & ---            & ---      \\
    2MASS~J10440384$-$5934344       & B2     & V       & ---       & ---            & ---    & ---     & ---       & ---            & ---      & B      & I       & ---       & ---            & D17      \\
\end{tabular}
\addtolength{\tabcolsep}{1mm}
}
\end{table}

%% file: non_members.tex
\begin{table*}
\caption{Stars of the census whose distance is not compatible with Car~OB1 and are located in the foreground and in the background. See Table~\ref{table_spclas} for reference acronyms.}
\label{non-members}
\centerline{
\tiny
\addtolength{\tabcolsep}{-1mm}
\begin{tabular}{llllllllllllllll}
\hline
\hline 
Name &\multicolumn{4}{c}{\rule[2pt]{0.5cm}{0.4pt} Gaia-ESO \rule[2pt]{0.5cm}{0.4pt}}&\multicolumn{5}{c}{\rule[2pt]{1.0cm}{0.4pt} GOSSS \rule[2pt]{1.0cm}{0.4pt}}&\multicolumn{5}{c}{\lili\ + STIS + literature} & group\\
&\mci{ST}& \mci{LC}&\mci{qual.}&   \mci{sec.}   &\mci{ST}& \mci{LC}&\mci{qual.}&   \mci{sec.}   &\mci{ref.}&\mci{ST}& \mci{LC}&\mci{qual.}&   \mci{sec.}   &\mci{ref.}\\ 
\hline 
    HD~93\,420                      & ---    & ---     & ---       & ---            & ---    & ---     & ---       & ---            & ---      & M4     & Ib      & ---       & ---            & H72   & foreground   \\
    HD~93\,342                      & ---    & ---     & ---       & ---            & ---    & ---     & ---       & ---            & ---      & B1     & Ia      & ---       & ---            & A16   & background   \\
    HD~93\,501                      & ---    & ---     & ---       & ---            & ---    & ---     & ---       & ---            & ---      & B0     & V       & ---       & ---            & A16   & foreground   \\
    CPD~$-$59~2592                  & ---    & ---     & ---       & ---            & ---    & ---     & ---       & ---            & ---      & B1     & Ib      & ---       & ---            & A16   & background   \\
    HDE~305\,439~A                  & B0     & Ia     & ---       & ---            & ---    & ---     & ---       & ---            & ---      & B0   & Ia     & ---       & ---            & V93   & background   \\
    HDE~305\,439~B                  & ---     & ---     & ---       & ---            & ---    & ---     & ---       & ---            & ---      & B0.7   & Ib     & ---       & ---            & M23b   & background   \\
\hline
\end{tabular}   
\addtolength{\tabcolsep}{1mm}
}
\end{table*}

%% file: six_members.tex
\begin{table*}
\caption{Stars identified in the Car~OB1 region as of WR or non-O supergiant (II to Ia) spectral class. See Table~\ref{table_spclas} for reference acronyms.}
\label{six-members}
\centerline{
\tiny
\addtolength{\tabcolsep}{-1mm}
\begin{tabular}{llllllllllllllll}
\hline
\hline 
Name &\multicolumn{4}{c}{\rule[2pt]{0.5cm}{0.4pt} Gaia-ESO \rule[2pt]{0.5cm}{0.4pt}}&\multicolumn{5}{c}{\rule[2pt]{1.0cm}{0.4pt} GOSSS \rule[2pt]{1.0cm}{0.4pt}}&\multicolumn{5}{c}{\lili\ + STIS + literature} & group\\
&\mci{ST}& \mci{LC}&\mci{qual.}&   \mci{sec.}   &\mci{ST}& \mci{LC}&\mci{qual.}&   \mci{sec.}   &\mci{ref.}&\mci{ST}& \mci{LC}&\mci{qual.}&   \mci{sec.}   &\mci{ref.}\\ 
\hline 

    $\eta$~Car                      & ---    & ---     & ---       & ---            & LBV    & ---     & ---       & ---            & M22a     & F:     & I       & ---       & ---            & W77    & \bf{O-025}  \\
    WR~24                           & ---    & ---     & ---       & ---            & WN6    & ---     & ha        & ---            & TW       & WN6    & ---     & ha-w...   & ---            & H06    & O-028  \\
    HD~93\,281                      & ---    & ---     & ---       & ---            & ---    & ---     & ---       & ---            & ---      & M1.5     & Iab:    & ---       & ---              & M23b    & O-028  \\
    HDE~303\,310                    & ---    & ---     & ---       & ---            & ---    & ---     & ---       & ---            & ---      & M3     & Iab      & ---       & ---            & M23b    & O-027  \\
    CPD~$-$58~2605                  & B0.5   & II      & ---       & ---            & ---    & ---     & ---       & ---            & ---      & B0     & III-IV: & ---       & ---            & M88    & O-002  \\
    CPD~$-$59~2504                  & B7     & II:     & nn        & ---            & ---    & ---     & ---       & ---            & ---      & B2/5   & ---     & ---       & ---            & L76    & O-028  \\

    \hline
\end{tabular}   
\addtolength{\tabcolsep}{1mm}
}
\end{table*}

%% file: obinaries.tex
\begin{table*}
\caption{Spectroscopic binary systems identified in the census containing at least one O-type star. See Table~\ref{table_spclas} for reference acronyms.}
\label{obinaries}
\centerline{
\tiny
\addtolength{\tabcolsep}{-1mm}
\begin{tabular}{lllllllllllllllll}
\hline
\hline 
Name &\multicolumn{4}{c}{\rule[2pt]{0.5cm}{0.4pt} Gaia-ESO \rule[2pt]{0.5cm}{0.4pt}}&\multicolumn{5}{c}{\rule[2pt]{1.0cm}{0.4pt} GOSSS \rule[2pt]{1.0cm}{0.4pt}}&\multicolumn{5}{c}{\lili\ + STIS + literature} & group & Notes\\
&\mci{ST}& \mci{LC}&\mci{qual.}&   \mci{sec.}   &\mci{ST}& \mci{LC}&\mci{qual.}&   \mci{sec.}   &\mci{ref.}&\mci{ST}& \mci{LC}&\mci{qual.}&   \mci{sec.}   &\mci{ref.}\\ 
\hline 

   QZ~Car~Aa,Ac                    & ---    & ---     & ---       & ---            & O9.7   & Ib      & n         & ---            & S14      & O9.7   & Ib      & ---       & O9 II:         & M22b   & \bf{O-028} & new SB2\\
    HD~93\,129~Aa                   & ---    & ---     & ---       & ---            & O2     & I       & f*        & ---            & S14      & O2     & I       & f*        & OB?            & M17    & O-002 \\
    HD~93\,403                      & ---    & ---     & ---       & ---            & O5     & I       & fc        & O7.5 V         & M22b     & O5     & I       & fc        & O7.5 V         & M22b   &  \bf{Car~OB1} & new SB2\\
    HD~93\,205                      & ---    & ---     & ---       & ---            & O3.5   & V       & ((f))     & O8 V           & S14      & O3.5   & V       & ((f))     & O8 V           & M20    &  O-003\\
    HD~93\,162                      & ---    & ---     & ---       & ---            & O2.5   & I       & f*/WN6    & OB             & S14      & WN6    & ---     & h         & O4             & V01    &  O-003\\
    HD~93\,161~A                    & O7     & V       & ((f))     & O9 IV          & O7.5   & V       & ---       & O9 V           & S14      & O7     & V       & ((f))     & O9 IV          & M22b   &  O-002\\
\hline
\end{tabular}   
\addtolength{\tabcolsep}{1mm}
}
\end{table*}

%% file: bbinaries.tex
\begin{table*}
\caption{Spectroscopic binary systems identified in the census formed by early B-type stars. See Table~\ref{table_spclas} for reference acronyms. }
\label{bbinaries}
\centerline{
\tiny
\addtolength{\tabcolsep}{-1mm}
\begin{tabular}{lllllllllllllllll}
\hline
\hline 
Name &\multicolumn{4}{c}{\rule[2pt]{0.5cm}{0.4pt} Gaia-ESO \rule[2pt]{0.5cm}{0.4pt}}&\multicolumn{5}{c}{\rule[2pt]{1.0cm}{0.4pt} GOSSS \rule[2pt]{1.0cm}{0.4pt}}&\multicolumn{5}{c}{\lili\ + STIS + literature} & group & Notes\\
&\mci{ST}& \mci{LC}&\mci{qual.}&   \mci{sec.}   &\mci{ST}& \mci{LC}&\mci{qual.}&   \mci{sec.}   &\mci{ref.}&\mci{ST}& \mci{LC}&\mci{qual.}&   \mci{sec.}   &\mci{ref.}\\ 
\hline 
    HDE~305\,534                    & ---    & ---     & ---       & ---            & ---    & ---     & ---       & ---            & ---      & B0     & V       & ---       & B0 V           & A16   & O-028   \\
    HDE~305\,522                    & B0.2   & V       & ---       & B1: V          & ---    & ---     & ---       & ---            & ---      & B0     & V       & ---       & B1 V           & A16   & O-028   \\
    HDE~305\,543                    & ---   & ---      & ---       & ---           & ---    & ---     & ---       & ---            & ---      & B0.2     & V       & (n)       & B1: V(n)          & M23b & O-028 & new SB2  \\    
    HDE~303\,300                    & B0.5   & V       & ---       & B              & ---    & ---     & ---       & ---            & ---      & B1     & V       & ---       & ---            & A16   & Car~OB1 & new SB2  \\
    HDE~305\,516                    & B0.5   & V       & (n)       & B2: V          & ---    & ---     & ---       & ---            & ---      & B0.5   & V       & ---       & ---            & G11b  & O-028  & new SB2\\
    HDE~303\,313                    & B2     & V       & ---       & B2 V           & ---    & ---     & ---       & ---            & ---      & B2     & V       & ---       & B2 V           & A16   & Car~OB1  \\
\hline
\end{tabular}   
\addtolength{\tabcolsep}{1mm}
}
\end{table*}

%% file: runaways.tex
\begin{table*}
\caption{ Proper motions from Gaia EDR3 for the identified runaway candidates.}
\label{runaways}
\centerline{
\tiny
\addtolength{\tabcolsep}{-1mm}
\begin{tabular}{lcccccc}
\hline
\hline 
Name                      & Gaia Source  & $\mu_{\alpha *}$        &  $\mu_{\delta}$          &\small{RUWE}     & group & n-qualifer\\
                          &              & [mas a$^{-1}$] & [mas a$^{-1}$]    &                 &       &           \\ 
\hline 

2MASS~J10440866$-$5933488 & 5350363085623074432 &   -7.181 $\pm$ 0.015   &   3.419 $\pm$ 0.014   &   0.972    & O-002 &    nnn       \\
CPD~$-$58~2657            & 5350395383780746496 &   -6.622  $\pm$  0.329  &  0.385 $\pm$ 0.300   &  12.985    & O-027 &  (n)            \\
HDE~303\,310              & 5350389095946525568 &   -7.432  $\pm$  0.031  &  2.851 $\pm$ 0.026   &  0.704     & O-027 &             \\
CPD~$-$59~2541            & 5254271056429426688 &  -5.647  $\pm$  0.032   &  2.545 $\pm$ 0.031   &  1.582     & O-028 &   (n)          \\
HD~93\,281               & 5350302097082780416 &  -8.028  $\pm$  0.022   &  2.029 $\pm$ 0.020   &  1.008     & O-028 &                 \\
QZ~Car~Aa,Ac              & 5350346970905044480 &  -5.750  $\pm$  0.123   &  2.455 $\pm$ 0.108   &  3.471     & O-028 &                 \\
HDE~305\,523              & 5350347520660979840 &  -5.804  $\pm$  0.025   &  2.165 $\pm$ 0.023   &  1.083     & O-028 &                 \\
2MASS~J10451588$-$5929563 & 5350384629180484224 &  -6.893  $\pm$  0.011   &  4.030 $\pm$ 0.010   &  0.797     & O-029 &   n              \\

\hline

\end{tabular}   
\addtolength{\tabcolsep}{1mm}
}
\end{table*}